\documentclass[a4paper,12pt]{article}
\usepackage[utf8x]{inputenc}
\usepackage[T1]{fontenc}        		
\usepackage{color}              		
\usepackage{amsmath,amsfonts,amssymb}   	
\usepackage{bm}
\usepackage{ae}                 %
\usepackage{longtable}          		
\usepackage{multirow}  				
\usepackage{multicol}      			
\usepackage{url}
\usepackage{paralist}				
\usepackage{graphicx}           		
\usepackage{epsfig}
\usepackage{hyperref}
\usepackage{epstopdf}
\usepackage{listings} 				
\usepackage{booktabs} 				
\usepackage{tabularx}
\usepackage{bm}
\usepackage{natbib}
\usepackage{morefloats}
\usepackage{rotating}
\usepackage{mathrsfs} 
\usepackage[capposition=top]{floatrow}
\usepackage{caption}
\usepackage{subcaption}
\usepackage{pdflscape}
\usepackage{array}
\usepackage{mdframed}
\usepackage{tikz}
\usetikzlibrary{decorations.markings}
\usetikzlibrary{angles,quotes}

\begin{document}
\hypersetup{pageanchor=false}

\pagenumbering{roman}   

\begin{titlepage}       

\thispagestyle{empty}   

\begin{center}
\vspace*{2.5cm}
{\large \textbf{Closed-form approximations in derivatives pricing: \\
The Kristensen-Mele approach}}
\vspace*{3cm}

{Degree Dissertation \\ for the \\ Master Examination in Economics \& Finance \\
at the \\ Faculty of Economics and Social Sciences \\ of the \\
Eberhard Karls Universit\"at \\ T\"ubingen}

\end{center}

\vfill

\hfill
\begin{tabular}{@{}l@{}}
Submitted by: \\
Michael Kurz \\
\\~\\
Date of submission: 14/09/2015
\end{tabular}

\end{titlepage}

\newpage                
\textbf{ABSTRACT}\\
~~~\\
Kristensen and Mele (2011) developed a new approach to obtain closed-form approximations to continuous-time derivatives pricing models. The approach uses a power series expansion of the pricing bias between an intractable model and some known auxiliary model. Since the resulting approximation formula has closed-form it is straightforward to obtain approximations of greeks. In this thesis I will introduce Kristensen and Mele's methods and apply it to a variety of stochastic volatility models of European style options as well as a model for commodity futures. The focus of this thesis is the effect of different model choices and different model parameter values on the numerical stability of Kristensen and Mele's approximation.\\
~~~\\
~~~\\

\newpage

~\\~\\
\\~\\
\\~\\
\begin{center}
	\textit{'As far as the laws of mathematics refer to reality, they are not certain; and as far as they are certain, they do not refer to reality.'}
\end{center}
\begin{flushright}
	- Albert Einstein
\end{flushright}

\newpage
\hypersetup{pageanchor=true}


\tableofcontents   
\newpage
\newpage
\listoffigures

\newpage
\listoftables

\newpage


\pagenumbering{arabic}      
\setcounter{page}{1}        


\section{Introduction}

In the field of financial engineering, the continuous-time framework had become the golden standard in the modeling of derivative securities. Continuous-time models for the prices of these securities are build on the notion of stochastic differential equations (SDEs) to describe the dynamics of the market of the assets or the risks on which the derivative is written on. The price of a specific derivative security then usually arises from the solution of a partial differential equation (PDE) which can be constructed from the markets SDEs. One of the most celebrated of such solutions is the \citet{BS} model, which resulted from the pioneering works of Robert C. Merton, Fischer Black and Myron Scholes, and yields the price of an European Call option written on a financial asset. While the solution to the Back-Scholes PDE can be derived in closed-form, this is not possible for the vast majority of the more realistic and thus more complex models nowadays used by practitioners and in academia. The continuous-time framework, in fact, owes its success at least partially to this very issue, as it opens convenient ways to cope with security prices that are not available in closed-form.\\
One widely used approach are Fourier transform methods, which allow for so called \textit{semi} closed-form solutions. Where the word \textit{semi} is attached since the solution is only given in closed-form for the models characteristic functions, which are complex valued functions of the underlying and time. The price of the derivative security is then computed by numerically integrating over the characteristic functions, what might be a mathematically non-trivial task and may introduce sources of numerical instability which are not easy to spot.\footnote{See the example on complex valued logarithms in Appendix \ref{ComplexLog} of this thesis.} Two classical approaches to obtain approximations in case (semi-)closed-form solutions are unavailable, are Finite Difference schemes and Monte Carlo simulations. While both approaches are widely applicable they might be computationally burdensome, as Finite Differences requires the repeated solution of linear systems of equations, and Monte Carlo requires the simulation of huge numbers of sample paths of the underlying based on discretized SDEs over a fine grid of time increments.\\
Even though these approaches can be described as standards to cope with PDEs for which closed-form solutions are not available, there appeared also one branch of academic literature attempting to develop approaches to derive \textit{closed-form approximations} to PDEs. These closed-form approximations are usually built on Taylor series expansions of conditional expectations. The approach suggested by \citet{KM2011} (henceforth KM) fits into this branch of literature. However, KM do not focus on approximating the asset price directly, but target on developing an approximation of the expected bias which arises if an oversimplifying baseline model (or "auxiliary" in KM's terms) would be used to value some derivative, e.g. when the Black-Scholes model would be used while the true market dynamics are assumed to obey the Heston model. The derivatives price is then computed as the sum of the price resulting from the baseline model and the expected bias. Where the bias is approximated by a power series expansion around the baseline model, what is among the main features distinguishing KM's approach from other closed-form approximations available in the literature. In a nutshell, KM's approximation consists of two steps: (i) Firstly the Feynman-Kac representation of the solution to a PDE which the bias obeys is derived. This Feynman-Kac representation consists of two conditional expectations, one describing the difference in the payoffs and one describing the difference in the market dynamics between the true and the baseline model. (ii) The second step is to approximate these expectations through a series expansion around the baseline model. One important restriction regarding the choice of a suitable model arises with the differentiability of the payoff function. KM's approach can only accommodate non-differentiable payoffs if one chooses a baseline model that has a payoff identical to the one of the model of interest.\\ 
The general idea of approximating the pricing bias instead of the asset price directly is not entirely new to the financial engineering literature. \citet{Hull88} already suggested an approach to approximate the price of a derivative in the Heston model through a series expansion of the bias between the Heston and the Black-Scholes model.\footnote{Note that in 1988 there were no solutions available for such models for non-zero correlation cases. (See \citet{Hull88}, pp. 30 - 31.) The model received the name \textit{Heston model} after Steve Heston published his influential work in 1991, showing how such models can be solved in (semi-)closed-form through Fourier transform methods.} However, while Hull/White and KM are using similar PDEs to determine the pricing bias their series expansions are very different. Specifically, Hull/White use a series of the bias in the model's volatility of variance parameter, whereby the elements of which the series consists are functions of the state variables and time which are defined through a system of differential equations. Hull/White provide closed-form solutions for up to three elements in their series.\footnote{See \citet{Hull88}, equations (10) and (16) to (18). Note that Hull/White's and KM's PDEs for the bias are only consistent if one assumes that the Black-Scholes constant volatility parameter is identical to Heston's spot volatility parameter.}\\
Despite its merit in approximating derivative prices, the approach of focusing on computing a pricing bias is also interesting in its own right. Using the same illustrative example as KM, assume that the true market dynamics would be given by the Heston model but a trader uses the Black-Scholes delta to determine the amount of stocks he needs to buy to hedge his short position in an option. KM claim that the pricing bias obtained through their approach can be interpreted as the expected total hedging costs from using a wrong model, such as in the described situation. Such an interpretation of the bias is confirmed by \citet{El13} who show that the expectation of the costs of a hedging strategy is equal to the difference between the price obtained through the true model and some simpler models used for valuation and portfolio management.\\
\citet{Younesian} tests the empirical option valuation performance of the CEV and the Heston model both approximated by KM's approach, whereby he uses data on prices of S\&P 500 call index options for the period of years from 2002 to 2011. Since the approximation has closed-form, \citet{Younesian} is able to estimate the model's parameters directly from cross-sectional option data. Nevertheless, \citet{Younesian} reports that the obtained parameter estimates are quite imprecise. Especially the correlation parameter $\rho$ in the CEV model seems to be difficult to estimate if the model is approximated via KM's approach. As he notes this may also seriously flaw the estimates of the other model parameters.\\
\citet{Catalan} investigates how KM's approximation performs when applied to jump processes. Specifically, he compares two alternative approaches to compute KM's pricing bias for the CEV-Merton model using the Black-Scholes model as baseline. \citet{Catalan} reports to be able to only compute the first two elements in the series expansion such that the obtain approximations are comparatively imprecise and it is difficult to infer whether the series expansion converges.\\
The implementation of KM's series expansion requires the computation of derivatives of the baseline model up to higher orders. If e.g. one wishes to implement KM's approximation for the Heston model, while using the Black-Scholes model as baseline and using five terms in the series expansion, this requires taking derivatives of the Black-Scholes model up to an order of 10 in stock price direction as well as a number cross-derivatives in stock price and time direction. While KM themselves do report how they implemented their approximation, \citet{Catalan} reports to have all baseline model derivatives implemented through finite difference approximations. Whereas this appears to be a convenient and flexible solution, it substantially increases the computation times and also creates an approximation that does not have a fully closed-form. Hence, I believe that this approach undermines the main advantages offered by KM's method. \citet{Younesian} uses Maple to implement the whole approximations, but reports difficulties in saving the obtained expressions in Matlab files if expansions using more than two terms are computed. Implementing KM's approach by using Maple calls inside Matlab avoids the need of saving the expressions to some file but slows down computation time substantially.\\  
In this thesis I will compute most parts of the series expansion manually. I only use Maple to compute time-stock price cross-derivatives of the Black-Scholes model, which I could successfully save to Matlab files. By using this approach all of my KM approximations are given in closed-form completely. Additionally, I am able to compute expansion using up to five or six terms without any loss in computation speed.\\
In terms of models I will focus on stochastic volatility models, which provide a good chance to test the performance and convergence behavior of KM's approximation when varying different model parameters. In particular, I will apply the method to approximate option prices in the Heston and CEV model as well as the Sch\"obel/Zhu model and to approximate Commodity Future prices in a stochastic volatility model which is due to \citet{Lutz09}. Technically it is straightforward to define a suitable baseline model for each stochastic volatility model by simply taking the same price process but replacing stochastic by constant volatility. In case of the option pricing examples this leads to the Black-Scholes model as baseline, while I use \textit{model 1} from \citet{Schwartz97} as baseline for the commodity future example.\\
The remainder of this thesis develops as follows: In section \ref{KMbase} KM's asset price approximation as well as KM's approximation for hedge ratios are introduced and the links to similar approximation approaches available in the literature are shown. Section \ref{HestonModel} shows the application to KM's approximation for option prices as well as different hedge ratios in the Heston model. This is followed by a discussion on the performance in approximating a slight generalization the Heston model - \citet{Jones03}'s CEV model - in section \ref{sec.CEVexpansion}. Section \ref{SZModel} continues the discussion by showing an application of KM's approach to option prices and hedge ratios under the non-affine \citet{Schobel99} model. In section \ref{CommModel} the focus is shifted away from option prices towards prices of commodity futures, where the price process itself is assumed to be mean-reverting. Section \ref{Conclude} summarizes and concludes the discussion.

\newpage
\section{Closed-form approximations to asset prices and greeks} \label{KMbase}
In the next two subsections I will, closely following \citet[sect. 3.1 \& 3.2]{KM2011}, state the KM asset price representation, its approximation formula, as well as an approximation to partial derivatives of the asset price.

\subsection{Preliminaries: Infinitesimal generators}
The approximation approach of KM uses heavily the notion of infinitesimal operators. Hence, I will start the description of KM's approximation  by briefly introducing these operator.\\
Let $z(t)$ be a $(d \times 1)$ vector of state variables and assume that under the risk-neutral probability measure $z(t)$ satisfies the following vector SDE,
\begin{align}
\begin{bmatrix}
dz(t)_1\\
\vdots\\
dz(t)_d
\end{bmatrix} &= 
\begin{bmatrix}
\mu_1\\
\vdots\\
\mu_d
\end{bmatrix} dt +
\begin{bmatrix}
\sigma_{11} & \cdots & \sigma_{1d} \\
\vdots & \ddots & \vdots \\
\sigma_{d1} & \cdots & \sigma_{dd}
\end{bmatrix}
\begin{bmatrix}
dW(t)_1 \\
\vdots \\
dW(t)_d
\end{bmatrix} \nonumber \\
dz(t) &= \mu\left(z(t)\right)dt + \sigma\left(z(t)\right)dW(t),~~t \in \left[0,\infty\right) \label{SDE1}
\end{align}
where $dW(t)_i, ~ i \in \left[1,d\right]$ are the increments of arbitrarily correlated standard Brownian motion processes and $\mu\left(z(t)\right)$ and $\sigma\left(z(t)\right)$ define some time-homogenous drift and diffusion terms. Further, let $f(z(t))$ be a continuously twice differentiable function of $z(t)$. Then e.g. \citet{Neftci} define the expected rate of change in $f(z(t))$ through the operator $\mathscr{A}$, such that
\begin{align}
\mathscr{A}f(z(t)) = \lim_{\Delta \rightarrow 0}\frac{\mathbb{E}_{z,t}\left[f(z(t+\Delta))\right]-f(z(t))}{\Delta} \label{Llim}
\end{align}
\citet{Neftci} state that since the expected change in $f(z(t))$ is a smoother function than $f(z(t))$ itself, the limit in (\ref{Llim}) can be defined even as a Brownian motion process itself is non-differentiable. This \textit{derivative-like} notion of $\mathscr{A}$ relates the rate change directly to It\^o's Lemma. Applying It\^o's Lemma to $f(z(t))$ and replacing the $dW(t)$ terms in resulting expression by it's drift, which is zero for $dW(t)$, and finally dividing the remainder by $dt$ yields the same result as taking the limit in (\ref{Llim}).\footnote{See e.g. \citet{Neftci}, p. 356.} Hence, the expected rate of change can be expressed as
\begin{align}
\mathscr{A}f(z(t)) = \sum^d_{i = 1}\mu_i(z)\dfrac{\partial f(z(t))}{\partial z_i} + \dfrac{1}{2}\sum^d_{i = 1}\sum^d_{j = 1}\sigma^2_{ij}(z)\dfrac{\partial^2 f(z(t))}{\partial z_i \partial z_j} \label{GenL}
\end{align}
where $\sigma^2(z(t)) = \sigma(z(t))\sigma(z(t))' \in \mathbb{R}^{d \times d}$ is the covariance matrix of (\ref{SDE1}).\\
The representation in (\ref{GenL}) also shows clearly the connection to PDEs. Using the definition of $\mathscr{A}$ the PDE for the derivative $f(z(t))$ can be constructed as\footnote{Note that this is not a derivation of the PDE, but simply shows the technical connection between the concepts of an infinitesimal generator and the PDE. While  $\mathscr{A}f(z(t))$ can be interpreted, as discussed before, as an expected rate of change, the PDE determines the intertemporal change in the derivatives value (See e.g. \citet{Fries}, pp. 46 - 47). It is easy to see that both concepts are mathematically related. However, generally PDEs are derived by arbitrage arguments, such that their economic meaning becomes clear. Throughout this thesis I will not derive the PDEs of the considered models by arbitrage arguments in order to keep the focus of this work on the method of developing closed-form approximations.}
\begin{align}
\underbrace{\dfrac{\partial f(z,t)}{\partial t} + \mathscr{A}f(z,t)}_{\equiv ~ \mathscr{L}f(z,t)} + c(z,t) = R(z,t)f(z,t) \label{fundaPDE}
\end{align}
Where $R(z,t)$ is the instantaneous short-term rate and $c(z,t)$ denotes the instantaneous coupon rate. Following KM it will be assumed that $c(z,t)$ is identical zero and hence will be dropped in the following.
Based on this notion of the infinitesimal generator, KM define a new operator $\mathscr{L}$ as shown in the above expression.\\
In the subsequent section the operator $\mathscr{L}$ will be repeatedly applied to different functions. Hence, I found it useful to also define the operator independent from any specific function as 
\begin{align}
\mathscr{L} &= \dfrac{\partial }{\partial t} + \mathscr{A}  \nonumber \\
            &= \dfrac{\partial }{\partial t} + \sum^d_{i = 1}\mu_i(z)\dfrac{\partial}{\partial z_i} + \dfrac{1}{2}\sum^d_{i = 1}\sum^d_{j = 1}\sigma^2_{ij}(z)\dfrac{\partial^2}{\partial z_i \partial z_j} \label{L}
\end{align}
The notation in (\ref{L}) emphasizes that applying the operator $\mathscr{L}$ to any function practically means constructing a sum of the derivatives and cross-derivatives of that function in direction of all of its state variables as well as in time direction and weigh these derivatives with the elements of the drift vector and the covariance matrix of the underlying system of SDEs.\\
Note that KM name the operator $\mathscr{L}$ an infinitesimal generator. However, in the literature an infinitesimal generator often is defined as $\mathscr{A}$ only.
\subsection{Asset price approximations} \label{AssetApprox}
KM's approximation approach essentially builds on a power series expansion around a 'baseline' model for which an exact closed-form solution is available. Hence, the approximation requires the formulation of two models.\\
First, assume a d-dimensional vector SDE as in (\ref{SDE1}) and assume that $f(z,t)$ denotes the price of a derivative written on $z(t)$.\footnote{For notational convenience I will write $z(t)$ simply as $z$ and $f(z(t))$ as $f(z,t)$ to emphasis that $f$ is also a function of time.} Using the definition of the operator $\mathscr{L}$ in (\ref{L}), $f(z,t)$ satisfies the partial differential equation
\begin{align}
\mathscr{L}f(z,t) - R(z,t)f(z,t) &= 0 \label{ActPDE} \\
\text{s.t.}~~ f(z,T) &= b(z)~  \nonumber
\end{align}
Where $b(z)$ defines the payoff function at some maturity date $T > t$ and $R(z,t)$ again denotes the instantaneous short-term interest rate. Further, assume that there is no closed-form solution to (\ref{ActPDE}) available. \\
As baseline model consider the price of another derivative $f_0(y,t)$, which is written on a $(m \times 1)$ vector of state variables $y(t)$, where $m \leq d$ and $y(t)$ follows the (vector) SDE,
\begin{align}
dy(t) = \mu_y\left(y(t)\right)dt + \sigma_y\left(y(t)\right)dW_y(t),~~t \in \left[0,\infty\right) \label{SDE2}
\end{align}
Where $\mu_y$, $\sigma_y$ are time-homogenous drift and diffusion terms and $dW_y(t)$ denotes the increments of a m-dimensional standard Brownian motion. If the baseline model is of lower dimension as the model we desire to approximate, it is necessary to perform a correction in order to fit the dimensions of the two models.\\
KM suggest to perform this correction in the following way: Let $z_0(t)$ be a $(d \times 1)$ vector process, such that $z_0(t) = y(t)$ and $dz_0(t) = dy(t)$ if $m = d$ and $z_0(t) = \left[y(t)' ~ \aleph\right]'$ if $m < d$. Where $\aleph$ is a $(d-m \times 1)$ vector of zeros.
Further, if $m < d$ define the SDE for $z_0(t)$ such that,
\begin{align}
dz_0(t) &= \mu_0\left(z_0(t)\right)dt + \sigma_0\left(z_0(t)\right)dW_y(t),~~t \in \left[0,\infty\right) \label{SDE3} \\
\mu_{0,i}\left(z_0(t)\right) &=
\begin{cases}
\mu_{y,i}\left(y(t)\right)  & \quad \text{if } 1 \leq i \leq m \\
0  & \quad \text{otherwise} \\
\end{cases} \\
\sigma_{0,ij}\left(z_0(t)\right) &=
\begin{cases}
\sigma_{y,ij}\left(y(t)\right)  & \quad \text{if } 1 \leq i,j \leq m \\
0  & \quad \text{otherwise} \\
\end{cases}
\end{align}
Hence, the drift and diffusion terms of the SDEs in (\ref{SDE1}) and (\ref{SDE3}) will be of same dimension, what later on, will be necessary to obtain an expression for the pricing difference between the two models. 
Note that $f_0(\cdot)$ satisfies the partial differential equation
\begin{align}
\mathscr{L}_0f_0(z_0,t) &= R(z_0,t)f(z_0,t) \label{AuxPDE} \\
\text{s.t.}~~ f(z_0,T) &= b_0(z_0) \nonumber
\end{align}
Where $b_0(z_0)$ denotes the payoff of the derivative $f_0(\cdot)$ at maturity. $\mathscr{L}_0$ is essentially the same operator as in (\ref{L}), but defined for the m-dimensional system (\ref{SDE3}). \\
If a closed-form solution to (\ref{AuxPDE}) exists, then $f_0(\cdot)$ can be used as a baseline model to derive an approximate closed-form expression for $f(z,t)$.
Such a closed-form approximation can be derived in two steps.\\
First, denote the difference in the price between the unknown model $f(z,t)$ and the auxiliary model $f_0(\cdot)$ as $\Delta f(z,t) = f(z,t) - f_0(z,t)$. If $f(z,t)$ and $f_0(z,t)$ are well-defined solutions to their respective PDEs given in (\ref{ActPDE}) and (\ref{AuxPDE}), then the pricing difference $\Delta f(z,t)$ will satisfy the PDE\footnote{See \citet{KM2011}, p. 394.}
\begin{align}
\mathscr{L}\Delta f(z,t) + \delta(z,t) &= R(z,t)\Delta f(z,t) \label{DiffPDE} \\
\text{s.t.}~~ \Delta f(z,T) &= d(z) \nonumber
\end{align}
Where $d(z) = b(z)-b_0(z)$ adjusts the boundary condition for possible mismatches in the payoff functions of the two models. The second adjustment term $\delta(z,t)$ corrects for the differences in the underlying driving forces of the two models, i.e. for the differences between the SDEs given in (\ref{SDE1}) and (\ref{SDE3}) which KM define as 
\begin{align}
\delta(z,t) = (\mathscr{L}-\mathscr{L}_0)f_0(z,t) 
\end{align}
Recalling the definition of the operator $\mathscr{L}$ (and $\mathscr{L}_0$) it is straightforward to derive an expression for $\delta(z,t)$.
\begin{align}
\left(\mathscr{L}-\mathscr{L}_0\right)f_0(z,t) =&~ \dfrac{\partial f_0}{\partial t} + \mathscr{A}f_0 - \dfrac{\partial f_0}{\partial t} - \mathscr{A}_0f_0 \nonumber \\
						  =&~ \sum^d_{i = 1} \mu_i\dfrac{\partial f_0}{\partial z_i} + \dfrac{1}{2}\sum^d_{i = 1}\sum^d_{j = 1}\sigma^2_{ij}(z)\dfrac{\partial^2 f_0}{\partial z_i \partial z_j} - \sum^d_{i = 1} \mu_{0,i}\dfrac{\partial f_0}{\partial z_i} + \dfrac{1}{2}\sum^d_{i = 1}\sum^d_{j = 1}\sigma^2_{0,ij}(z)\dfrac{\partial^2 f_0}{\partial z_i \partial z_j} \nonumber \\
						  =&~ \sum^d_{i = 1} \left(\mu_i(z) - \mu_{0,i}(z)\right)\dfrac{\partial f_0}{\partial z_i} + \dfrac{1}{2}\sum^d_{i = 1}\sum^d_{j = 1}\left(\sigma^2_{ij}(z) - \sigma^2_{0,ij}(z)\right)\dfrac{\partial^2 f_0}{\partial z_i \partial z_j} \nonumber \\
 =&~ \sum^d_{i = 1} \Delta\mu_i\dfrac{\partial f_0(z_t,t)}{\partial z_i} + \dfrac{1}{2}\sum^d_{i = 1}\sum^d_{j = 1}\Delta\sigma^2_{ij}(z)\dfrac{\partial^2 f(z,t)}{\partial z_i \partial z_j}
\label{initdel}
\end{align}
Where $\Delta\mu_i(z) = \mu_i(z) - \mu_{0,i}(z)$ and $\Delta\sigma^2_{ij}(z) = \sigma^2_{ij}(z) - \sigma^2_{0,ij}(z)$. Recalling (\ref{Llim}) and (\ref{GenL}) the first line in (\ref{initdel}) suggests that KM's $\delta(z,t)$ represents the difference in expected rate of change of $f_0$ under the true and the baseline dynamics.\\ 
Since $f_0(\cdot)$ and $d(\cdot)$ are known, both adjustment terms are straightforward to obtain and a solution to the PDE in (\ref{DiffPDE}) can be derived as a function of the two adjustments $d(z)$ and $\delta(z,t)$. KM do this by applying Theorem 7.6 in \citet{Shreve}, which states the Feynman-Kac representation of the solution to (\ref{DiffPDE}) \\
\begin{align}
\Delta f(z,t) = \mathbb{E}_{z,t}\left[e^{\left(- \int_{t}^{T}R\left(z(s),s\right)ds\right)}d(z(T)) + \int_{t}^{T}e^{\left(- \int_{t}^{T}R\left(z(u),u\right)du\right)}\delta(z(s),s)ds\right] \label{Theo7.6}
\end{align}
 Recalling that $\Delta f(z,t) = f(z_t,t) - f_0(z_t,t)$ and rearranging yields KM's \textit{Theorem 1}:
\begin{align}
f(z,t) = f_0(z,t) &+ \mathbb{E}_{z,t}\left[e^{\left(- \int_{t}^{T}R\left(z(s),s\right)ds\right)}d(z(T))\right] \nonumber \\ &+ \int_{t}^{T}\mathbb{E}_{z,t}\left[e^{\left(- \int_{t}^{T}R\left(z(u),u\right)du\right)}\delta(z(s),s)\right]ds \label{Theo1}
\end{align}
The expression in (\ref{Theo7.6}) yields an exact formulation of the pricing bias between the baseline and the true model. Hence, if $f(z,t)$ represents e.g. the \citet{Heston93} model and this is assumed to reflect the true market dynamics while $f_0(z,t)$ represents e.g. the Black-Scholes model and this is used to manage a portfolio of derivative contracts, then (\ref{Theo7.6}) yields the bias from stochastic volatility. KM note that this bias can be interpreted as the expected hedging costs arising from the use of the wrong model. By using a different approach \citet{El13} show that this interpretation of the bias or pricing difference $\Delta f(\cdot)$ indeed applies. Note that the right hand side in (\ref{Theo7.6}) depends only on the solution known solution to the baseline.\\
Since the aim of KM's method is to derive asset price approximations, it is more useful to consider the formulation in (\ref{Theo1}). However, the expression in (\ref{Theo1}) relies on the evaluation of two conditional moments, which have no trivial solution. Generally there would be several possibilities to evaluate (\ref{Theo1}). As KM note, it would be possible to compute the two conditional moments by Monte Carlo integration. In such a case the chosen baseline model would have a similar role as a control variate. Hence, KM conclude that there might be no obvious advantage to directly estimating $f(z,t)$ using Monte Carlo with $f_0(z,t)$ as control variate.\\
This directly leads to the second step of KM's approach, which is to approximate the conditional moments in (\ref{Theo1}) by a series expansions. The expansion suggested for this purpose is given in KM's \textit{Definition 1}:
\begin{align}
f_N(z,t) = f_0(z,t) + \sum_{n = 0}^{N}\dfrac{1}{n!}d_n(z,t)\left(T-t\right)^n + \sum_{n = 0}^{N}\dfrac{1}{(n+1)!}\delta_n(z,t)\left(T-t\right)^{n+1}
\label{Dif1}
\end{align}
Where $\delta_0(z,s)$ is defined as in (\ref{initdel}) and $d_0(z,t) = b(z) - b_0(z)$. While for $ n > 0$
\begin{align}
\delta_n(z,s) =& \mathscr{L}\delta_{n-1}(z,t) - R(z,t)\delta_{n-1}(z,t) \label{It1}\\
d_n(z,s) =& \mathscr{L}d_{n-1}(z,t) - R(z,t)d_{n-1}(z,t) \label{It2}
\end{align}
The number $N$ denotes the order of the approximation, such that $f_N(z,t)$ denotes the $N^{th}$ order approximation to $f(z,t)$. Under some regulatory conditions, which I will discuss in a moment, $f_N(z,t) \rightarrow f(z,t)$ as $N \rightarrow \infty$. KM note that, in order to compute the iterations on the adjustment terms, it is must be ensured that both, $\delta(z,t)$ and $d(z,t)$, are $2N$ times differentiable with respect to $z$ and additionally that $\delta(z,t)$ is $N$ times differentiable with respect to $t$. Hence, if the payoff function $b(z)$ is non-differentiable, as it is the case e.g. for a plain vanilla options, the choice of a baseline model is restricted to models with $b_0(z) = b(z)$, such that $d_n(z,t) = 0 ~ \forall~ n$. \\
The formula in (\ref{Dif1}) provides a method to obtain a closed-form approximation of the asset price or the expected hedging costs even for cases in which the true model is not known in closed-form. Hence, computation will be fast. This feature provides the greatest advantage of KM's method over approaches like \citet{El13}, which is computationally costly as it is based on simulations of hedging strategies.\footnote{See \citet{El13}, p. 1016.}\\
Both steps of KM's approximation approach rely on a number of assumption. These are given in KM's Proposition A.1 \& A.2, which I summarize as follows: 
\begin{enumerate}
\item Since (\ref{Theo1}) is basically just an application of the Feynman-Kac formula, the expression holds under the same regulatory conditions under which this formula holds. These condition are given in \citet[Theorem 5.7.6]{Shreve} or in \citet[condition (A.3)-(A.5), p. 412]{KM2011}. Among these conditions only the linear growth condition which is imposed on the drift and diffusion terms of the SDEs may not hold in some models considered in finance. However, this condition only serves to ensure the existence of the solution to the model's SDEs. Hence, if existence can be ensured for both the baseline and the true model by other means then the linear growth condition can be ignored.\footnote{For the last statement see the footnote in KM, p. 394.}
\item In order to ensure that the expansion in (\ref{Dif1}) converges assumptions on the operator $\mathscr{L}$ need to be imposed. Specifically, $\mathscr{L}$ must have a transition density $p_t(y|x)$ with respect to Lebesgue measure. Hence, the considered diffusion model must have a transition density, what is generally true for most diffusion models. Further, there must be a measure $\pi$, such that $\pi(x)p_t(y|x) = \pi(y)p_t(x|y)$. This can be understood as a generalization of time-reversibility, which e.g. is always satisfied for stationary, univariate processes. 
\item The last set of assumption concerns the corrective terms $\delta(z,t)$ and $d(z,t)$. Assume that for a function space $H$, which contains some function norm $||\cdot ||_H$, there exists some $\hat{\tau}>0$ and some functions $\psi_{\delta},\psi_d \in H$ such that $\delta : \mathbb{R}^n \times \mathbb{R}_+ \rightarrowtail \mathbb{R}$ and $d : \mathbb{R}^n \times \mathbb{R}_+ \rightarrowtail \mathbb{R}$ satisfy $\mathbb{E}\left[\psi_{\delta}\left(x(\hat{\tau})\right)\mid x(0) = x\right] = \delta(x,\hat{\tau})$ and $\mathbb{E}\left[\psi_{d}\left(x(\hat{\tau})\right)\mid x(0) = x\right] = d(x,\hat{\tau})$, where additionally $t \rightarrowtail \delta(x,t)$ is analytic uniformly in $||\cdot ||_H$. In words this means that it must be possible to match the two adjustment terms $\delta$ and $d$ by conditional moments. Further, it is required that the instantaneous short-term rate is time homogeneous and $sup_x\mid R(x,t) = R(x)\mid < \infty$.
\end{enumerate}
These assumptions are difficult to verify for a specific model.\footnote{See \citet{KM2011}, p. 411.} In fact, KM do not verify them in their experiments but rather show empirically that their expansion converges for the models they consider.



\subsection{Approximating greeks} \label{sec.ApproxGreeks}
So far the KM approximation was introduced as a method to obtain prices of financial derivatives or hedging costs. Besides this the computation of hedge ratios for a derivative contract, the so called greeks, is of equal importance in many applications, such as risk management. The greeks are defined as partial derivatives of the respective pricing formula with respect to relevant parameters and hence can be interpreted as the sensitivity of the derivatives price to changes in these parameters. I will consider the following greeks in the numerical examples below:
\begin{enumerate}
	\item $\Delta~=~\dfrac{\partial C}{\partial S}$ : Sensitivity of the derivatives price to changes in the underlying
	\item $\Gamma~=~\dfrac{\partial^2 C}{\partial S^2}$ : Sensitivity of the derivatives price to changes in $\Delta$
	\item $\mathscr{V}~=~\dfrac{\partial C}{\partial v}$ : Sensitivity of the derivatives price to changes in spot variance
\end{enumerate}
From the definition of the greeks it is obvious that their computation is more or less trivial as long as a closed-form solution for the derivatives price is available. For cases in which a closed-form solution is not available the literature developed a number of numerical approaches which are often based on Monte Carlo simulation. However, these approach often increase estimation risk and, if the payoff function is non-differentiable, straightforward procedures such as finite differences perform poorly.\footnote{See \citet{LehPre}, p. 4 - 6, for an example on the estimation of the $\Delta$ of a Knock-in option via finite differences.} \citet{LehPre} provides a good overview of publications on the computation of greeks. \\
KM argue that their approximation as given in (\ref{Dif1}) can be used to also estimate the greeks of the derivative by differentiating the approximation formula with respect to the parameter of interest. Hence, KM state the following expression to compute the above mentioned greeks
\begin{align}
\dfrac{\partial^k f_N(z,t)}{\partial z^k_i} = \dfrac{\partial^k f_0(z,t)}{\partial z^k_i} + \sum_{n = 0}^{N}\dfrac{\left(T-t\right)^n}{n!}\dfrac{\partial^k d_n(z,t)}{\partial z^k_i} + \sum_{n = 0}^{N}\dfrac{\left(T-t\right)^{n+1}}{(n+1)!}\dfrac{\partial^k \delta_n(z,t)}{\partial z^k_i} \label{GrApprox}
\end{align}
where the first and second partial derivatives of the adjustment terms are given by $k = 1$ and  $k = 2$. KM provide the results for the analytic derivatives of first and second order for these adjustment terms.\\
Alternatively, KM suggest that these derivatives could be evaluated numerically, e.g. by finite-differences. Note, that the before mentioned poor performance of finite difference methods in the case of non-differentiable payoffs does not apply here. If the payoff function is non-differentiable, then a baseline model must be chosen such that $d_n(z,t) = 0~\forall~n$, whereby the function $\delta_n(z,t)$ necessarily needs to be differentiable in any region relevant for the computation of greeks. Since non-differentiability of the payoff function is given for many derivatives I will restrict the attention to that case.\\
Since the approximation expressions obtained from (\ref{Dif1}) becomes quite lengthy even for small numbers of corrective terms analytic differentiation might be cumbersome. Hence, numerical differentiation seems to be a more efficient choice. \\
In order to obtain numerical first and second derivatives of in (\ref{GrApprox}) I apply central differences to each of the corrective terms, while using the analytic derivative of the baseline model. Hence, the derivatives of the corrective terms in (\ref{GrApprox}) are estimated as
\begin{align}
\dfrac{\partial \delta_n(z,t)}{\partial z_i}~ \approx&~ \frac{\delta_n(z_i + h,t) - \delta_n(z_i - h,t)}{2h} \label{FirstD} \\
\dfrac{\partial^2 \delta_n(z,t)}{\partial z_i^2}~ \approx&~ \frac{\delta_n(z_i + h,t) + 2\delta_n(z_i,t) - \delta_n(z_i - h,t)}{h^2} \label{SecD}  \\
\text{for } n =& 1, \hdots ,N \nonumber
\end{align}
The numerical derivatives of $d_n(z,t)$ could be computed analogously. Using central differences yields an error of order $\mathscr{O}\left(h^2\right)$ for the approximation of the derivatives. This error comes additionally to the overall approximation error of KM's method. However, my numerical results shown in the later sections suggest that this approach nevertheless achieves sufficient accuracy. Following the suggestion in \citet[p. 257 \& Ch. 8.1.3]{Hermann} the step size $h$ is chosen to be a function of the differentiation variable and the machine precision of the computer used to perform the computations, such that
\begin{align}
h = \left(z_i + 1\right)10^{\frac{log_{10}(eps)}{3} - 1} \label{StepS}
\end{align}
Where $eps$ denotes the machine precession. On the system I used to run all Matlab scripts related to this thesis, $eps$ had a value of $2.2204\cdot 10^{-16}$. Note that, while in general the step size should be chosen as small as possible to minimize truncation errors, choosing too small steps would lead to round-off errors.\footnote{See e.g. \citet{Gill}, p. 11 - 12.} Hence, (\ref{StepS}) attempts to balance these two effects. Note that the specific functional form in (\ref{StepS}) does not necessarily lead to an optimal step size. Better solutions for $h$ might be available in the literature. Also, one might try the complex-step approach suggested by \citet{Martins}. Therein the real function of which we want to estimate the derivatives is transformed into the complex plane. Since this yields a situation in which no difference operation is required to estimate the derivative the previously mentioned trade-off between truncation and round-off error is circumvented.\footnote{See \citet[Eq. 6]{Martins}}  However, as I will show in section \ref{HestGreeks}, the specification in (\ref{StepS}) turned out to yield precisely the same results as were reported in KM's original publication, so I decided to keep this definition. \citet[Programm 5.2, p. 257]{Hermann} provides Matlab code to estimate the Jacobi-Matrix of a function based on forward differences and the step size in (\ref{StepS}). I adapted this code to estimate the four greeks previously defined.

\subsection{Connection to other approaches}
In the finance literature there are two closed-form approximation approaches bearing an especially close relationship to KM's approach, that have not been mentioned in the introduction. In the following I will provide a short overview of these approaches by summarizing KM's treatment of the two methods. 
\subsubsection{Risk-neutral densities and saddlepoint approximations}
The closed-form approximation approach from KM closely relates to saddlepoint approximations of risk-neutral transition densities and cumulative distribution functions, which also can be used to approximate prices of derivatives and, as claimed by KM, arise as a special case of their expansion. The notion of saddlepoint approximations to probability distribution dates back to work of \citet{Daniels54} and had been applied in the context of option pricing e.g. by \citet{Rogers99} and \citet{Ait06}. Whereas \citet{Goutis99} provide a good introduction on how this method can be applied.
As noted by \citet{Ait06}, saddlepoint approximations can be applied to models for which a characteristic function of the transition density can found in closed-form, but the transition density itself is unknown. Hence, the method can be seen as an alternative to Fourier inversion, but comes with the advantage of avoiding numerical integration.\footnote{See \citet{Rogers99}, p. 494.} To see the connection to KM's approach recall from the definition of the Feynmann-Kac theorem that the price of a derivative equals the discounted exception of its payoff, i.e. $f(z,t)=\mathbb{E}\left[ e^{\left(- \int_{t}^{T}R(z(s),s)\right)ds} b(z(T)) \right]$ . Where $b(z(T))$, as before, denotes the assets payoff. Following KM, the instantaneous short-term rate is assumed to be identical zero within this section, such the exponential term in the expectation drops out. Again considering the model of interest $f(z,t)$ and a suitable baseline model $f_0(z,t)$, both asset prices can be expressed through the Feynmann-Kac formula as\footnote{See \citet{KM2011}, p. 396.}
\begin{align}
f(z,t) =& \int_{\mathbb{R}^d}b(z(T))p\left(z(T),T\mid z,t\right)dz \\
f_0(z,t) =& \int_{\mathbb{R}^d}b_0(z(T))p_0\left(z(T),T\mid z,t\right)dz
\end{align}
Where the expectation is expressed as an Riemann-integral and $p$ and $p_0$ denote the risk-neutral transition densities of the two models. If the baseline model is chosen to match the payoff of the actual model at maturity, i.e. $b(z(T)) = b_0(z(T))$, the pricing bias between the models is
\begin{align}
 f(z,t) - f_0(z,t) = \int_{\mathbb{R}^d}b(z(T))\Delta p\left(z(T),T\mid z,t\right)dz \label{Transdis}
\end{align}
Where $\Delta p = p - p_0$ denotes the transition discrepancy, the difference between the two transition densities. Comparing (\ref{Transdis}) with (\ref{Theo7.6}) and recalling that the short-term rate is assumed to be zero suggests that\footnote{See also \citet{KM2011}, p. 397.} 
\begin{align}
\int_{\mathbb{R}^d}b(z(T))\Delta p\left(z(T),T\mid z,t\right)dz = \int_{t}^{T}\mathbb{E}\left[\delta(z(s),s) \mid z,t\right]ds \label{SaddleEqu}
\end{align}
Such that applying KM's approximation would be equivalent to computing a derivatives price through an estimate of its transition discrepancy. Indeed, \citet[Appendix B]{KM2011} provide a proof that the above equation holds. Nevertheless, the mere equality in (\ref{SaddleEqu}) is not of practical use until it is specified how the integrals on the left and the right hand side should be evaluated. For the left-hand side, saddlepoint approximations can be used to first estimate the transition discrepancy and then the pricing error can be computed through the evaluation of the Riemann-integral. KM's approach, as described at the beginning, directly approximates the pricing error by means of a series expansion around the baseline model. However, KM note that the equality in (\ref{SaddleEqu}) hinges on the assumption that the instantaneous short-term rate is identical zero. In section \ref{CommModel}, I will show that KM's approximation applied to a future indeed yields a representation of the pricing bias as on right hand side of (\ref{SaddleEqu}). Such that the use of saddlepoint approximations for transition discrepancies and KM's series expansion of the pricing bias indeed are equivalent in this case. However, this is not true for e.g. option pricing models.

\subsubsection{A similar expansion: Yang's approach}
\citet{Yang06} suggests a closed-form approximation for option prices in the context of stochastic volatility models, which appears to be very similar to the approach suggested by KM. They provide a good comparison between their approximation and the one in \citet{Yang06}, which I attempt to summarize in this section.\footnote{The approach is a part of Yang's doctoral dissertation, which unfortunately was not accessible. Hence, while I still include the original reference to Yang's dissertation it should be noted that this subsection solely summarizes the treatment of the method as described in KM.}\\
Yang's approximation is also built on the notion of a pricing bias between some baseline model $f^{0}(z,t)$ known in closed-form and some unknown model $f(z,t)$. Following the treatment of the method in KM the instantaneous short-term rate $R(z,t)$ is assumed to be identical zero, and the true model and the baseline model are assumed to have the same payoff at maturity. The difference between the two models, denoted by $\Delta f(z,t) = f(z,t) - f^{(0)}(z,t)$, satisfies the PDE
\begin{align}
\mathscr{L}_0\Delta f(z,t) + \tilde{\delta}(z,t) = 0 \label{YangPDE} \\
\text{with } \tilde{\delta}(z,t) = \left(\mathscr{L} - \mathscr{L}_0\right)f(z,t) \label{YangErr}
\end{align}
The two operators $\mathscr{L}$ and $\mathscr{L}_0$ are defined the same as for KM's approximation. However, KM note two important differences between their and Yang's approximation. First, in (\ref{YangPDE}) the operator associated with the baseline model, instead the one of the true model, is applied to the difference in the two asset prices. Second, the operator $\left(\mathscr{L} - \mathscr{L}_0\right)$, even though it is identical to the one used in KM's approximation, is applied to the true model instead of the baseline model. Hence, $\tilde{\delta}$ can not be expressed in closed-form, as directly as KM's $\delta$. By applying the Feynmann-Kac formula to (\ref{YangPDE}) a solution similar to the one in (\ref{Theo1}) is obtained
\begin{align}
f(z,t) = f^{(0)}(z,t) + \int_{t}^{T}\mathbb{E}^0\left[\tilde{\delta}(z,t) \mid z,t\right]
\end{align}
Where $\mathbb{E}^0$ indicates that the expectation is taken under the probability measure of the baseline model. Note that in the case of KM's approach this expectation is taken under the probability of the true model. Again similar to KM, Yang suggests a series expansion to approximate the integral expression
\begin{align}
f(z,t) = f^{(0)}(z,t) + \sum_{m = 1}^{M}f^{(m)}(z,t) 
\end{align}
Where each $f^{(m)}(z,t)$ is a solution to the PDE $\mathscr{L}_0 f^{(m)} + \left(\mathscr{L} - \mathscr{L}_0\right)f^{(m-1)} = 0$ with boundary condition $f^{(m)}(z,T) = 0$. Such that the $f^{(m)}(z,t)$ can be computed recursively starting from the known $f^{(0)}(z,t)$ until $f^{(M)}(z,t)$. Yang suggests to do so by applying the Feynmann-Kac formula to each of the PDEs
\begin{align}
f^{(m)} = \int_{t}^{T}\mathbb{E}^0\left[\left(\mathscr{L} - \mathscr{L}_0\right)f^{(m-1)}\mid z,t\right]
\end{align}
Note that for $m = 1$ the expression is, besides the expectation operator, the same as in (\ref{Theo1}) of KM's approximation and thus the expression inside the expectation can be expressed in closed-form. The whole recursion can be solved e.g. using standard symbolic software. KM compare the performance of their method to the performance of Yang's expansion for the \citet{Heston93} as well as the more general CEV model. They find that, while both models perform reasonably well, Yang's expansion is slightly more accurate in approximating a non-affine formulation of the CEV model, but is slightly less accurate for the Heston model.   
\clearpage

\section{Approximating the Heston model} \label{HestonModel}
The abstract description of KM's expansion in section \ref{AssetApprox} already showed that the implementation of an approximation can be cumbersome if a large number of corrective terms would be required to achieve precise approximations. Hence, for practical applications, the approximation should converge fast to the true price of the derivative contract of interest, such that only a small number of corrective terms need to be computed. In order to assess the convergence behavior of the approximation I follow KM and apply their expansion to the well-known \citet{Heston93} model. Since for this model (semi-)closed form solutions are available it is straightforward to analyze the accuracy of the approximation. Hence, in this section I first attempt to replicate the results already shown in KM, but also to go further and analyze in greater depth the effect of different parameter values on the convergence behavior of the approximation.\\
The Heston model is defined by the following system of SDEs
\begin{align}
	dS(t)~ =&~ rSdt + \sqrt{v(t)}SdW_1(t) \label{HestSDE1}\\
	dv(t)~ =&~ \kappa\left(\theta - v(t)\right)dt + \omega \sqrt{v(t)}dW_2(t) \label{HestSDE2}\\
	dW_1(t)dW_2(t)~ =& ~\rho dt \nonumber
\end{align}
Where $\kappa$ denotes the speed of mean-reversion parameter, $\theta$ and $\omega$ the long-run variance and the instantaneous volatility of variance respectively. The instantaneous short-term rate is constant over time and all states, such that $R(S,t) = r ~\forall ~ S,t$ in the notation of section \ref{AssetApprox}. This implies, i.e. for equation (\ref{Theo1}), that $\int_{t}^{T}R(S,u)du = r\left(T-t\right)$.


\subsection{KM expansion for the Heston model} \label{HestonApproxKM}
In this section I will use the principles outlined in section \ref{AssetApprox} to derive the KM approximation of the Heston model. As a first step one needs to derive the vector $\mu(S(t),v(t))$ and the matrix $\sigma(S(t),v(t))$ as described in (\ref{SDE1}).\\ 
Trivially, $\mu(S(t),v(t)) = \left[rS ~~\kappa\left(\theta - v(t)\right)\right]'$. $\sigma(S(t),v(t))$ is derived by decomposing $dW_1(t),~dW_2(t)$ into independent Wiener processes denoted by $\hat{dW_1(t)},~\hat{dW_2(t)}$, such that
\begin{align}
\begin{bmatrix}
dW(t)_1\\
dW(t)_2
\end{bmatrix} = 
\begin{bmatrix}
1 & 0\\
\rho & \sqrt{1 - \rho^2}
\end{bmatrix} \times
\begin{bmatrix}
\hat{dW(t)_1}\\
\hat{dW(t)_2}
\end{bmatrix}
\end{align}
Where $\hat{dW_1(t)}\hat{dW_2(t)} = 0$. From this $\sigma(S(t),v(t))$ as in (\ref{SDE1}) is given by
\begin{align}
\begin{bmatrix}
\sqrt{v(t)}S & 0 \\
0 & \omega \sqrt{v(t)}
\end{bmatrix} \times
\begin{bmatrix}
1 & 0\\
\rho & \sqrt{1 - \rho^2}
\end{bmatrix} = 
\begin{bmatrix}
\sqrt{v(t)}S & 0 \\
\rho \omega \sqrt{v(t)} & \sqrt{1 - \rho^2} \omega \sqrt{v(t)}
\end{bmatrix} \label{CovHest}
\end{align} 
Hence, the covariance matrix $\sigma^2(\cdot) = \sigma(\cdot)\sigma(\cdot)'$ of the \citet{Heston93} model is
\begin{align}
\begin{bmatrix}
\sqrt{v(t)}S(t) & 0 \\
\rho \omega \sqrt{v(t)} & \sqrt{1 - \rho^2} \omega \sqrt{v(t)}
\end{bmatrix} &\times
\begin{bmatrix}
\sqrt{v(t)}S(t) & \rho \omega \sqrt{v(t)} \\
0 & \sqrt{1 - \rho^2} \omega \sqrt{v(t)}
\end{bmatrix} = \nonumber \\ ~ \nonumber \\
&=	\begin{bmatrix}
v(t)S(t)^2 & \rho \omega v(t) S(t) \\
\rho \omega v(t) S(t) & \omega^2 v(t)
\end{bmatrix}
\end{align}
Denoting the price of a European style call option written on $S(t)$ by $C\left(S(t),t\right)$ and applying the operator $\mathscr{L}$ to this price one obtains,
\begin{align}
\mathscr{L}C\left(S(t),t\right) = \dfrac{\partial C}{\partial t} + rS\dfrac{\partial C}{\partial S} +& \nonumber  \\ + \kappa\left(\theta - v\right)\dfrac{\partial C}{\partial v} +& \dfrac{1}{2} \left( v S^2\dfrac{\partial^2 C}{\partial S^2} + \omega^2 v \dfrac{\partial^2 C}{\partial v^2} \right) + \rho \omega v S \dfrac{\partial^2 C}{\partial S \partial v} \label{OpHest}
\end{align}
Such that the PDE of a  European style call in the Heston model can be constructed from this expression as  
\begin{align}
rC\left(S,t\right) = \dfrac{\partial C}{\partial t} + rS\dfrac{\partial C}{\partial S} +& \nonumber  \\ + \kappa\left(\theta - v\right)\dfrac{\partial C}{\partial v} +& \dfrac{1}{2} \left( v S^2\dfrac{\partial^2 C}{\partial S^2} + \omega^2 v \dfrac{\partial^2 C}{\partial v^2} \right) + \rho \omega v S \dfrac{\partial^2 C}{\partial S \partial v} \label{PDEHest} \\ ~ \nonumber \\
\text{s.t. } b(S,T) ~ =& ~ max\left[S - K,0\right] \nonumber
\end{align}
In a next step one needs to define a suitable baseline model at which KM's series expansion can be developed. As mentioned earlier the Black-Scholes model would be a convenient choice in the set up of stochastic volatility models. Assume that the underlying follows the same process as in (\ref{HestSDE1}), but replacing $\sqrt{v(t)}$ with a constant $\eta_0$, i.e.
\begin{align}
dS(t) ~=&~ rSdt + \eta_0SdW_1(t) \label{BS1}
\end{align}
By applying the principles for defining a baseline model described in section \ref{AssetApprox} the drift vector and the covariance matrix of the Black-Scholes model are given by
\begin{align}
\mu_0 = \left[rS~~0\right]' \quad \text{,} \quad \sigma^2_0(\cdot) = \begin{bmatrix}
											\eta_0^2S^2 & 0 \\
											0 & 0
											\end{bmatrix}
\end{align} 
Denoting by $C^{BS}(S,t)$ the price of a European call option under the Black-Scholes model the models PDE can be again written by using the definition of the operator $\mathscr{L}$
\begin{align}
rC^{BS}\left(S,t\right) &= \dfrac{\partial C^{BS}}{\partial t} + rS\dfrac{\partial C^{BS}}{\partial S} + \dfrac{1}{2} \eta_0^2S^2\dfrac{\partial^2 C^{BS}}{\partial S^2}  \label{PDEBS} \\ ~ \nonumber \\
\text{s.t. } b(S,T) ~ =& ~ max\left[S - K,0\right] \nonumber
\end{align}
The solution to this PDE is the well-known Black-Scholes formula
\begin{align}
C^{BS}(S,t) &= \mathbb{N}(d_1)S - \mathbb{N}(d_2)Ke^{-r(T-t)}\\
d_1 &= \frac{log(S/K) + \left(r + \sigma^2_0/2\right)\left(T-t\right)}{\eta_0\sqrt{T-t}}\nonumber \\
d_2 &= d_1 - \sigma_0\sqrt{T-t}\nonumber
\end{align} 
Where $\mathbb{N}(\cdot)$ denotes a standard normal distribution.\\
Using the PDEs in (\ref{PDEHest}) and (\ref{PDEBS}) as well as the definition in (\ref{DiffPDE}) the PDE of the pricing difference $\Delta C(S,t) = C\left(S,t\right) - C^{BS}\left(S,t\right)$ is defined as
\begin{align}
\mathscr{L}\Delta C(S,t) + \delta(S,t) = r\Delta C(S,t)
\end{align}
As both model specifications, Heston and Black-Scholes, share identical boundary conditions $d_n(S) = 0 ~\forall~n$ in (\ref{Dif1}). As previously mentioned it is necessary to match the boundary condition if the approximated model does have a non-differentiable boundary.\\
Applying (\ref{initdel})  yields the initial pricing error
\begin{align}
\delta_0(S,t) =& \left(rS - rS\right)\dfrac{\partial C^{BS}}{\partial S} + \left(\kappa\left(\theta - v(t)\right) - 0\right)\dfrac{\partial C^{BS}}{\partial v} + \nonumber\\ &\dfrac{1}{2}\left(\left(v(t)S^2 - \eta_0^2S^2\right)\dfrac{\partial^2C^{BS}}{\partial S^2} + \left(\omega^2v(t) - 0\right)\dfrac{\partial^2C^{BS}}{\partial v^2}\right) + \left(\rho \omega S v(t)^ - 0\right)\dfrac{\partial^2C^{BS}}{\partial S\partial v} \nonumber \\
=& \dfrac{1}{2}\left(v(t) - \eta_0\right)S^2\dfrac{\partial^2C^{BS}}{\partial S^2} \label{Hestd0}
\end{align} 
Where the last equality in (\ref{Hestd0}) results because $\partial C^{BS}/\partial v = \partial^2C^{BS}/\partial v^2 = \partial^2C^{BS}/\partial S\partial v = 0$. Below Table \ref{tab.Hestit} shows how the series of corrective terms develops with the order of the approximation, starting from (\ref{Hestd0}) up to any positive integer $N$. Each of the rows in Table \ref{tab.Hestit} is developed according to the rule in (\ref{It1}). KM note that, through the initial pricing error only a convexity adjustment is added to Black-Scholes price, while for $N = 1$ and $N = 2$ (see also Table \ref{tab.Hestit}) more information about the variance process is included. \\
Using the equations from Table \ref{tab.Hestit} in (\ref{Dif1}) yields the $N^{th}$ order KM approximation for a Heston call option
\begin{align}
C_N(S,t;\eta_0) = C^{BS}(S,t,\eta_0) +& (T-t)\delta_0 + \frac{(T-t)^2}{2}\delta_1 + \nonumber \\ ~ \nonumber \\ &+ \frac{(T-t)^3}{6}\delta_2 +  \hdots + \frac{(T-t)^{N+1}}{(N+1)!}\delta_N
\label{CorrecHest}
\end{align}
Since $C_N(S,t;\sigma_0) \rightarrow C(S,t)$ as $N \rightarrow \infty$ each additional corrective term $\delta_n$ should increase the accuracy of the approximation. However, such convergence should occur already after only a view corrective terms since otherwise the approach would loose it tangibility. \\
A feature of the KM approximation is that the approximated price $C_N$ depends on the constant volatility parameter of the Black-Scholes model while the Heston model does not. Such a dependence on one or more nuisance parameters originating from the chosen baseline model will usually arise in the application of the KM approximation. Hence, it is necessary to make a choice regarding the Black-Scholes volatility. 
\newpage
\begin{landscape}
	\newcolumntype{C}{>{\centering\arraybackslash} m{18cm} } 
	\renewcommand{\arraystretch}{2.0}
	\begin{table}[H]
		\centering
		\caption{Iterations of the pricing error for the \citet{Heston93} model}
		\begin{tabular}{m{1cm}|C}
			\hline
			\textbf{n} & \textbf{Pricing error $\delta_n(S,t)$} \\\hline\hline
			\textbf{0} & $\dfrac{1}{2}\left(v - \eta_0\right)S^2\dfrac{\partial^2C^{BS}}{\partial S^2}$ \\[5pt]\hline
			\textbf{1} & $\dfrac{\partial \delta_0}{\partial t} + rS\dfrac{\partial \delta_0}{\partial S} + \kappa\left(\theta - v\right)\dfrac{\partial \delta_0}{\partial v} + \frac{1}{2}vS^2\dfrac{\partial^2 \delta_0}{\partial S^2} + \rho \omega S v\dfrac{\partial^2\delta_0}{\partial S\partial v} - r\delta_0$~~~ (*)\\[5pt]\hline
			\textbf{2} & $\dfrac{\partial \delta_1}{\partial t} + rS\dfrac{\partial \delta_1}{\partial S} + \kappa\left(\theta - v\right)\dfrac{\partial \delta_1}{\partial v} + \frac{1}{2}\left(vS^2\dfrac{\partial^2 \delta_1}{\partial S^2} + \omega^2v\dfrac{\partial^2\delta_1}{\partial v^2}\right) + \rho \omega S v\dfrac{\partial^2\delta_1}{\partial S\partial v} - r\delta_1$\\[5pt]\hline
			\textbf{$\vdots$} & $\vdots$ \\[5pt]\hline
			\textbf{N} & $\dfrac{\partial \delta_{N-1}}{\partial t} + rS\dfrac{\partial \delta_{N-1}}{\partial S} + \kappa\left(\theta - v\right)\dfrac{\partial \delta_{N-1}}{\partial v} + \frac{1}{2}\left(vS^2\dfrac{\partial^2 \delta_{N-1}}{\partial S^2} + \omega^2v\dfrac{\partial^2\delta_{N-1}}{\partial v^2}\right) + \rho \omega S v\dfrac{\partial^2\delta_{N-1}}{\partial S\partial v} - r\delta_{N-1}$ \\[5pt]\hline
		\end{tabular} \\
		\tiny{\textit{(*) Note that $\partial^2 \delta_0/\partial v^2 = 0$.}}
		\label{tab.Hestit}
	\end{table}
\end{landscape}
\newpage

\subsubsection{Digression on optimal nuisance parameters}
\begin{figure}[H]
	\centering
	\includegraphics[scale=.9]{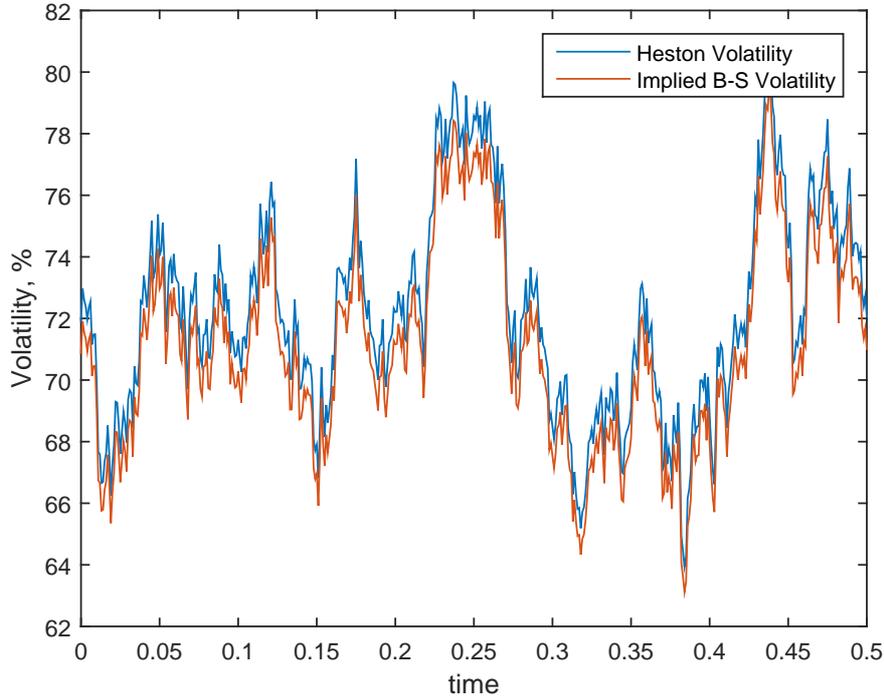}
	\caption{Simulated vs. B-S implied volatility}
	\floatfoot{\textit{Notes: The variances series had been a single sample path generated by an Euler discretazation of (\ref{CEV1}) \& (\ref{CEV2}) with 500 time increments. Option prices had been computed by Fourier inversion using numerical integration, implied Black-Scholes volatilities by Matlabs bsimpl() function. Parameters: $\kappa = 2.00$, $\theta = 0.04$, $\omega = 0.10$, $\rho = -0.5$ and $r = 0.10$.}}
	\label{fig.impV}
\end{figure}

For the present case of the Heston model, KM suggest to choose either $\eta_0 = \sqrt{v(t)}$ or $\eta_0 = \sqrt{\theta}$. As I will show in the next sections, the specific choice of $\eta_0$ has no big influence on the result of the approximation. However, setting $\eta_0 = \sqrt{v(t)}$ appears to be superior regarding accuracy compared to other choices. KM themselves also use this specification in all their experiments based on the Heston model\footnote{And its generalization the CEV model.}, but do not further explain the reason for this particular choice. However, since $v(t)$ denotes the current spot volatility in the market, this choice also makes intuitive sense. Recall that (\ref{Theo7.6}) showed clearly that the KM approximation mainly provides a correction term if one e.g. would compute an option price by the Black-Scholes model when the true market dynamics are determined by the Heston model. However, both models require the definition of a "spot volatility" parameter. The only difference is that in the Black-Scholes case this value is assumed to be a constant over time, while in the Heston case this value is interpreted as the initial value of the stochastic variance process. Since $v(t)$ denotes a variance in (\ref{HestSDE2}), the choice $\eta_0 = \sqrt{v(t)}$ implies that one would use the same spot volatility estimate to compute Black-Scholes and Heston option prices. Since such spot volatilities are usually estimated from observed prices of options and their respective underlying this seems to be reasonable. KM suggest that one might also use an optimal nuisance parameter, defined via the optimization problem:
\begin{align}
\eta^{Opt}_N(S,v,t) = \text{arg}~\underset{\eta_0}{\text{min}} \left(C_N(S,t;\eta_0) - C^{BS}(S,t,\eta_0)\right)^2
\label{OptNu}
\end{align}  
Note that this is equivalent to minimizing the square of the pricing error. As KM point out, $\eta^{Opt}_N(S,v,t)$ convergences to the implied volatility of the Black-Scholes model as $N \rightarrow \infty$. Consider the experiment shown in Figure \ref{fig.impV}, there the blue line shows one possible sample path of the variance process as defined in (\ref{HestSDE2}). The red line in the same figure shows Black-Scholes implied volatilities, where the option prices had been computed under the Heston model using the respective value $v(t)$ from the blue line in each time step. Hence, the red line shows the implied volatilities one would estimate at each point in time if working under the assumption of the Black-Scholes model, while the true market dynamics are governed by the Heston model. Figure \ref{fig.impV} shows clearly that the spot volatility of the Heston model and the implied volatility of the Black-Scholes model are highly correlated, where the implied Black-Scholes volatility always slightly underestimates the true volatility. However, this admittedly quite heuristic example shows that the choice $\eta_0 = \sqrt{v(t)}$ is almost equivalent to using the optimization criterion in (\ref{OptNu}), explaining why this specific choice is superior to the suggested alternative regarding accuracy of the approximation if a finite and low number of corrective terms is used.   

\subsection{Accuracy of the approximation}
\begin{figure}[H]
	\centering
	\begin{subfigure}{.7\textwidth}
		\centering
		\includegraphics[width=.9\linewidth]{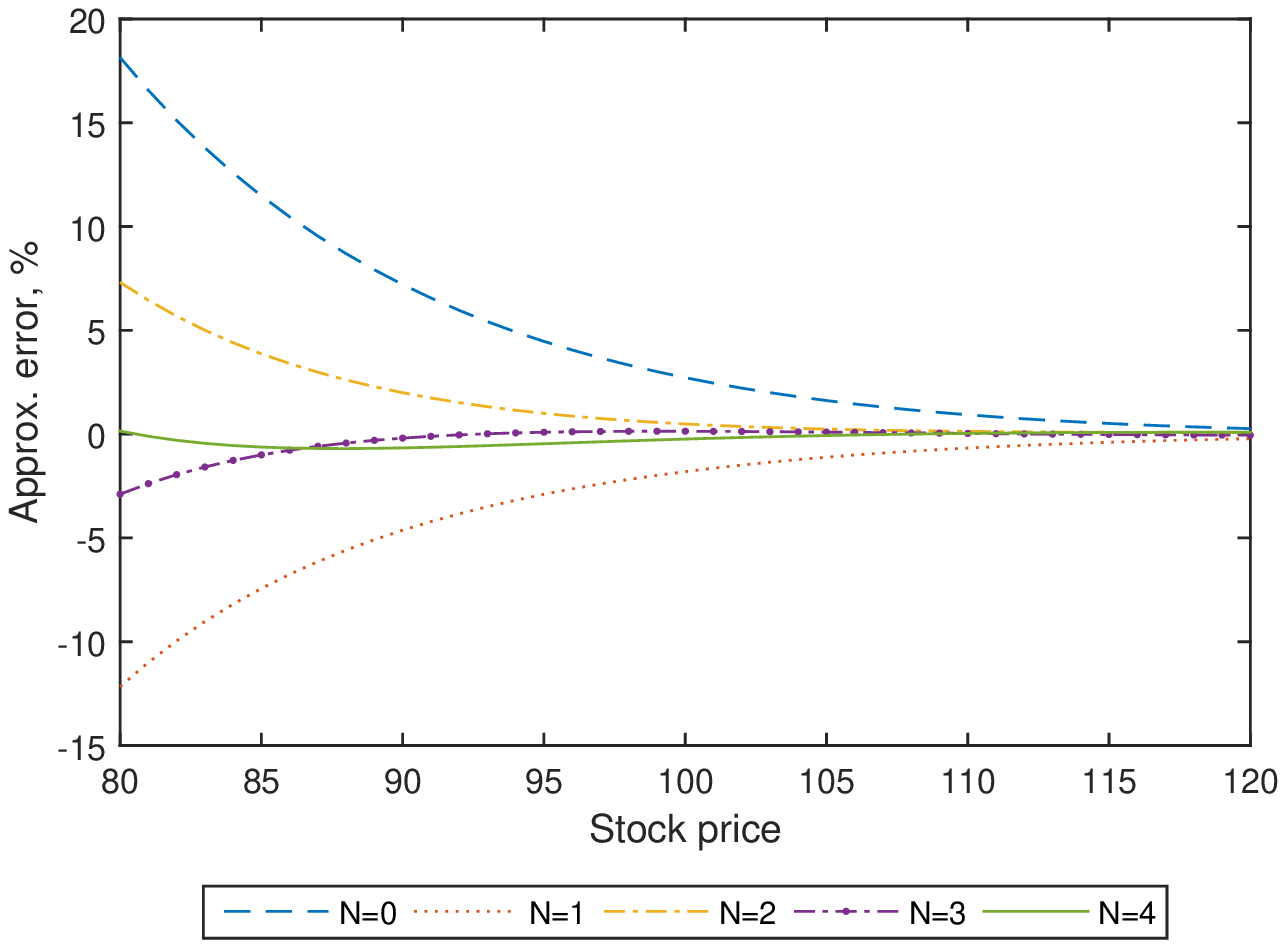}
		\caption{Using $\sigma_0 = \sqrt{v(t)}$}
		\label{KM1A}
	\end{subfigure}\\
	\begin{subfigure}{.7\textwidth}
		\centering
		\includegraphics[width=.9\linewidth]{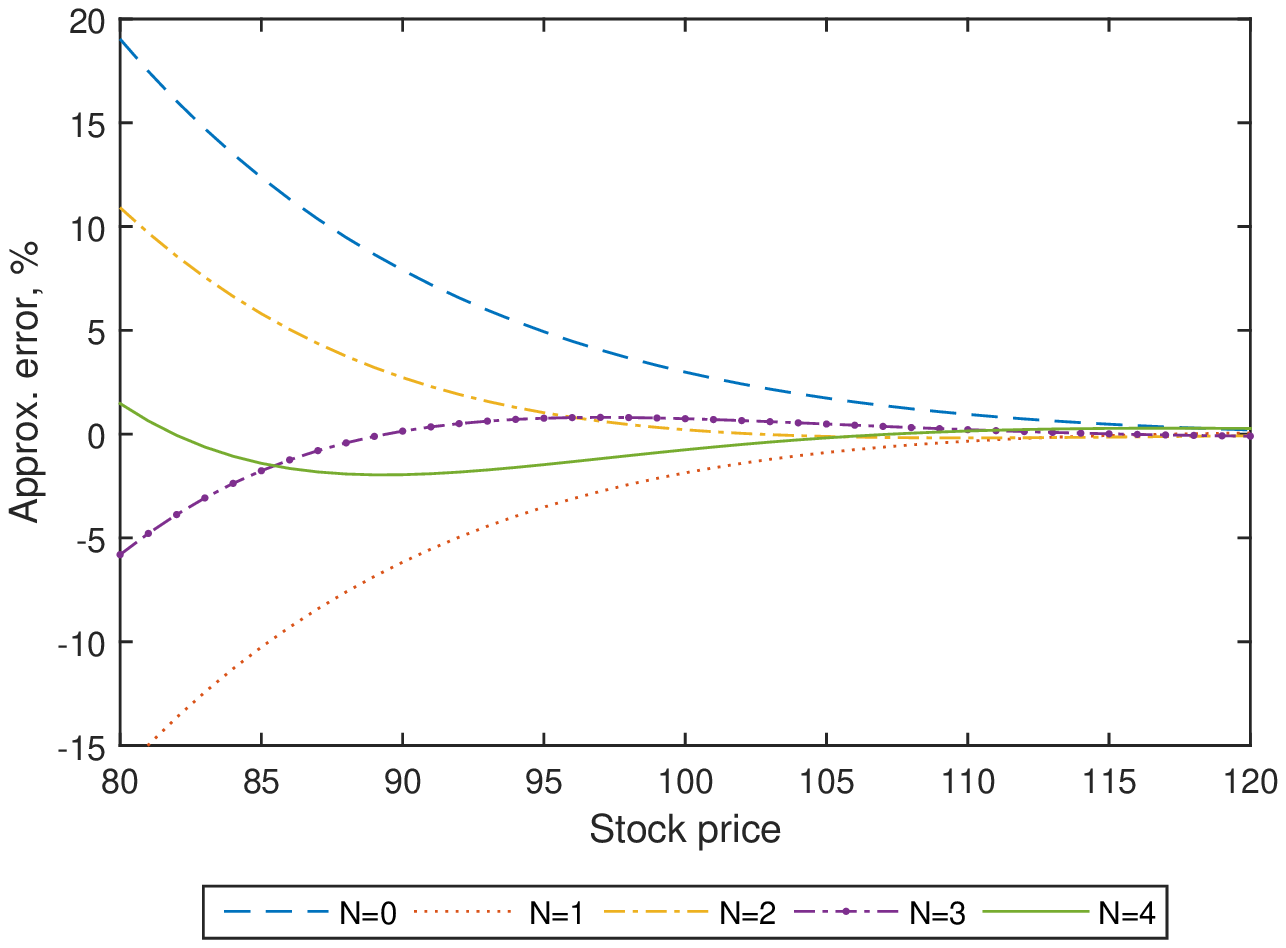}
		\caption{Using $\sigma_0 = \sqrt{\theta}$}
		\label{KM1B}
	\end{subfigure}
	\caption{KM approximation of the \citet{Heston93} model}
	\label{fig.KM1}
	\floatfoot{\textit{Notes: Analytic prices had been computed via Fourier transforms. $N$ = numer of corrective terms. Strike price = 100, time to maturity = 1 year, $\kappa = 2.00$, $\theta = 0.04$, $\omega = 0.10$, $\rho = -0.5$, $r = 0.10$, $v(t) = 0.04$. All B-S parameters that coincide with the \citet{Heston93} parameters are set equal.}}
\end{figure}

In order to investigate the accuracy of the approximations I will follow KM and express approximation errors relative to the analytic solution as
\begin{align}
\%Diff = \frac{C_{Approximation} - C_{Analytic}}{C_{Analytic}} \cdot 100 \nonumber
\end{align}
I will use this definition of the approximation error for all models considered in this thesis and not only for the Heston model. If an analytic solution is not available I use prices obtained via Monte Carlo Simulation like analytic prices in the above expression.\\
Firstly I use the same parameters as KM did in their initial example, which are also roughly the same as in \citet{Heston93}. Such that $\kappa = 2.00$, $\theta = 0.04$, $\omega = 0.10$, $\rho = -0.5$ and $r = 0.10$. The spot volatility is $v(t) = 0.04$, the strike price is set to $100$ and time to maturity is one year. In Figure \ref{KM1A} the Black-Scholes volatility is set equal to $\eta_0 = \sqrt{v(t)}$, such that the resulting figure replicates fig. 3 in KM. For Figure \ref{KM1B} the nuisance parameter is set equal to  $\eta_0 = \sqrt{\theta}$. In both cases the approximation is convergent, achieving sufficient precession after only five corrective terms. As already indicated before choosing the nuisance parameter as in Figure \ref{KM1A} appears to be superior to the alternative. Note that setting $\eta_0 = \sqrt{v(t)}$ implies for the initial corrective term $\delta_0 = 0$. Hence, in Figure \ref{KM1A} the $N = 0$ line effectively shows just the percentage difference between the Black-Scholes and the Heston price. This is not the case in Figure \ref{KM1B}, where $\eta_0 = \sqrt{\theta} \neq \sqrt{v(t)}$. Since in this case the approximation looses accuracy, this loss seems to be related to the occurrence of the $\delta_0$ term in the series expansion.

\begin{figure}[H]
	\centering
	\begin{subfigure}{.7\textwidth}
		\centering
		\includegraphics[width=.7\linewidth]{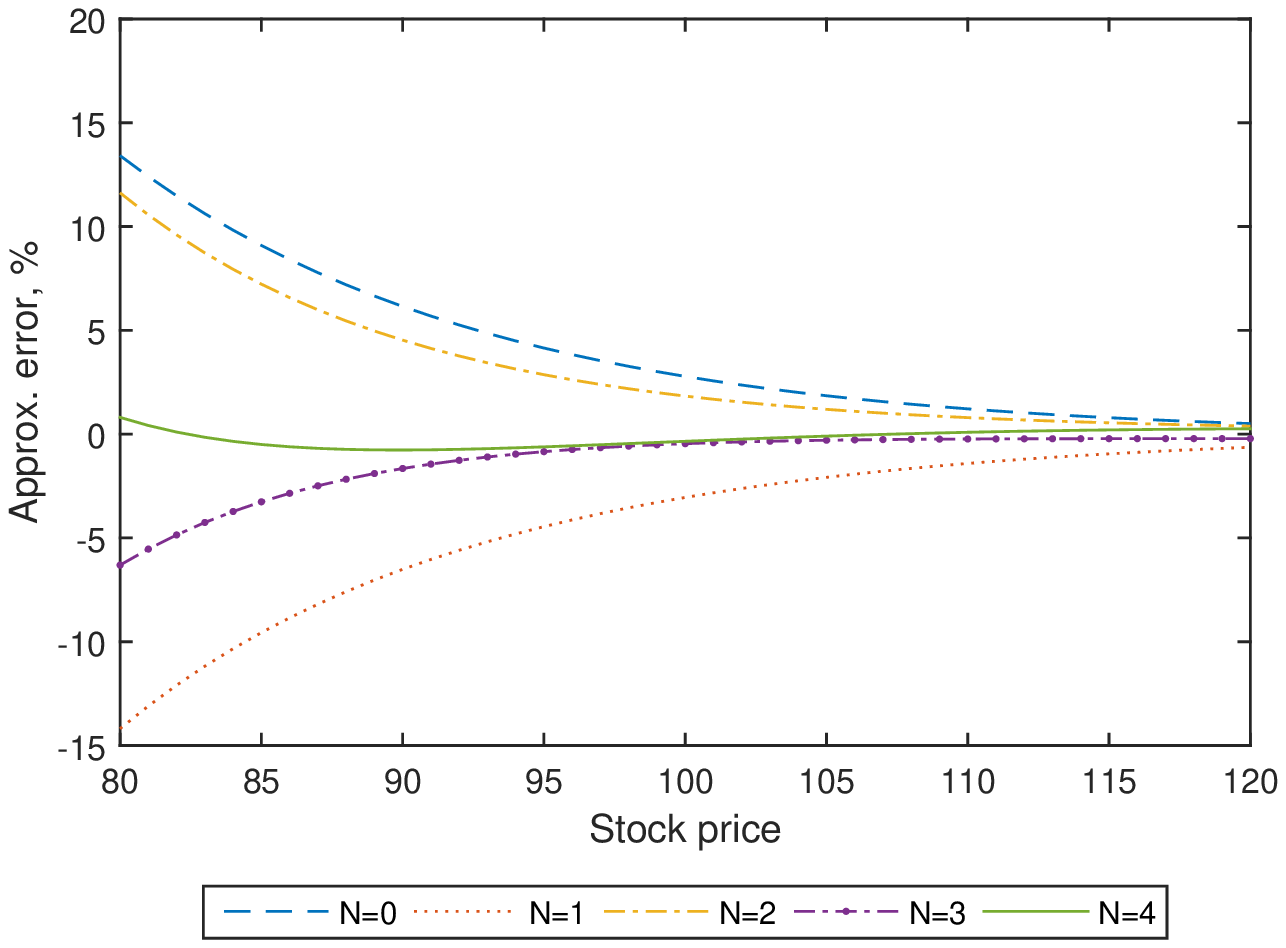}
		\caption{Time to maturity 1.5 years}
		\label{HestLONG1}
	\end{subfigure}\\
	\begin{subfigure}{.7\textwidth}
		\centering
		\includegraphics[width=.7\linewidth]{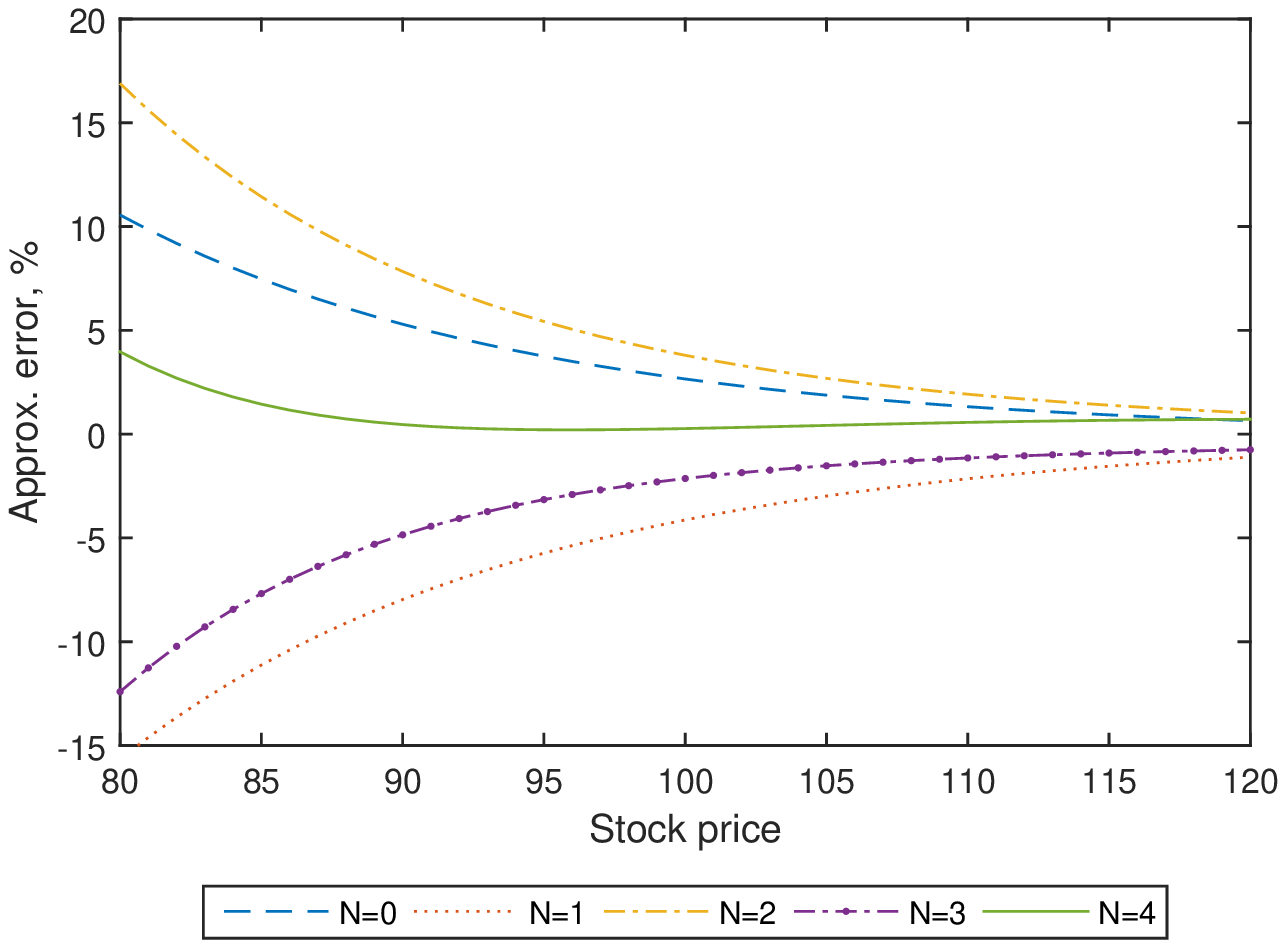}
		\caption{Time to maturity 2.0 years}
		\label{HestLONG2}
	\end{subfigure}
	\begin{subfigure}{.7\textwidth}
		\centering
		\includegraphics[width=.7\linewidth]{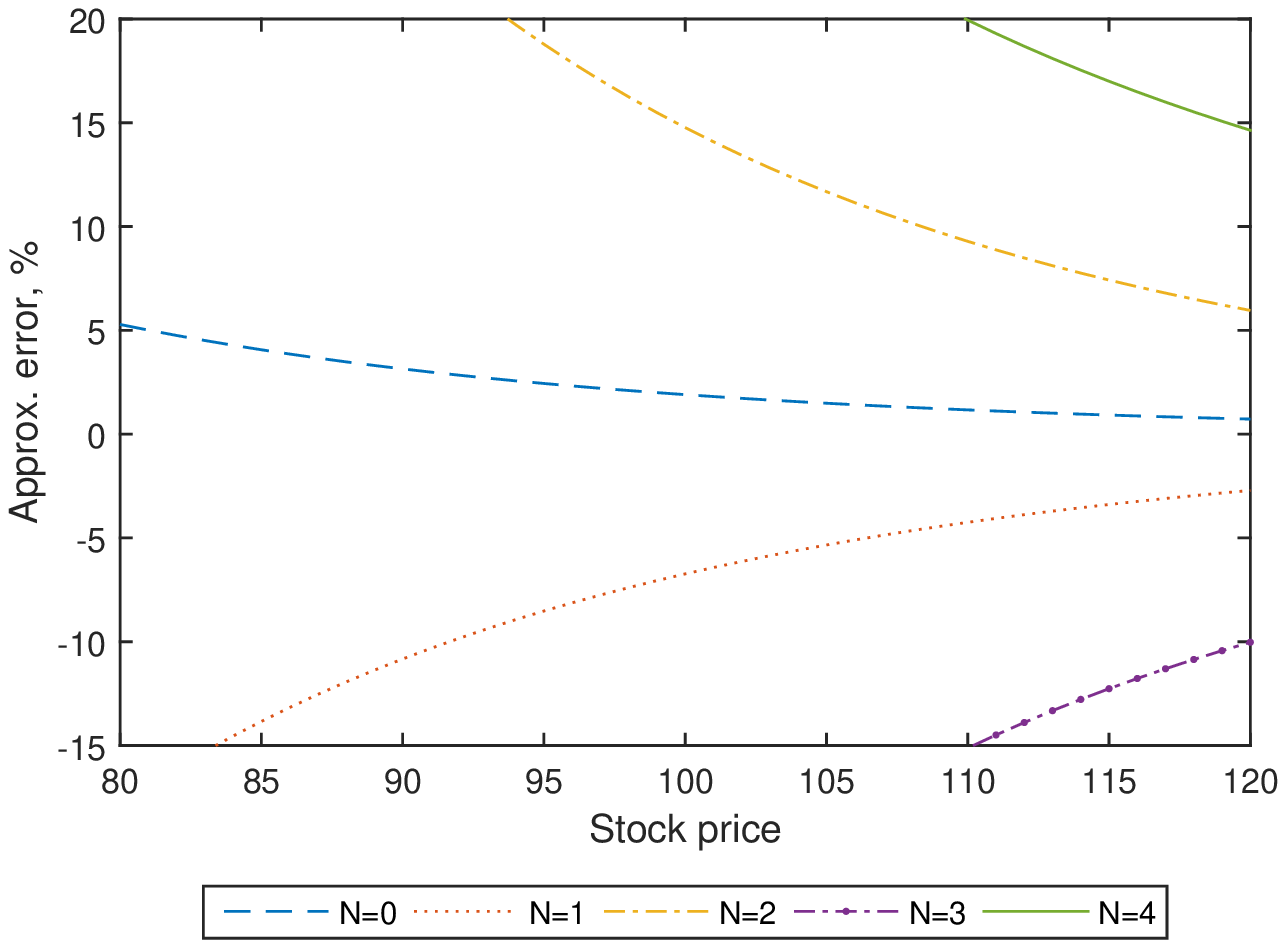}
		\caption{Time to maturity 4.0 years}
		\label{HestLONG3}
	\end{subfigure}
	\caption{KM approximation of the \citet{Heston93} model}
	\label{fig.HestLONG}
	\floatfoot{\textit{ Analytic prices had been computed via Fourier transforms. $N$ = numer of corrective terms. Strike price = 100, $\kappa = 2.00$, $\theta = 0.04$, $\omega = 0.10$, $\rho = -0.5$, $r = 0.10$, $v(t) = 0.04$. All B-S parameters that coincide with the \citet{Heston93} parameters are set equal.}}
\end{figure}

\newpage
Note that in both cases there is a range of stock prices for which the approximation involving four corrective terms performs slightly better than the one using five. However, the maximum absolute error is lowest for $N = 4$. Hence, I will use $\eta_0 = \sqrt{v(t)}$ and $N = 4$ throughout this section. Figures \ref{HestLONG1} to \ref{HestLONG3} show the errors of the same approximation for different maturities. These examples show an interesting pattern. If time to maturity is increased the accuracy of all approximations except the $N = 0$ approximation reduces. For a very long maturity of four years the convergence even seems to turn into divergence, since the $N = 0$ approximation yields the most precise result. Recall that the nuisance parameter of the approximation was set such that the $N = 0$ approximation consists only of the Black-Scholes price. When analyzing volatility smiles in options data it is a well documented fact that the smile is far more pronounced for short maturities than for long maturities, i.e. the volatility smile tends to flatten with increasing volatility. Since the Black-Scholes model predicts a completely flat "smile" the true option price will be closer to the Black-Scholes price for long maturities compared to short maturities. This suggests that KM's approach, which simply adds non-zero corrective terms to the Black-Scholes price in Figure \ref{HestLONG3}, gets increasingly imprecise with increasing time to maturities. Hence, convergence of KM's series expansion might not be uniform over time. This issue had been also detected in KM for the approximation of prices of long-run bonds. However, in case of equity or FX options such long maturities are rare. 


\newpage
Figure \ref{fig.TimeMoney} below shows how the accuarcy of the approximation changes when time to maturity and moneyness are varied, whereas now also maturities below one year are considered. As expected the performance of the approximation increases for lower maturities. The pattern that the approximation of ITM (\textit{in-the-money}, moneyness > 1) is better than the approximation of OTM (\textit{out-of-the-money}, moneyness < 1) calls remains over all maturities. Interestingly for very long maturities the approach overestimates the call price while for the shortest maturity the price is underestimated. It should be noted that, while for maturities below one year the approximation error of far-ITM calls virtually becomes zero, it remains slightly above zero for all longer maturities. Overall Figure \ref{fig.TimeMoney} suggests a high accuracy of the approximation for most maturities.

\begin{figure}[H]
	\centering
	\includegraphics[scale=.9]{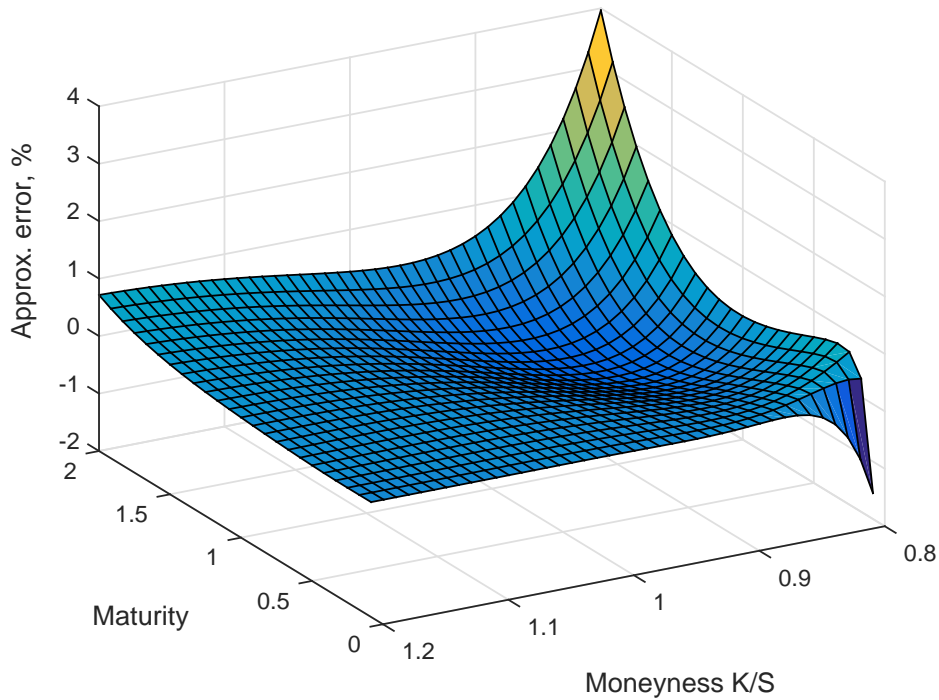}
	\caption{Effect of maturity and moneyness on KM's expansion}
	\floatfoot{\textit{Notes:  Analytic prices had been computed via Fourier transforms. KM's approximation uses five corrective terms. $\kappa = 2.00$, $\theta = 0.04$, $\omega = 0.10$, $\rho = -0.5$, $r = 0.10$, $v(t) = 0.04$. All B-S parameters that coincide with the \citet{Heston93} parameters are set equal.}}
	\label{fig.TimeMoney}
\end{figure}

\begin{figure}[H]
	\centering
	\begin{subfigure}{.9\textwidth}
		\centering
		\includegraphics[width=.9\linewidth]{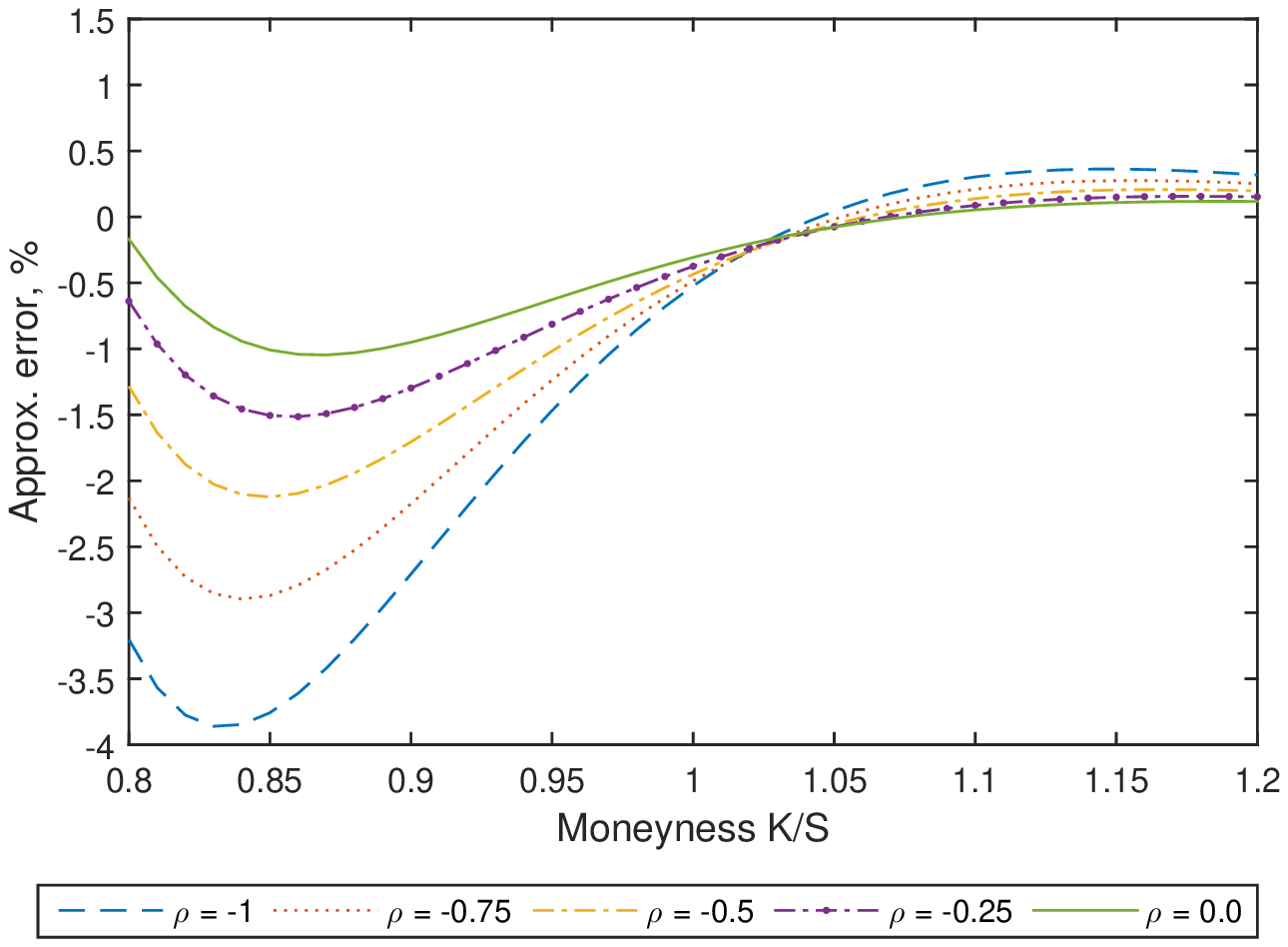}
		\caption{Time to maturity one year}
		\label{HestCorr3d1}
	\end{subfigure}\\
	\begin{subfigure}{.9\textwidth}
		\centering
		\includegraphics[width=.9\linewidth]{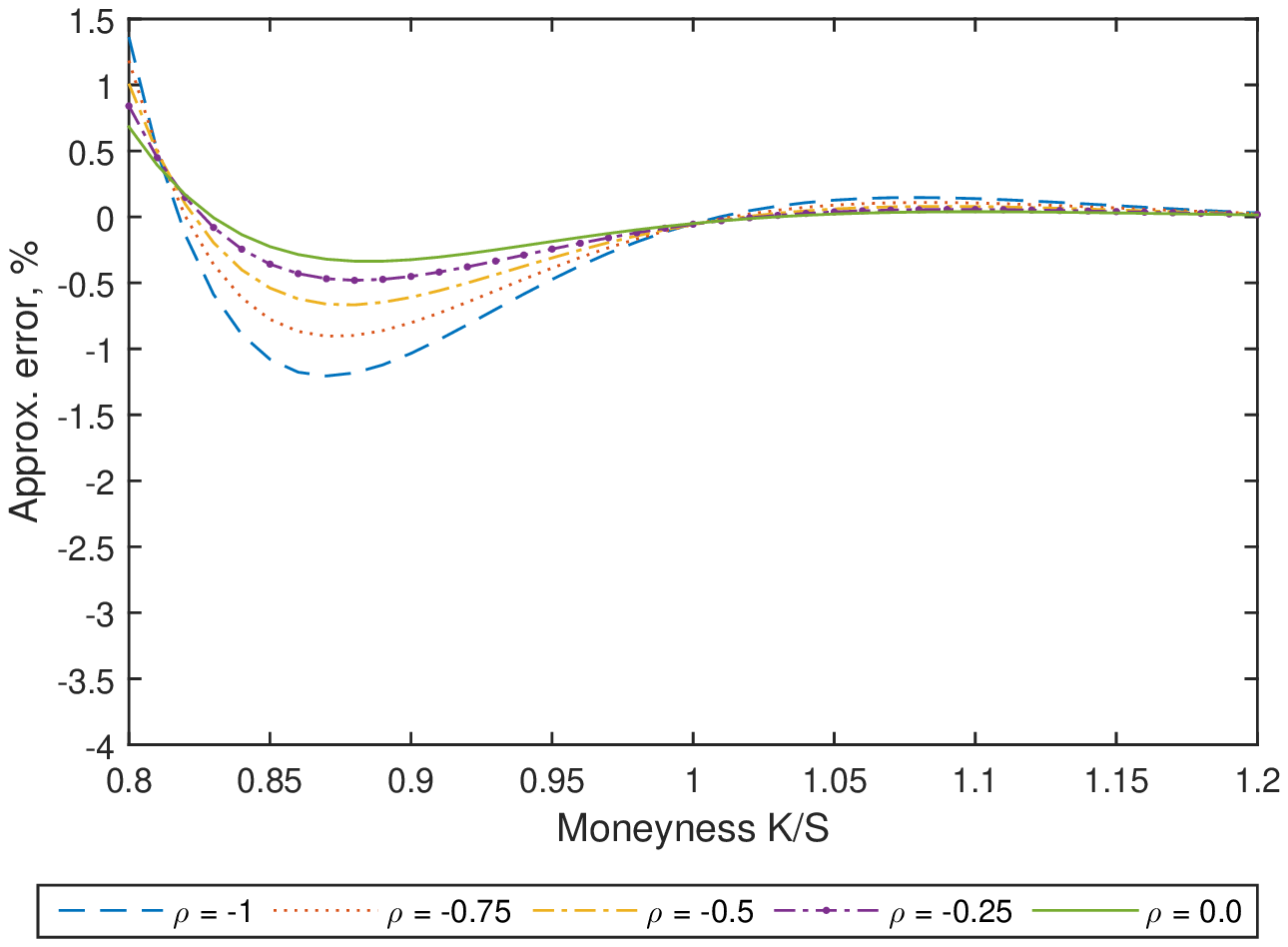}
		\caption{Time to maturity 0.5 year}
		\label{HestCorr3d2}
	\end{subfigure}
	\caption{Influence of correlation on KM's approximation}
	\label{fig.HestCorr3d}
	\floatfoot{\textit{Notes: Analytic prices had been computed via Fourier transforms. KM's approximation uses five corrective terms. Time to maturity = 1 year (Panel A) or = 0.5 year (Panel B), $\kappa = 2.00$, $\theta = 0.04$, $\omega = 0.10$, $r = 0.10$, $v(t) = 0.04$. All B-S parameters that coincide with the \citet{Heston93} parameters are set equal.}}
\end{figure}

\begin{figure}[H]
	\centering
	\includegraphics[scale=.9]{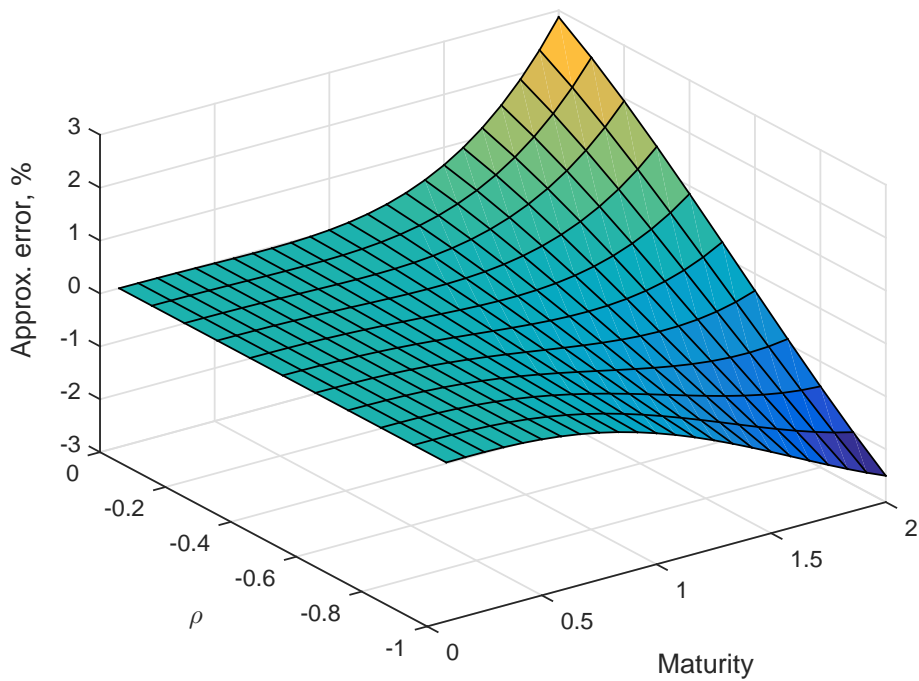}
	\caption{Effect of maturity and correlation on KM's expansion}
	\floatfoot{\textit{Notes: Analytic prices were obtained via Fourier inversion. KM's approximation uses five corrective terms. $\kappa = 2.00$, $\theta = 0.04$, $\omega = 0.10$, $r = 0.10$, $v(t) = 0.04$. All Black-Scholes parameters that coincide with the \citet{Heston93} parameters are set equal.}}
	\label{fig.TimeCorr}
\end{figure}

Next I analyze the effect of correlation between the stock price and variance process. In all previous figures the coefficient of correlation $\rho$ was set equal to $-0.5$. Figure \ref{fig.HestCorr3d} shows KM's approximation of the Heston model when varying $\rho$ as well as the stock price simultaneously, whereas the number of corrective terms is fixed to five (i.e. $N = 4$). Panel (a) shows the result for a maturity of one year, Panel (b) for a maturity of 0.5 year. Both plots show that the approximation loses precision fast for OTM calls as correlation increases. With increasing correlation accuracy of the approximation of OTM calls in Figure \ref{HestCorr3d2} is significantly reduced. However, the precision of the approximation of ITM calls appears to be almost unaffected by correlation. Figure \ref{HestCorr3d1} clearly shows a lower overall accuracy of the approximation for a longer time to maturity. There correlation always affects accuracy over the range of different moneyness. Nevertheless, a similar pattern as in Figure \ref{HestCorr3d2} applies. For low levels of correlation the difference in accuracy between ITM and OTM calls is far less pronounced than for high correlations. Overall, Figures \ref{fig.TimeMoney} and \ref{fig.HestCorr3d} show that KM's approximation is strongly affected by correlation, moneyness and time to maturity. Since Figure \ref{fig.HestCorr3d} also indicates a connection between time to maturity and correlation, Figure \ref{fig.TimeCorr} shows explicitly the effect on the approximation error of KM's expansion for an ATM call (\textit{at-the-money}, moneyness = 1), if these two quantities are varied. For a time to maturity blow 6 months the level of correlation has no significant effect on the accuracy. However, for longer times to maturity correlations between $0.0$ and $-0.5$ lead to an overestimation of the call price, while correlations between $-0.5$ and $-1.0$ lead to an underestimation. The highest approximation errors in either case are reached at the extreme ends of the considered spectrum of correlation and the longest considered time to maturity of two years.  

All figures that appeared in this section so far had in common that the accuracy of KM's approximation had it's maximum absolute error in the range of OTM options. The errors increase fast the deeper the option is in the OTM region and decreased as the option price moves in the ITM region. Hence, it seems like the moneyness of the option has a significant influence on the accuracy of the approximation. For short maturities and/or low levels of correlation the approximation appears to be quite accurate over a wide range of moneyness. However, if maturity and/or correlation increases precision deteriorates strongest for OTM options.\\  
KM also analyze their approximation of the Heston model for another set of parameters. Specifically KM used the following values: Strike $= 1000$, $T-t= 1/12$, $\kappa = 0.1465$, $\theta = 0.5172$, $\omega = 0.5786$, $\rho = -0.0243$, $r = 0.00$, $v(t) = \theta$ and $\gamma = 0.5$. These parameters had been estimated by \citet{Boll02} in the context of FX options.\footnote{As also observed by \citet{Younesian} this parameter choice violates the Feller condition, since $4\kappa\theta / v = 0.5860 < 2$. However, \citet{Cizek11} note that this commonly happens when calibrating the Heston model to market data in the FX context. It merely implies that the variance process hits the zero boundary recurrently but leaves it immediately, i.e. the time spent at the boundary is zero. (See \citet{Cizek11}, p. 144). Hence, the violation of the Feller condition should not affect the results of KM's approximation.} The results of KM's approximation can be seen in Table \ref{tab.Hest1}, which replicates the results already shown in KM. Panel A shows how spot moneyness influences the accuracy of the approximation, whereas Panel B shows the influence of different values for the spot variance $v(t)$ for an ATM call. Again accuracy of the approximation is very high in both Panels. The extreme high accuracy in this case might also be related to the very low level of correlation and the short time to maturity. The previous examples indicated that the both is likely to improve the approximation. 

\begin{table}[H]
	\centering
	\caption{Approximating the \citet{Heston93} model}
	\begin{tabular}{c|c c c}
		\hline
		\multicolumn{4}{c}{\textbf{Panel A:}} \\
		\textbf{Stock price}&\textbf{Fourier Transform}&\textbf{KM approx.}&\textbf{\%Diff}\\\hline
		\textbf{ 950} & 57.8425&57.8449&0.00418\\
		\textbf{ 960}&62.3711&62.3738&0.0042574\\
		\textbf{ 970}&67.1005&67.1033&0.0042447\\
		\textbf{ 980}&72.0291&72.0321&0.0041553\\
		\textbf{ 990}&77.1553&77.1584&0.0040021\\
		\textbf{1000}&82.4766&82.4797&0.003797\\
		\textbf{1010}&87.9903&87.9934&0.0035513\\
		\textbf{1020}&93.6933&93.6964&0.003275\\
		\textbf{1030}&99.5822&99.5852&0.0029773\\
		\textbf{1040}&105.6532&105.656&0.0026663\\
		\textbf{1050}&111.9021&111.9048&0.0023492\\\hline
	\end{tabular}
	\begin{tabular}{c|c c c}
		\multicolumn{4}{c}{\textbf{Panel B:}} \\
		~~~\textbf{$v(t = 0)$}~~~~~~~&\textbf{Fourier Transform}&\textbf{KM approx.}&\textbf{\%Diff}\\\hline
		\textbf{0.1}&36.4488&36.4854&0.10045\\
		\textbf{0.2}&51.4125&51.4255&0.025319\\
		\textbf{0.3}&62.8997&62.9068&0.011276\\
		\textbf{0.4}&72.5792&72.5838&0.0063472\\
		\textbf{0.5}&81.1007&81.104&0.0040628\\
		\textbf{0.6}&88.7981&88.8006&0.002821\\
		\textbf{0.7}&95.8702&95.8721&0.002072\\
		\textbf{0.8}&102.4465&102.4481&0.0015857\\
		\textbf{0.9}&108.6171&108.6184&0.0012524\\
		\textbf{  1}&114.4477&114.4488&0.001014\\
		\textbf{1.1}&119.9878&119.9888&0.00083766\\\hline
	\end{tabular}
	\floatfoot{\textit{Comparsion of European Call option prices under the dynamics of the \citet{Heston93} model. The Fourier transform prices had been obtained via numerical integration using Matlab's \texttt{integral()} function. KM's approximation uses the Black-Scholes model as baseline model. Panel A: Accuracy for differing spot moneyness of the option. Strike price = 1,000. Panel B: At-the-money options (S(t) = strike = 1,000) for differing spot variance. In both Panles: Time to maturity $= 1/12$, $\kappa = 0.1465$, $\theta = 0.5172$, $\omega = 0.5786$, $\rho = -0.0243$, $r = 0.00$, $v(t) = \theta$ and $\gamma = 0.5$. The volatility for the Black-Scholes model is $\sigma_0 = \sqrt{v(t)}$. All B-S parameters that coincide with the \citet{Heston93} parameters are set equal.}}
	\label{tab.Hest1}
\end{table}

\subsubsection{Approximating Put Prices}
\begin{figure}[H]
	\centering
	\begin{subfigure}{.7\textwidth}
		\centering
		\includegraphics[width=.7\linewidth]{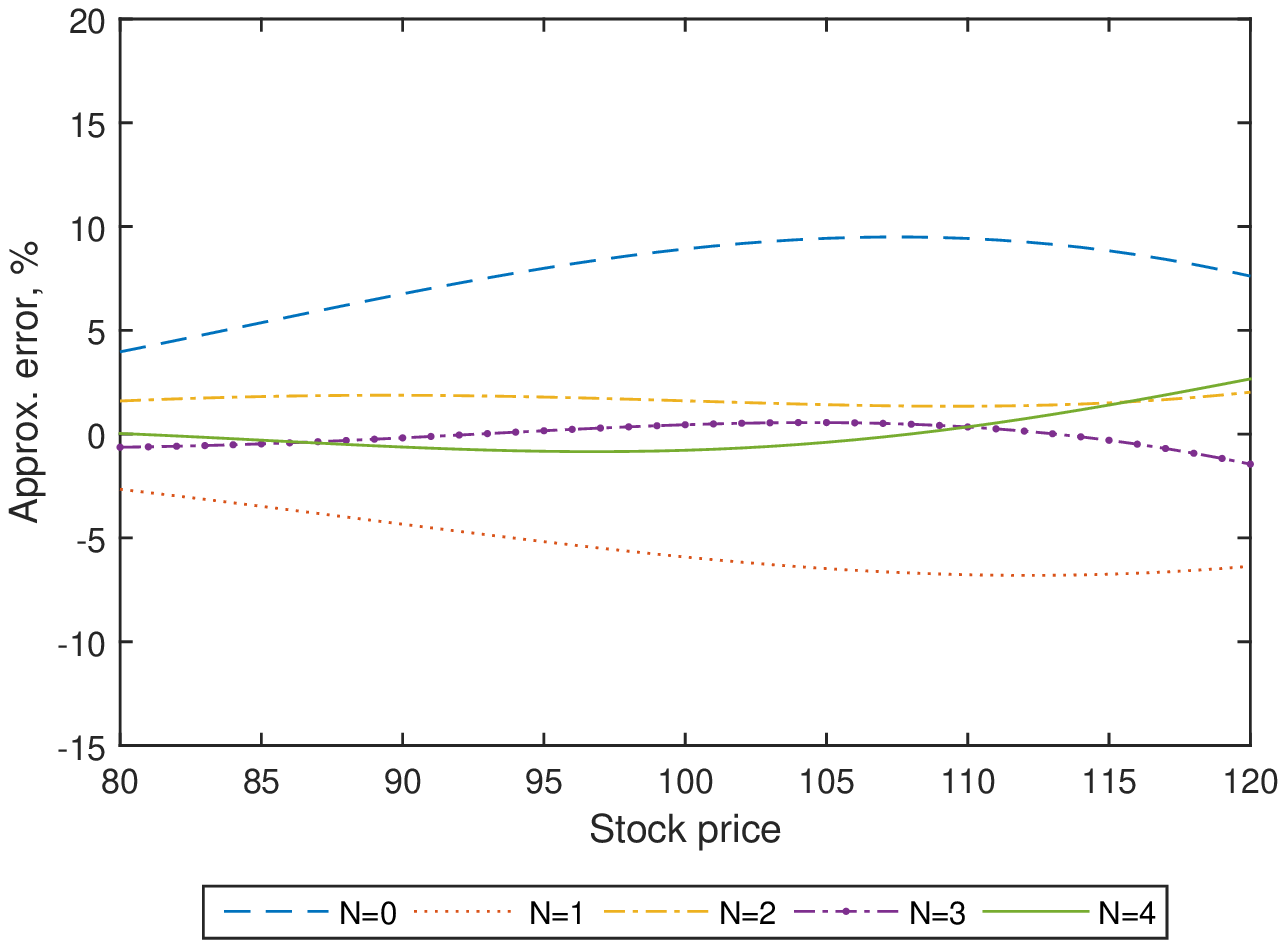}
		\caption{Time to maturity one year}
		\label{HestPut1}
	\end{subfigure}\\
		\begin{subfigure}{.7\textwidth}
			\centering
			\includegraphics[width=.7\linewidth]{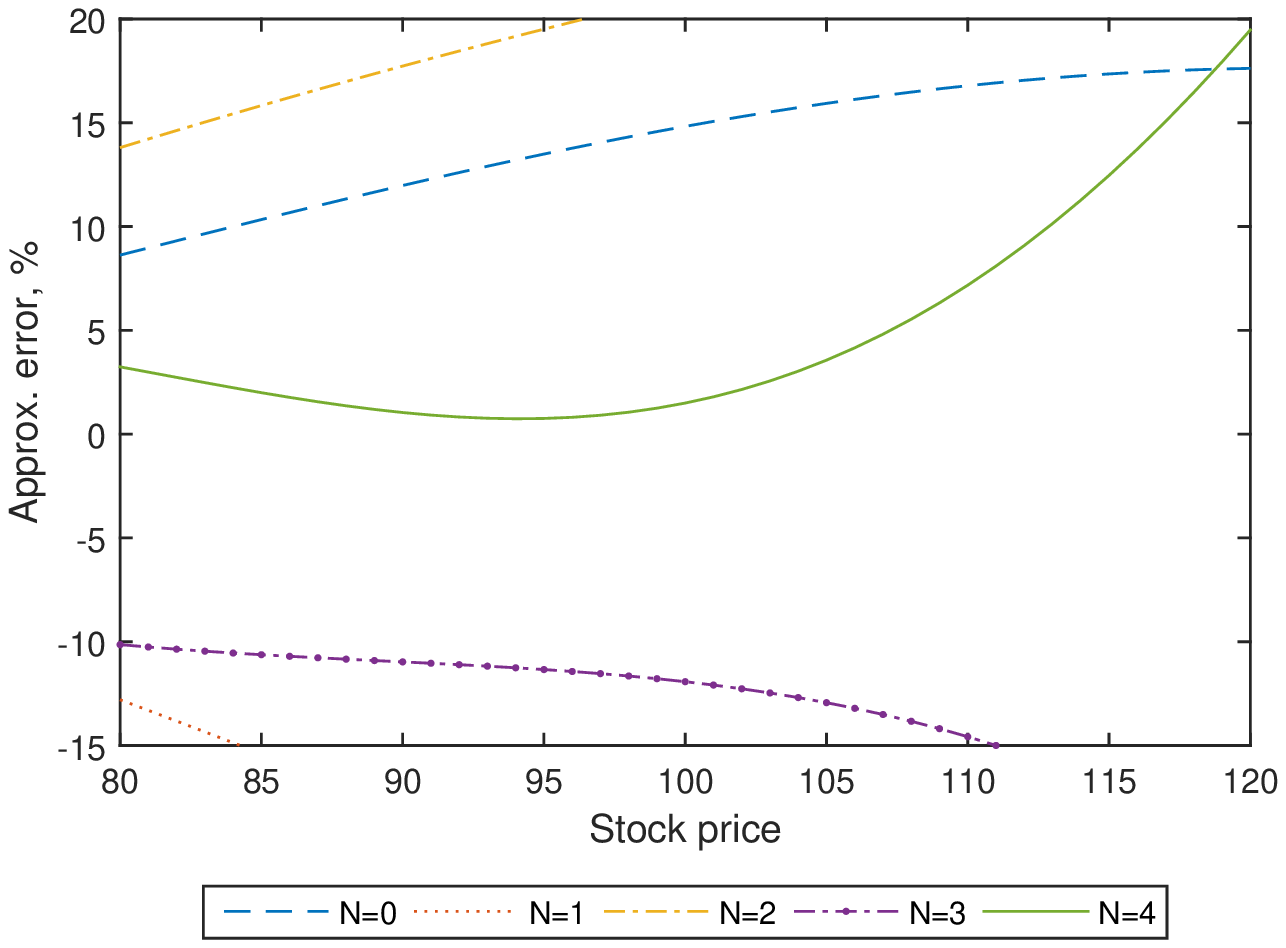}
			\caption{Time to maturity two year}
			\label{HestPut2}
		\end{subfigure}
	\begin{subfigure}{.7\textwidth}
		\centering
		\includegraphics[width=.7\linewidth]{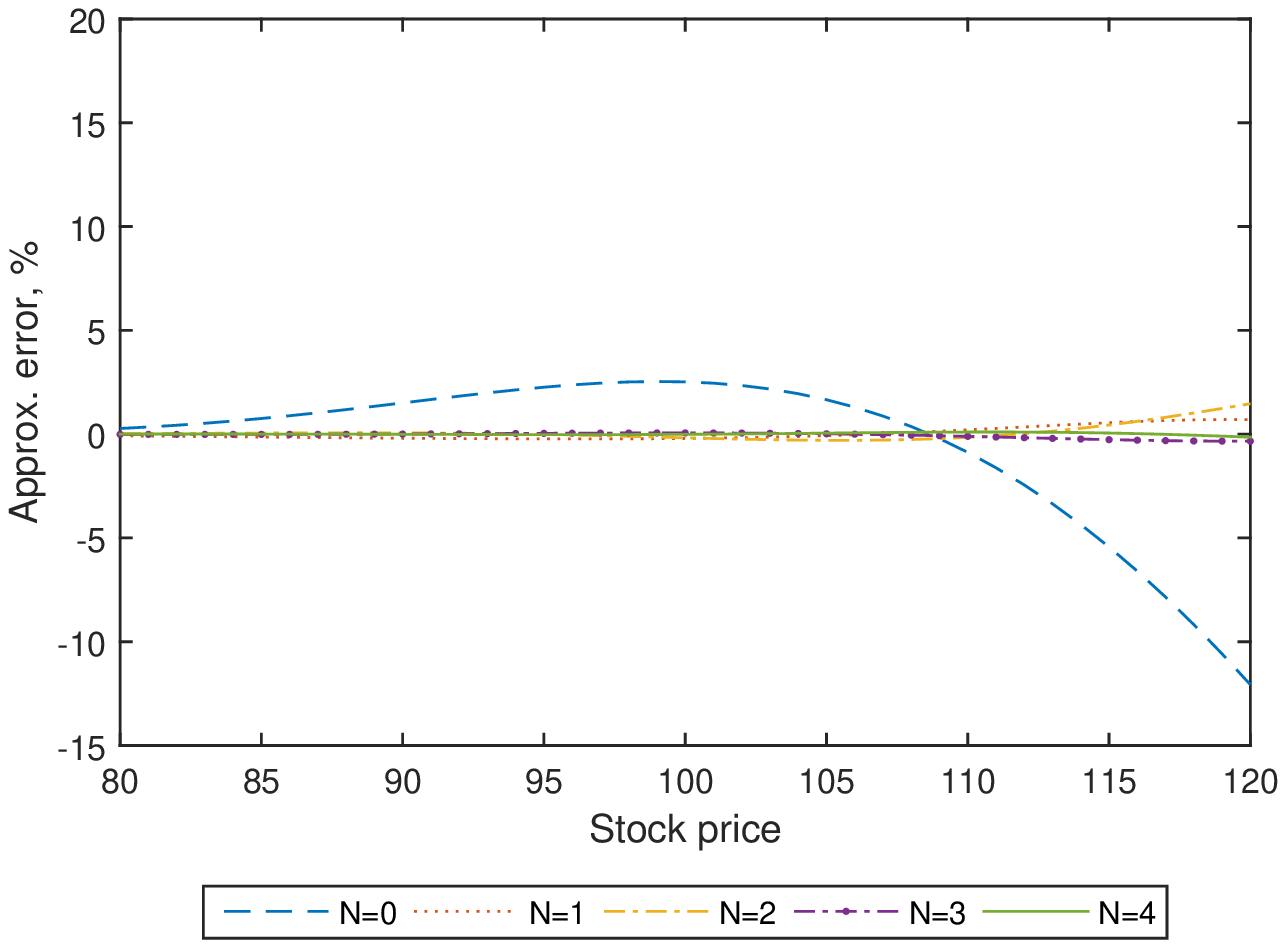}
		\caption{Time to maturity 0.25 year}
		\label{HestPut3}
	\end{subfigure}
	\caption{KM approximation of European Put option in the \citet{Heston93} model}
	\label{fig.HestPut}
	\floatfoot{\textit{Notes: Convergance behavior of the KM approximation for a European Put. True model is \citet{Heston93}, baseline model is the Black-Scholes model. Percentage errors are in pct. of the analytic Heston price. Analytic prices had been computed via Fourier transforms and numerical integration using Matlab's \texttt{integral()} function. $N$ = numer of corrective terms. Strike price = 100, time to maturity = 1 year in Panel (a) and 0.25 year in Panel (b), $\kappa = 2.00$, $\theta = 0.04$, $\omega = 0.10$, $\rho = -0.5$, $r = 0.10$, $v(t) = 0.04$ and $\gamma = 0.5$. The nuisance parameter is set to $\sigma_0 = \sqrt{v(t)}$.}}
\end{figure}

Independent from the chosen parameter values the approximation always performed comparatively poor for OTM calls and comparatively good for ITM calls. In the following I will analyze whether this effect also appears in the case of European Put options. KM do not provide an example for the approximation of a put option. However, the series expansion of a put is straightforward to derive. Recall the definitions of the corrective terms $d(\cdot)$ and $\delta_n(\cdot)$ in the series expansion in (\ref{CorrecHest}). $d(\cdot)$ was defined as the difference between the final payoffs of the true model and the baseline model. If one uses a Black-Scholes put option as baseline model to approximate a European put option in the Heston framework, the term $d(\cdot)$ remains equal to zero. Recall that it is necessary to choose a baseline model with an identical final payoff, since the payoff function of a plain vanilla option is non-differentiable. The other corrective term $\delta_n$ accounts for the difference in the driving forces of the true and the baseline market. Technically this correction is built on the difference of the coefficients in the two model's systems of SDEs and the derivatives of the baseline model (see Table \ref{tab.Hestit}). \\
The initial pricing error and all of the higher order corrective terms given in Table \ref{tab.Hestit} remain unchanged if one switches from call to put options. In order to see this recall the following: The coefficients are the same in either case and the derivatives of the baseline model appearing in the corrective terms are also the same. In the Black-Scholes model only the first derivative in stock price direction differs between call and put, while the second derivative is identical in both cases. In the initial pricing error $\delta_0$ only the second derivative of the Black-Scholes formula appears and all other derivatives of the baseline model that are used in the approximation are taken from this second derivative. Hence, the whole series of corrective terms $\delta_n$ is identical for a call and put option.\\
Hence, one can switch from call to put prices in the KM approximation simply by using the Black-Scholes price of a Put option as the first element in the series expansion in (\ref{CorrecHest}) and leaving the rest unchanged. This result can be checked by applying the put-call parity relationship\footnote{See e.g. \citet{Hull15}, p. 242.} $C(S,t) + Ke^{-r(T-t)} = Put(S,t) + S_0$ to the call prices approximated via KM's series expansion, i.e. Put prices obtained from Call prices through put-call parity should be identical to the Put prices obtained through the series expansion using the Black-Scholes Put prices as its first element. Applying both approaches I could confirm that this is indeed true.\\
Figure \ref{HestPut1} shows the convergence of the KM approximation for a Put option. Regarding the parameter values I again use the values from the first part of the last section, i.e. $\kappa = 2.00$, $\theta = 0.04$, $\omega = 0.10$, $\rho = -0.5$, $r = 0.10$, $v(t) = 0.04$, with a strike of 100 and a time to maturity of one year. Overall, each of the KM approximations show lower absolute maximum errors than the approximation of the call price with the same number of corrective terms. Each additional corrective term appears to increase the accuracy of the approximation. However, as before there is a range of option prices for which the expansion including four terms performs slightly better than the one including five terms. The difference in accuracy in the approximation of the OTM and ITM options is less clear in the case of a put option. While the effect, that the precision of the approximation of OTM options is higher than for ITM options, is present in all the approximations except for the $N = 2$ expansion, it is far less pronounced than for the Call option. Figures \ref{HestPut2} and \ref{HestPut3} repeat the analysis for a time to maturity of two and $0.25$ years respectively. As expected the accuracy increases significantly for the short maturity and decreases for the long maturity. Also the effect of better approximations of ITM options becomes more pronounced. As before, in the long maturity case the convergence pattern starts to dissolve.


\subsubsection{Approximating Greeks in the Heston model} \label{HestGreeks}
So far I analyzed the performance of KM's series expansion in approximating the price of European call and put options in the framework of the Heston model. Next, I will assess the accuracy of the KM approximation for the computation of hedge ratios in the Heston framework.\\ 
In order to compute the KM approximation of the hedge ratios I use the series expansion in (\ref{GrApprox}) and the finite difference approximation to compute the derivatives therein. I calculate theses greeks using the same set of parameters as for the computation of Table \ref{tab.Hest1}. This example is identical to the one conducted in KM. However, KM do not explain whether they used numerical or analytic approximations to obtain their series expansion values. By using finite differences I am able to obtain the same values for all Greeks as reported in KM for the case of varying monyness. For the case of varying spot volatility, both my approximations as well as my analytic hedge ratios deviate slightly from the values reported in KM. The results can been seen in Table \ref{tab.HestDelta} and \ref{tab.HestGamma}. As I represent the options $\Delta$ in $\Gamma$ in percent the approximation errors are reported in percentage points.\\
Table \ref{tab.HestVega} shows the result when approximating $\mathscr{V}$ by KM's series expansion. The approximation errors are given as absolute deviations here. For $\mathscr{V}$ I obtain the most significant deviation of my results from KM's, when the spot volatility is varied (see Panel B). This is somewhat surprising since I obtained exactly KM's results for varying moneyness, using the same method of computation in both cases.\\
Note that in the case of $\mathscr{V}$ there is no corresponding Greek of the baseline model. Even though the value of the Black-Scholes volatility parameter is set equal to spot volatility parameter in the Heston model, the variables are not identical. This was also emphasized when deriving the initial pricing error in (\ref{Hestd0}), where the corresponding derivatives of the Black-Scholes model were considered to be equal to zero. Hence, when considering the series expansion in (\ref{GrApprox}) the first element representing the derivative of the baseline model equals zero in the approximation of $\mathscr{V}$.

\begin{table}[H]
	\centering
	\caption{Approximating Heston's $\Delta$}
	\begin{tabular}{c|c c c}
		\hline
		\multicolumn{4}{c}{\textbf{Panel A:}} \\
		\textbf{Stock price}&\textbf{Fourier Transform}&\textbf{KM approx.}&\textbf{Diff., pp.}\\\hline
		\textbf{ 950}&44.2794&44.2819&-0.0025864\\
		\textbf{ 960}&46.2918&46.294&-0.0021588\\
		\textbf{ 970}&48.2928&48.2945&-0.0016931\\
		\textbf{ 980}&50.2776&50.2788&-0.001201\\
		\textbf{ 990}&52.2414&52.2421&-0.00069418\\
		\textbf{1000}&54.18&54.1801&-0.00018398\\
		\textbf{1010}&56.0893&56.089&0.00031874\\
		\textbf{1020}&57.9657&57.9649&0.00080396\\
		\textbf{1030}&59.8058&59.8046&0.0012626\\
		\textbf{1040}&61.6066&61.6049&0.0016869\\
		\textbf{1050}&63.3654&63.3633&0.00207\\\hline
	\end{tabular}
	\begin{tabular}{c|c c c}
		\multicolumn{4}{c}{\textbf{Panel B:}} \\
		~~~\textbf{$v(t = 0)$}~~~~~~~&\textbf{Fourier Transform}&\textbf{KM approx.}&\textbf{Diff., pp.}\\\hline
		\textbf{0.1}&51.9512&51.9516&-0.00044321\\
		\textbf{0.2}&52.6614&52.6509&0.010469\\
		\textbf{0.3}&53.2189&53.2149&0.0040575\\
		\textbf{0.4}&53.6929&53.6917&0.0012105\\
		\textbf{0.5}&54.1121&54.1121&0.0000348\\
		\textbf{0.6}&54.492&54.4932&-0.0012002\\
		\textbf{0.7}&54.8416&54.8419&-0.00021102\\
		\textbf{0.8}&55.1673&55.1588&0.0085098\\
		\textbf{0.9}&55.4732&55.4419&0.031343\\
		\textbf{  1}&55.7625&55.69&0.072555\\
		\textbf{1.1}&56.0376&55.9066&0.13101\\
	\end{tabular}
	\floatfoot{\textit{Notes: Comparsion of $\Delta_S$ for European Call option under the dynamics of the \citet{Heston93} model. All Greek are in pct, differences in percentage points. The Fourier transform prices had been obtained via numerical integration using Matlab's \texttt{integral()} function. Greeks for the KM approx. are obtained via finite differences. Panel A: Differing spot moneyness of the option. Strike price = 1,000. Panel B: At-the-money options (S(t) = strike = 1,000) for differing spot variance. In both Panles: Time to maturity $= 1/12$, $\kappa = 0.1465$, $\theta = 0.5172$, $\omega = 0.5786$, $\rho = -0.0243$, $r = 0.00$, $v(t) = \theta$. The volatility for the Black-Scholes model is $\sigma_0 = \sqrt{v(t)}$. All B-S parameters that coincide with the \citet{Heston93} parameters are set equal.}} \label{tab.HestDelta}
\end{table}

\begin{table}[H]
	\centering
	\caption{Approximating Heston's $\Gamma$}
	\begin{tabular}{c|c c c}
		\hline
		\multicolumn{4}{c}{\textbf{Panel A:}} \\
		\textbf{Stock price}&\textbf{Fourier Transform}&\textbf{KM approx.}&\textbf{Diff., pp.}\\\hline
		\textbf{ 950}&0.20165&0.20161&4.0526e-05\\
		\textbf{ 960}&0.20076&0.20071&4.4782e-05\\
		\textbf{ 970}&0.19937&0.19932&4.8141e-05\\
		\textbf{ 980}&0.1975&0.19745&5.0149e-05\\
		\textbf{ 990}&0.19519&0.19514&5.0908e-05\\
		\textbf{1000}&0.19246&0.19241&5.0701e-05\\
		\textbf{1010}&0.18935&0.1893&4.9582e-05\\
		\textbf{1020}&0.18588&0.18583&4.7187e-05\\
		\textbf{1030}&0.18209&0.18205&4.4327e-05\\
		\textbf{1040}&0.17802&0.17798&4.0407e-05\\
		\textbf{1050}&0.1737&0.17366&3.6158e-05\\\hline
	\end{tabular}
	\begin{tabular}{c|c c c}
		\multicolumn{4}{c}{\textbf{Panel B:}} \\
		~~~\textbf{$v(t = 0)$}~~~~~~~&\textbf{Fourier Transform}&\textbf{KM approx.}&\textbf{Diff., pp.}\\\hline
		\textbf{0.1}&0.44642&0.38533&0.061096\\
		\textbf{0.2}&0.31234&0.30392&0.0084156\\
		\textbf{0.3}&0.25395&0.25273&0.0012173\\
		\textbf{0.4}&0.21935&0.21918&0.0001669\\
		\textbf{0.5}&0.1958&0.19575&5.1236e-05\\
		\textbf{0.6}&0.17844&0.17843&1.3833e-05\\
		\textbf{0.7}&0.16496&0.1652&-0.00023702\\
		\textbf{0.8}&0.15409&0.15448&-0.00038083\\
		\textbf{0.9}&0.14509&0.1436&0.0014905\\
		\textbf{  1}&0.13748&0.1273&0.010179\\
		\textbf{1.1}&0.13092&0.096136&0.034784\\\hline
	\end{tabular}
	\floatfoot{\textit{Notes: Comparsion of $\Gamma_S$ for European Call option under the dynamics of the \citet{Heston93} model. All Greek are in pct, differences in percentage points. The Fourier transform prices had been obtained via numerical integration using Matlab's \texttt{integral()} function. Greeks for the KM approx. are obtained via finite differences. Panel A: Differing spot moneyness of the option. Strike price = 1,000. Panel B: At-the-money options (S(t) = strike = 1,000) for differing spot variance. In both Panles: Time to maturity $= 1/12$, $\kappa = 0.1465$, $\theta = 0.5172$, $\omega = 0.5786$, $\rho = -0.0243$, $r = 0.00$, $v(t) = \theta$. The volatility for the Black-Scholes model is $\sigma_0 = \sqrt{v(t)}$. All B-S parameters that coincide with the \citet{Heston93} parameters are set equal.}} \label{tab.HestGamma}
\end{table}

	\begin{table}
		\centering
		\caption{Approximating Heston's $\mathscr{V}$}
		\begin{tabular}{c|c c c}
		\hline
		\multicolumn{4}{c}{\textbf{Panel A:}} \\
		\textbf{Stock price}&\textbf{Analytic $\mathscr{V}$}&\textbf{KM approx.}&\textbf{Diff, abs.}\\
		\textbf{ 950}&74.9687&74.9679&0.00076581\\
		\textbf{ 960}&76.221&76.2212&-0.00023367\\
		\textbf{ 970}&77.2834&77.2847&-0.0013377\\
		\textbf{ 980}&78.1538&78.1563&-0.0025286\\
		\textbf{ 990}&78.8316&78.8354&-0.0037851\\
		\textbf{1000}&79.3178&79.3229&-0.005083\\
		\textbf{1010}&79.6148&79.6212&-0.0063961\\
		\textbf{1020}&79.7259&79.7336&-0.0076966\\
		\textbf{1030}&79.6561&79.6651&-0.0089566\\
		\textbf{1040}&79.4111&79.4213&-0.010148\\
		\textbf{1050}&78.9977&79.0090&-0.0112\\\hline
		\end{tabular}
		\begin{tabular}{c|c c c}
			\multicolumn{4}{c}{\textbf{Panel B:}} \\
			~~~\textbf{$v(t = 0)$}~~~~~~~&\textbf{Fourier Transform}&\textbf{KM approx.}&\textbf{Diff., abs.}\\\hline
			\textbf{0.1}&180.4329&151.6085&28.8244\\
			\textbf{0.2}&127.7884&122.9584&4.8299\\
			\textbf{0.3}&104.3134&103.4803&0.83308\\
			\textbf{0.4}&90.279&90.1519&0.12705\\
			\textbf{0.5}&80.6826&80.6861&-0.0035699\\
			\textbf{0.6}&73.5874&73.5304&0.056988\\
			\textbf{0.7}&68.0654&67.8674&0.19792\\
			\textbf{0.8}&63.6084&63.6146&-0.0061472\\
			\textbf{0.9}&59.9122&61.4241&-1.512\\
			\textbf{  1}&56.7816&62.6834&-5.9018\\
			\textbf{1.1}&54.0853&69.5145&-15.4293\\\hline
		\end{tabular}	
		\floatfoot{\textit{Notes: Comparsion of $\mathscr{V}$ for European Call option under the dynamics of the \citet{Heston93} model.  The Fourier transform prices had been obtained via numerical integration using Matlab's \texttt{integral()} function. Greeks for the KM approx. are obtained via finite differences. Panel A: Differing spot moneyness of the option. Strike price = 1,000. Panel B: At-the-money options (S(t) = strike = 1,000) for differing spot variance. In both Panles: Time to maturity $= 1/12$, $\kappa = 0.1465$, $\theta = 0.5172$, $\omega = 0.5786$, $\rho = -0.0243$, $r = 0.00$, $v(t) = \theta$. The volatility for the Black-Scholes model is $\sigma_0 = \sqrt{v(t)}$. All B-S parameters that coincide with the \citet{Heston93} parameters are set equal.}}\label{tab.HestVega}
	\end{table}

\clearpage
\section{Approximating the CEV model} \label{sec.CEVexpansion}
In this section I will continue to follow the path of KM and apply their series expansion method to another model of stochastic volatility. This also had been done by KM themselves, such that I also in this section intend to start my numerical analysis by replicating KM's results. However, as in the previous case of the Heston model, I will also consider some own examples.\\
The \textit{constant elasticity of variance} (CEV) model\footnote{I use the same nomenclature as in KM. However, in the literature the term CEV is often used to describe models in which variance is given by a deterministic function e.g. of the stock price.} was first introduced by \citet{chan} to model stochastic interest rates and had been further analyzed in the context of stochastic volatility, among others, by \citet{Jones03} or \citet{Lewis00} and is defined by the following system of SDEs 
\begin{align}
dS(t) ~=&~ rSdt + \sqrt{v(t)}SdW_1(t) \label{CEV1}\\
dv(t) ~=&~ \kappa\left(\theta - v(t)\right)~dt + \omega|v(t)|^\gamma~ dW_2(t) \label{CEV2} \\
dW_1(t)dW_2(t) ~=&~ \rho dt \nonumber
\end{align}
Where $\kappa$ denotes the speed of mean-reversion parameter, $\theta$ and $\omega$ the long-run variance and the diffusion coefficient respectively and $r$ the constant instantaneous short-term rate. $\gamma$ denotes the CEV parameter. The representation in (\ref{CEV1}) and (\ref{CEV2}) is the same as in KM. However, usually the variance process includes $v(t)^\gamma$ instead of $|v(t)|^\gamma$. KM do not point out their reasons to choose this version of the CEV model, but it does not influence the series expansion in any way. The CEV model appears to be a generalization of the Heston model. By setting $\gamma = 0.5$ one obtains the square root process in (\ref{HestSDE2}).\\
Despite the popularity of the Heston model, cases of the CEV model with $\gamma \neq \frac{1}{2}$ are of empirical relevance as shown in \citet{Jones03}. While in the Heston model the volatility of instantaneous volatility is not level dependent, the volatility of instantaneous of variance appears to be level dependent.\footnote{See \citet{Jones03}} \citet{Jones03} concludes that, level dependence appears to be a feature of the data and hence the CEV model might perform better in replicating the observed volatility smiles. Hence, the process in (\ref{CEV2}) with choices for $\gamma \neq 1/2$ might be a sensible alternative to Heston's specification. Indeed, by analyzing large samples of daily return data from the S\&P 500 index with 3537 observations, \citet{Jones03} estimated values of $\gamma$ of $1.33$. As reported therein such values, indicating a high elasticity of variance, are consistent with other findings in empirical finance.\footnote{See the discussion in \citet{Jones03}, p. 196.} Using a more current data set on S\&P 500 call index options \citet{Younesian} estimates a CEV parameter of approximately $0.6$. Note that whereas \citet{Jones03} relies on Monte Carlo simulations and a Baysian estimation approach to obtain his estimates, \citet{Younesian} estimates the parameters directly from cross-sections of option prices using non-linear least square estimation together with KM's closed-form approximation formula. However, \citet{Younesian} notes that the estimate of the CEV parameter might seriously flawed through difficulties in the correct estimation of the correlation coefficient which he linked to KM's series approximation.\\
Since there are no analytic solutions available for any case of the CEV model other than $\gamma = \frac{1}{2}$, I will rely on Monte Carlo simulations to obtain reference values for the option prices estimated by the KM approach. Hence, all approximation errors reported in this section can be understood as percentage or absolute deviation from the price obtained via Monte Carlo (MC).

\subsection{KM's expansion for the CEV model} \label{CEVKM}
Since the CEV model is a simple generalization of the Heston model, it is straightforward to derive KM's series expansion for this model. One can simply repeat the steps taken in the case of the Heston model, just replacing the $\sqrt{v(t)}$ term by $v(t)^{\gamma}$ in all equations in section \ref{HestonApproxKM}. Following the approach of KM I again use the Black-Scholes model as baseline model. As I again intend to approximate European call options under the CEV model and also use calls in the Black-Scholes model, the corrective terms adjusting the boundary are equal to zero.\\ 
Applying (\ref{initdel})  yields an initial pricing error that is identical to the one obtained for the Heston model
\begin{align}
\delta_0(S,t) = \dfrac{1}{2}\left(v(t) - \eta_0\right)S^2\dfrac{\partial^2C^{BS}}{\partial S^2} \label{CEVd0}
\end{align} 
Where (\ref{CEVd0}) again results because $\partial C^{BS}/\partial v = \partial^2C^{BS}/\partial v^2 = \partial^2C^{BS}/\partial S\partial v = 0$. Below Table \ref{tab.CEVit} shows how the series of pricing errors develop with the order of the approximation, starting from (\ref{CEVd0}) up to any positive integer $N$. Comparing Table \ref{tab.CEVit} with Table \ref{tab.Hestit} shows that the only difference in the corrective terms lies in the variance term $v(t)$, which is taken to the power of $\gamma$ instead of $1/2$. The resulting series expansion is the same as in (\ref{CorrecHest}), just using the corrective terms form Table \ref{tab.CEVit} instead.\\
With respect to the assumptions underlying KM's approximation it might be interesting to investigate one property of the CEV process more carefully. Specifically, if $\gamma > 1$ then $v \rightarrow \infty$ and thus the growth condition fails to hold.\footnote{See \citet{Jones03}, p. 187.} Recalling the brief summary of the assumptions on which KM's approach is built on, the growth condition was mentioned in the context of the Feynman-Kac representation of the solution to the PDE which determines the difference between the true and the baseline model (see section \ref{AssetApprox}). However, this condition was only necessary to ensure existence of a solution to the SDEs of the true model. Hence, it would be sufficient to prove existence differently if the growth condition is violated. In order to show the existence of a solution to the CEV SDEs I follow closely the prove in \citet[Appendix C]{Jones03}.\footnote{\citet{Jones03} shows a proof of the existence of a solution but uses a different formulation of the CEV process. Hence, I am basically redoing the steps in \citet{Jones03} using the formulation of the CEV model as it is used in this thesis.}\\
Note that the price process of the CEV model essentially is a geometric Brownian motion. For the case of constant variance the existence of a solution to this kind of SDE had been proven elsewhere in the literature. For the present case of stochastic volatility \citet{Jones03} shows that a solution exists if there exists a solution to the SDE determining the stochastic behavior of the variance. Hence, in order to show that a solution to the CEV model exists and thus the  Feynman-Kac representation of the solution to the PDE determining the pricing bias in KM's approximation is well defined, it is sufficient to show the existence of a solution to (\ref{CEV2}).
To do so it is sufficient to show that the scale measure of (\ref{CEV2}) is unattainable in 0 and $+\infty$.\footnote{See \citet{Jones03}, p. 217 or the sources cited therein.} Whereby the scale measure of (\ref{CEV2}) is
\begin{align}
\Omega(v) &= \int_{m}^{n} \Theta(v)  dv \label{CEVscale}\\
\text{with } \Theta(v) &= exp\left(\frac{2\kappa\theta}{\omega^2\left(2\gamma - 1\right)}\frac{1}{v^{2\gamma - 1}} - \frac{\kappa}{\omega^2\left(\gamma - 1\right)}\frac{1}{v^{2\gamma - 2}}\right) \nonumber
\end{align}
Appendix \ref{AppendixScale} includes a brief derivation of the above scale measure.\\
\citet{Jones03} states that the upper bound $+\infty$ is unattainable if $\Omega(v) = \int_{m}^{+\infty} \Theta(v)  dv  = +\infty$. For $\gamma > 1$ this is true, since then $\frac{1}{v^{2\gamma - 1}} \rightarrow 0$ and $\frac{1}{v^{2\gamma - 2}} \rightarrow 0$ such that $\Theta(v) \rightarrow 1$ as $v \rightarrow \infty$. Hence, the integral fails to converge what causes $+\infty$ to be unattainable.\\
Accordingly \citet{Jones03} denotes the lower bound 0 as unattainable if $\Omega(v) = \int_{0}^{n} \Theta(v)  dv  = +\infty$. Note that for $\gamma > 1$, $\Theta(v) \approx exp\left(\frac{2\kappa\theta}{\omega^2\left(2\gamma - 1\right)}\frac{1}{v^{2\gamma - 1}}\right)$ as $v \rightarrow 0$.\footnote{See \citet{Jones03}, p. 216.} If additionally to $\gamma > 1$ also $\kappa > 0$ and naturally $\theta > 0$, then $\Xi \equiv \frac{2\kappa\theta}{\omega^2\left(2\gamma - 1\right)} > 0$ holds true. For small values of $v$ this implies the following relation\footnote{See \citet{Jones03}, p. 216.}
\begin{align}
exp\left(\frac{2\kappa\theta}{\omega^2\left(2\gamma - 1\right)}\frac{1}{v^{2\gamma - 1}}\right) = exp\left(\frac{\Xi}{v^{2\gamma - 1}}\right) > exp\left(\frac{\Xi}{v}\right) > \frac{\Xi}{v}
\end{align}
Since $1/v \rightarrow +\infty$ as $v \rightarrow 0$ the above inequality indicates that also $\Theta(v) \rightarrow +\infty$ as $v \rightarrow 0$.
Hence, for the scale measure of the CEV variance process both 0 and $+\infty$ are not attainable and thus a solution to (\ref{CEV2}) exists, such that the existence of a solution to the SDEs determining the CEV model can be ensured even though the growth condition is violated for $\gamma > 1$. Hence, at least the first step of KM's approximation, i.e. the Feynman-Kac representation of the solution of the PDE which the pricing error obeys, should be well defined.

\newpage
\begin{landscape}
	\newcolumntype{C}{>{\centering\arraybackslash} m{18cm} } 
	\renewcommand{\arraystretch}{2.0}
\begin{table}[H]
	\centering
	\caption{Iterations of the pricing error for the CEV model}
	\begin{tabular}{m{1cm}|C}
		\hline
		\textbf{n} & \textbf{Pricing error $\delta_n(S,t)$} \\\hline\hline
		\textbf{0} & $\dfrac{1}{2}\left(v - \sigma_0\right)S^2\dfrac{\partial^2C^{BS}}{\partial S^2}$ \\[5pt]\hline
		\textbf{1} & $\dfrac{\partial \delta_0}{\partial t} + rS\dfrac{\partial \delta_0}{\partial S} + \kappa\left(\theta - v\right)\dfrac{\partial \delta_0}{\partial v} + \frac{1}{2}vS^2\dfrac{\partial^2 \delta_0}{\partial S^2} + \rho \omega S v(t)^{\gamma + \frac{1}{2}}\dfrac{\partial^2\delta_0}{\partial S\partial v} - r\delta_0$~~~ (*)\\[5pt]\hline
		\textbf{2} & $\dfrac{\partial \delta_1}{\partial t} + rS\dfrac{\partial \delta_1}{\partial S} + \kappa\left(\theta - v\right)\dfrac{\partial \delta_1}{\partial v} + \frac{1}{2}\left(vS^2\dfrac{\partial^2 \delta_1}{\partial S^2} + \omega^2v^{2\gamma}\dfrac{\partial^2\delta_1}{\partial v^2}\right) + \rho \omega S v^{\gamma + \frac{1}{2}}\dfrac{\partial^2\delta_1}{\partial S\partial v} - r\delta_1$\\[5pt]\hline
		\textbf{$\vdots$} & $\vdots$ \\[5pt]\hline
		\textbf{N} & $\dfrac{\partial \delta_{N-1}}{\partial t} + rS\dfrac{\partial \delta_{N-1}}{\partial S} + \kappa\left(\theta - v\right)\dfrac{\partial \delta_{N-1}}{\partial v} + \frac{1}{2}\left(vS^2\dfrac{\partial^2 \delta_{N-1}}{\partial S^2} + \omega^2v^{2\gamma}\dfrac{\partial^2\delta_{N-1}}{\partial v^2}\right) + \rho \omega S v^{\gamma + \frac{1}{2}}\dfrac{\partial^2\delta_{N-1}}{\partial S\partial v} - r\delta_{N-1}$ \\[5pt]\hline
	\end{tabular} \\
	\tiny{\textit{(*) Note that $\partial^2 \delta_0/\partial v^2 = 0$.}}
	\label{tab.CEVit}
\end{table}
\end{landscape}
\newpage

 
\subsection{Numerical accuracy}
\begin{figure}[H]
	\centering
	\begin{subfigure}{.7\textwidth}
		\centering
		\includegraphics[width=.9\linewidth]{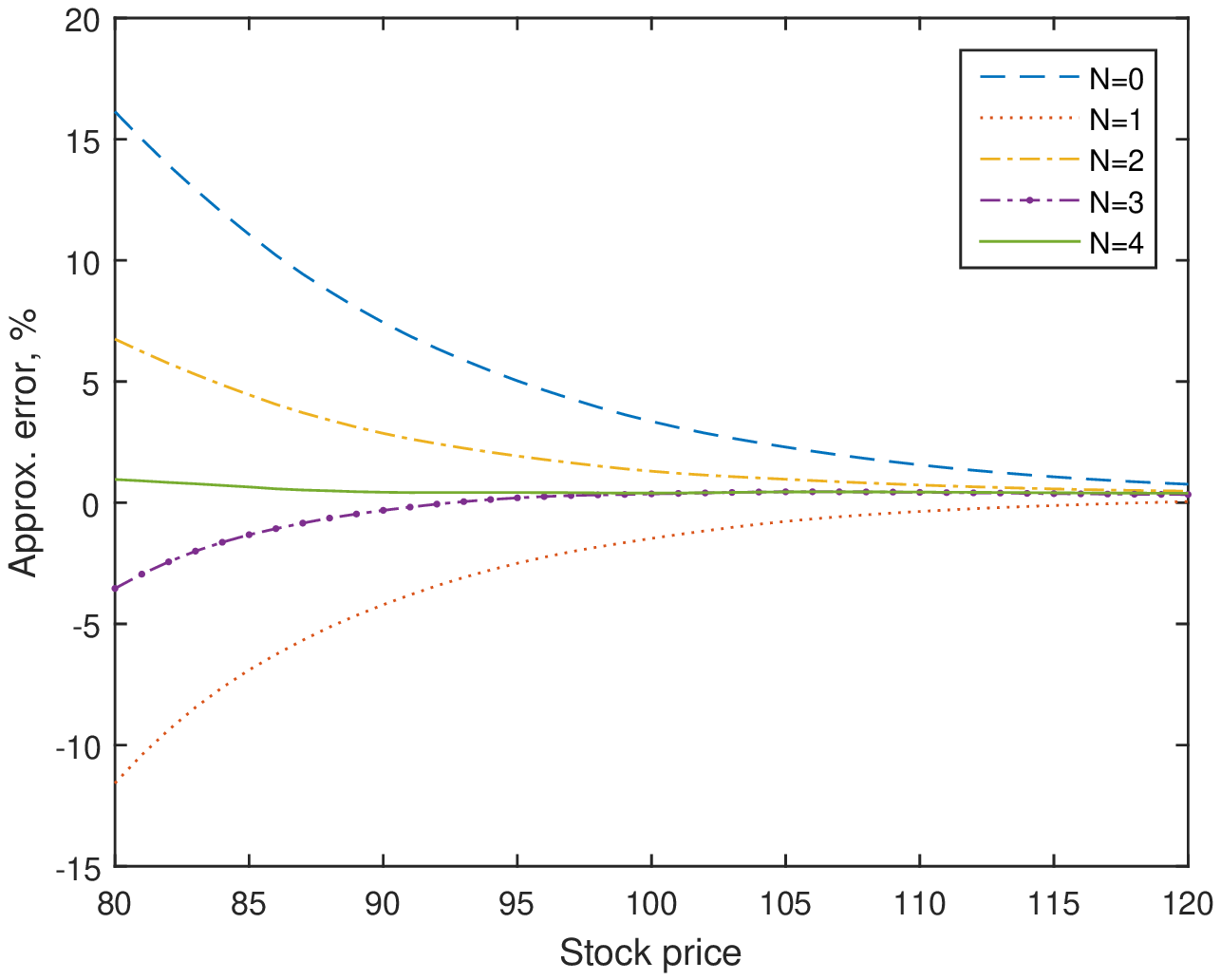}
		\caption{CEV with $\gamma = 0.6$}
		\label{CEV1A}
	\end{subfigure}\\
	\begin{subfigure}{.7\textwidth}
		\centering
		\includegraphics[width=.9\linewidth]{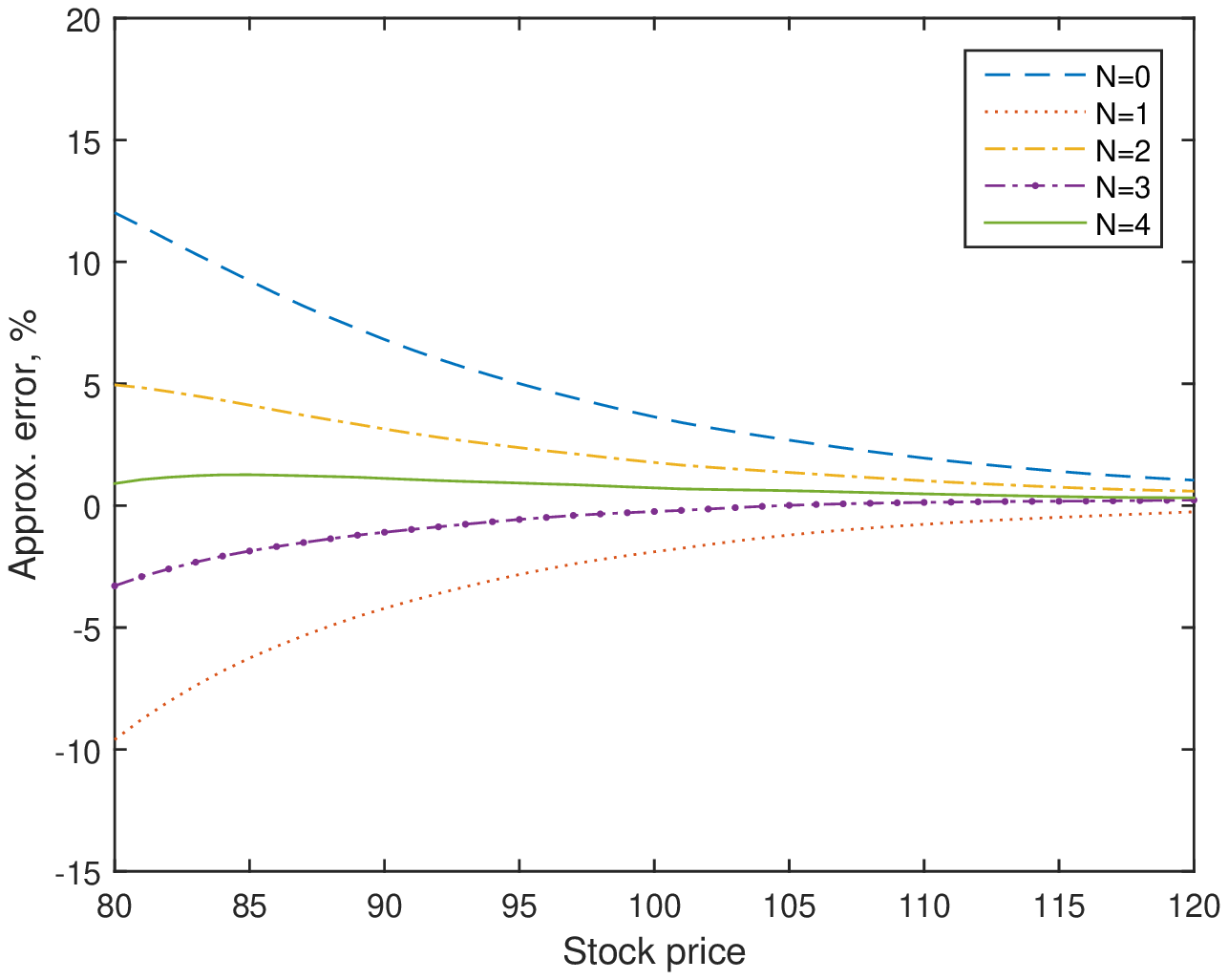}
		\caption{CEV with $\gamma = 1.33$}
		\label{CEV1B}
	\end{subfigure}
	\caption{Different CEV parameter values}
	\label{fig.CEV1}
	\floatfoot{\textit{Notes: Approximation of the CEV model. Panel (a): $\gamma = 0.6$, Panel (b): $\gamma = 1.33$. In both panels: $\theta = 0.04$, $\omega = 0.10$, $\rho = -0.5$, $r = 0.1$ and the spot variance $v = 0.05$, time to maturity is one year and the strike price is 100. Black-Scholes volatility is $\sqrt{v}$. Reference values are obtained via Monte Carlo simulation using the specification in Appendix \ref{CEVMCRef}.}}
\end{figure}

Figure \ref{fig.CEV1} shows KM's approximation of a European call in the CEV model for the two CEV parameter choices  $\gamma = 0.6$ (Panel a) and $\gamma = 1.33$ (Panel b). Time to maturity is one year and the strike price is set to 100. The other parameters are: $\theta = 0.04$, $\omega = 0.10$, $\rho = -0.5$, $r = 0.1$ and the spot variance $v = 0.05$, time to maturity is one year and the strike price is 100. The Black-Scholes volatility is $\sqrt{v}$. As in the previous section, $N$ denotes the order of the approximation. Since there is no analytic solution to the CEV model for these two choices of the CEV parameter, the reference values are obtained through Monte Carlo simulation. Both panels in Figure \ref{fig.CEV1} clearly show that KM's series expansion converges to the Monte Carlo solution. The approximation for $\gamma = 1.33$ appears to be slightly more precise than for $\gamma = 0.6$. Overall the convergence pattern pretty much resembles the pattern already observed for the Heston model. For both cases of the CEV parameter the approximation improves with each new corrective term, such that for both values of $\gamma$ clearly approximations of order $N = 4$ should be used. The results indicate a slightly higher accuracy for $\gamma = 1.33$. However, the error in both panels seems to not completely die out in the ITM region as this was the case for the Heston model.

KM analyze the accuracy of their approximation for the CEV model for $\gamma = 0.6$ and use the set of parameters estimated by \citet{Boll02} for FX options. Table \ref{tab.CEV1} below attempts to replicate parts of \citet[Table 2]{KM2011}. Time to maturity is one month, $\kappa = 0.1465$, $\theta = 0.5172$, $\omega = 0.5786$, $\rho = -0.0243$, $r = 0.00$, and $v(t) = \theta$. The strike of the Call option is set to 1,000. I also include 95\% confidence intervals in the table which are not reported in KM. Both prices, the MC as well as the KM prices, differ from the ones reported therein.  KM mention that they design their MC simulation in a way such that the MC price obtained with $\gamma = 0.6$ can not deviate more than 0.5\% from the analytic Heston price, i.e. from the price obtained via Fourier inversion with the same set of parameters but $\gamma = 0.5$.\footnote{See \citet{KM2011}, p. 401.} Nevertheless, when comparing the option prices reported in \citet[Table 1]{KM2011} and \citet[Table 2]{KM2011}, it turns out that the  MC prices in their Table 2 deviate from the Fourier prices in their Table 1 by more than 0.5\%. Additionally, the procedure of fixing the MC prices to the Fourier results might be questionable. Even though the parameter values are close, there are structural consequence on the resulting model. As mentioned already before, while the Heston model (i.e. $\gamma = 0.5$) is affine, the general CEV case with $\gamma = 0.6$ yields a non-affine model. Further, the barrier for the deviation of 0.5\% seems arbitrary. Hence, I will not follow KM's suggestion to restrict the MC prices. Consequently, the prices based on the MC simulation may deviate from the Heston prices in another way than reported in KM. However, KM's MC prices lie inside the confidence intervals of my own MC results.\\
The results obtained via KM's series expansion deviate from the values reported in \citet[Table 2]{KM2011}. The program which I employed to obtain the approximation in Table \ref{tab.CEV1} is the same as for Table \ref{tab.Hest1}. Since there the results matched the values reported in KM exactly, the deviation is surprising. While KM's original results overestimate their MC prices constantly by around 1\%, my own results constantly underestimate my MC prices. However, my MC and KM approximation prices differ by only about -0.55\%. Nevertheless, the conclusion would remain unchanged if I would use KM's original results. For the non-affine CEV case KM's series expansion seems to yield less precise approximations than it did for the affine Heston model.\\ 
Table \ref{tab.CEV2} uses the same parameter values as Table \ref{tab.CEV1}, but now setting $\gamma$ to $1.33$ in order to analyze the case of high elasticity of variance. The difference to the MC price remains below 1\%, but now the series expansion constantly yields prices that are higher than the MC prices. The approximation errors also show now an unexpected pattern. They increase when the option moves from OTM to ITM. This seems to be related to specific parameter values and the high elasticity of variance.  

\begin{table}[H]
	\centering
	\caption{Approximating the CEV model - $\gamma = 0.6$}
	\begin{tabular}{c|c c c c c}
		\hline
		\multicolumn{6}{c}{\textbf{Panel A:}} \\
		\textbf{Stock price}&\textbf{Monte Carlo}&\textbf{Conf. Lower}&\textbf{Conf. Upper}&\textbf{KM approx.}& \textbf{\% Diff}\\\hline
		\textbf{ 950}&58.178&56.5888&59.7672 & 57.8674&-0.53393\\\hline
		\textbf{ 960}&62.7283&61.0752&64.3814& 62.3967&-0.52869\\\hline
		\textbf{ 970}&67.4865&65.7693&69.2036& 67.1266&-0.5333\\\hline
		\textbf{ 980}&72.4486&70.6673&74.2298& 72.0555&-0.54255\\\hline
		\textbf{ 990}&77.6149&75.7696&79.4602& 77.1817&-0.55814\\\hline
		\textbf{1000}&82.9715&81.0622&84.8809& 82.5029&-0.56483\\\hline
		\textbf{1010}&88.5223&86.5491&90.4955& 88.0163&-0.57164\\\hline
		\textbf{1020}&94.2615&92.2246&96.2983& 93.7188&-0.57575\\\hline
		\textbf{1030}&100.1746&98.0743&102.2748& 99.6069&-0.56664\\\hline
		\textbf{1040}&106.2693&104.106&108.4326& 105.677&-0.55739\\\hline
		\textbf{1050}&112.5434&110.3175&114.7693& 111.9249&-0.54959\\\hline
	\end{tabular}
	\begin{tabular}{c|c c c c c}
		\multicolumn{6}{c}{\textbf{Panel B:}} \\
		~~~\textbf{$v(t = 0)$}~~~~~~~&\textbf{Monte Carlo}&\textbf{Conf. Lower}&\textbf{Conf. Upper}&\textbf{KM approx.}& \textbf{\% Diff}\\\hline
		\textbf{0.1}&36.7795&35.9868&37.5721&36.6167&-0.44249\\\hline
		\textbf{0.2}&51.7656&50.6285&52.9026&51.5021&-0.50903\\\hline
		\textbf{0.3}&63.2958&61.8824&64.7093&62.9573&-0.53486\\\hline
		\textbf{0.4}&73.0207&71.3661&74.6754&72.6188&-0.55043\\\hline
		\textbf{0.5}&81.5878&79.7144&83.4613&81.1286&-0.56283\\\hline
		\textbf{0.6}&89.3302&87.2538&91.4066&88.8177&-0.57367\\\hline
		\textbf{0.7}&96.4461&94.1786&98.7137&95.8836&-0.58329\\\hline
		\textbf{0.8}&103.0649&100.6155&105.5143&102.455&-0.59176\\\hline
		\textbf{0.9}&109.2765&106.6527&111.9003&108.6217&-0.59925\\\hline
		\textbf{  1}&115.1459&112.3539&117.9379&114.449&-0.60525\\\hline
		\textbf{1.1}&120.7237&117.7688&123.6786&119.9864&-0.61074\\\hline
	\end{tabular}
	\floatfoot{\textit{Notes: Comparsion of European Call option prices under the dynamics of the CEV model with $\gamma = 0.6$. MC prices are obtained using the Milstein scheme, 500 time steps and 20,000 sample paths. Confidence intervals are computed on the 95\% level. KM's approximation uses the Black-Scholes model as baseline model. Panel A: Accuracy for differing spot moneyness of the option. Strike price = 1,000. Panel B: At-the-money options (S(t) = strike = 1,000) for differing spot variance. In both Panles: Time to maturity $= 1/12$, $\kappa = 0.1465$, $\theta = 0.5172$, $\omega = 0.5786$, $\rho = -0.0243$, $r = 0.00$, and $v(t) = \theta$. The volatility for the Black-Scholes model is $\sigma_0 = \sqrt{v(t)}$.}}
	\label{tab.CEV1}
\end{table}

\begin{table}[H]
	\centering
	\caption{Approximating the CEV model - $\gamma = 1.33$}
	\begin{tabular}{c|c c c c c}
		\hline
		\multicolumn{6}{c}{\textbf{Panel A:}} \\
		\textbf{Stock price}&\textbf{Monte Carlo}&\textbf{Conf. Lower}&\textbf{Conf. Upper}&\textbf{KM approx.}& \textbf{\% Diff}\\\hline
		\textbf{ 950}&57.7911&56.2022&59.3801&57.9685&0.30688\\\hline
		\textbf{ 960}&62.2887&60.6359&63.9416&62.4995&0.33837\\\hline
		\textbf{ 970}&66.974&65.257&68.6911&67.2303&0.38266\\\hline
		\textbf{ 980}&71.8629&70.0816&73.6442&72.1595&0.4128\\\hline
		\textbf{ 990}&76.9577&75.1121&78.8032&77.2853&0.42574\\\hline
		\textbf{1000}&82.2442&80.3345&84.1539&82.6053&0.43906\\\hline
		\textbf{1010}&87.7206&85.7469&89.6943&88.1168&0.45168\\\hline
		\textbf{1020}&93.3737&91.3361&95.4113&93.8168&0.47455\\\hline
		\textbf{1030}&99.2136&97.1124&101.3148&99.7018&0.49215\\\hline
		\textbf{1040}&105.2331&103.0687&107.3975&105.7682&0.50848\\\hline
		\textbf{1050}&111.4358&109.2087&113.6629&112.0119&0.51693\\\hline
	\end{tabular}
	\begin{tabular}{c|c c c c c}
		\multicolumn{6}{c}{\textbf{Panel B:}} \\
		~~~\textbf{$v(t = 0)$}~~~~~~~&\textbf{Monte Carlo}&\textbf{Conf. Lower}&\textbf{Conf. Upper}&\textbf{KM approx.}& \textbf{\% Diff}\\\hline
		\textbf{0.1}&36.6629&35.8701&37.4558&36.8541&0.52153\\\hline
		\textbf{0.2}&51.4396&50.3018&52.5775&51.6922&0.49096\\\hline
		\textbf{0.3}&62.818&61.4036&64.2324&63.1147&0.47232\\\hline
		\textbf{0.4}&72.4188&70.7634&74.0743&72.7493&0.45633\\\hline
		\textbf{0.5}&80.8778&79.0039&82.7517&81.235&0.44158\\\hline
		\textbf{0.6}&88.5227&86.4464&90.5991&88.9015&0.42796\\\hline
		\textbf{0.7}&95.5489&93.2821&97.8157&95.9457&0.41536\\\hline
		\textbf{0.8}&102.0848&99.6369&104.5326&102.4961&0.40293\\\hline
		\textbf{0.9}&108.2186&105.5973&110.8399&108.642&0.39128\\\hline
		\textbf{  1}&114.0145&111.2261&116.8029&114.4488&0.38094\\\hline
		\textbf{1.1}&119.5232&116.573&122.4734&119.9658&0.37029\\\hline
	\end{tabular}
	\floatfoot{\textit{Notes: Comparsion of European Call option prices under the dynamics of the CEV model with $\gamma = 1.33$. MC prices are obtained using the Milstein scheme, 500 time steps and 20,000 sample paths. Confidence intervals are computed on the 95\% level. KM's approximation uses the Black-Scholes model as baseline model. Panel A: Accuracy for differing spot moneyness of the option. Strike price = 1,000. Panel B: At-the-money options (S(t) = strike = 1,000) for differing spot variance. In both Panles: Time to maturity $= 1/12$, $\kappa = 0.1465$, $\theta = 0.5172$, $\omega = 0.5786$, $\rho = -0.0243$, $r = 0.00$, and $v(t) = \theta$. The volatility for the Black-Scholes model is $\sigma_0 = \sqrt{v(t)}$.}}
	\label{tab.CEV2}
\end{table}



\begin{figure}[H]
	\centering
	\begin{subfigure}{.7\textwidth}
		\centering
		\includegraphics[width=.9\linewidth]{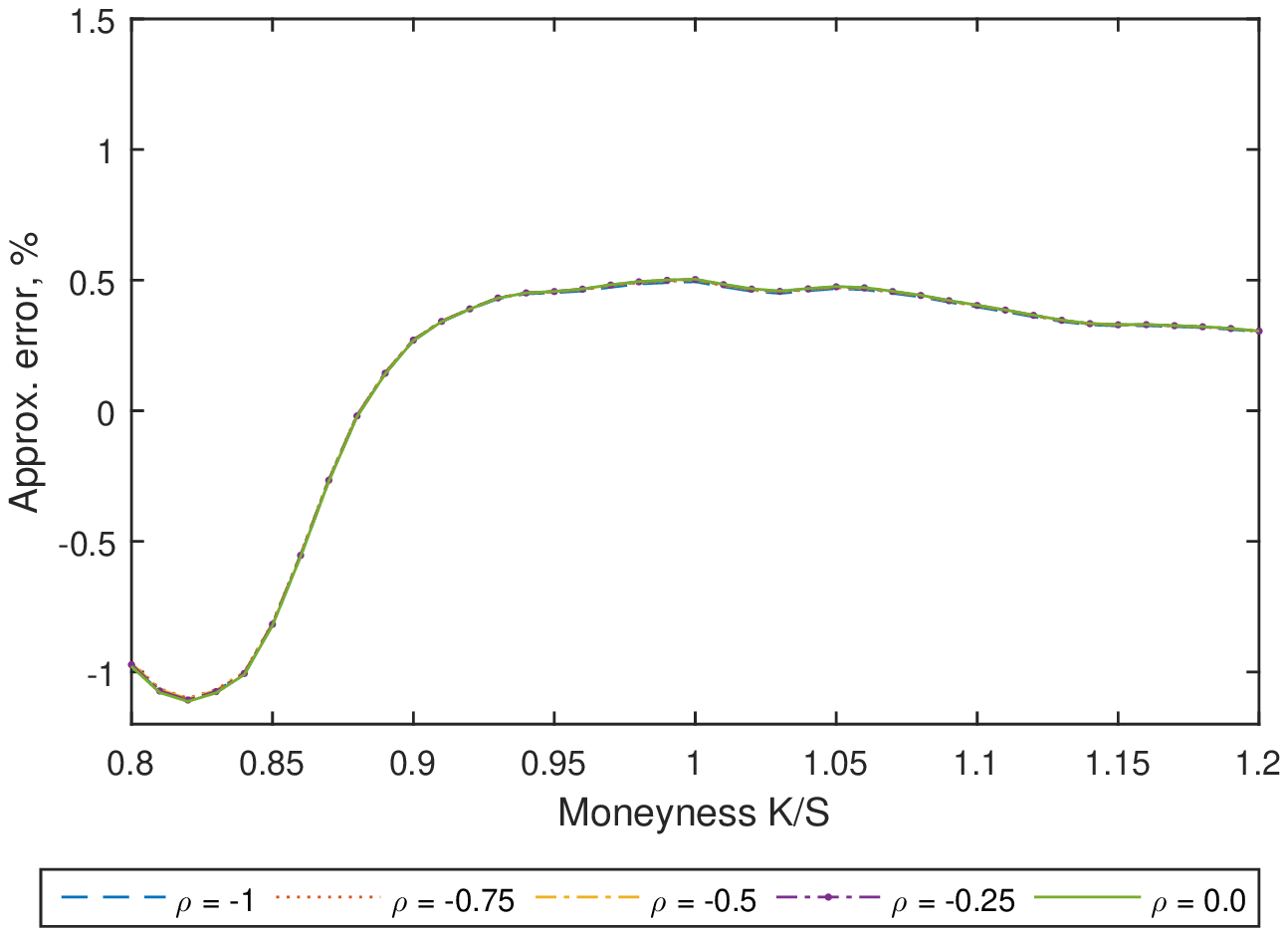}
		\caption{$T - t = 0.5$}
		\label{CEVcorr1}
	\end{subfigure}\\
	\begin{subfigure}{.7\textwidth}
		\centering
		\includegraphics[width=.9\linewidth]{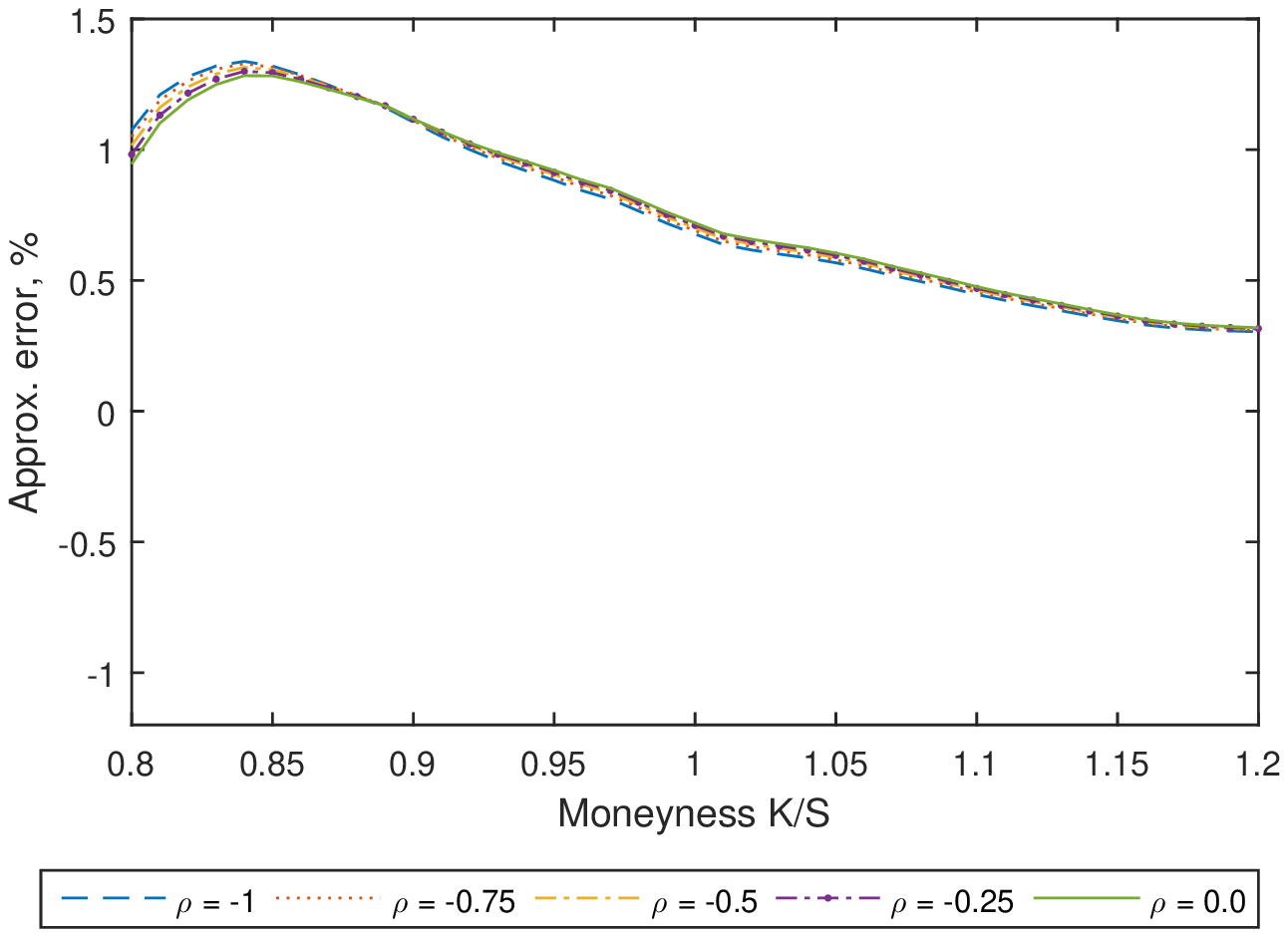}
		\caption{$T - t = 1.0$}
		\label{CEVcorr2}
	\end{subfigure}
	\caption{CEV: Correlation and Moneyness for $\gamma = 1.33$}
	\label{fig.CEVCorr}
	\floatfoot{\textit{Notes: Effect mooneyness and correlation on the KM approximation of the CEV model, using five corrective terms. Panel (a): Time to maturity is three months. Panel (b): Time to maturity is six months. $\gamma = 1.33$, $\theta = 0.04$, $\omega = 0.10$, $\rho = -0.5$, $r = 0.1$ and the spot variance $v = 0.05$. Black-Scholes volatility is set to $\sqrt{v}$. Reference values are obtained via Monte Carlo simulation using the specification in Appendix \ref{CEVMCRef}.}}
\end{figure}

\begin{figure}[H]
	\centering
	\begin{subfigure}{.7\textwidth}
		\centering
		\includegraphics[width=.9\linewidth]{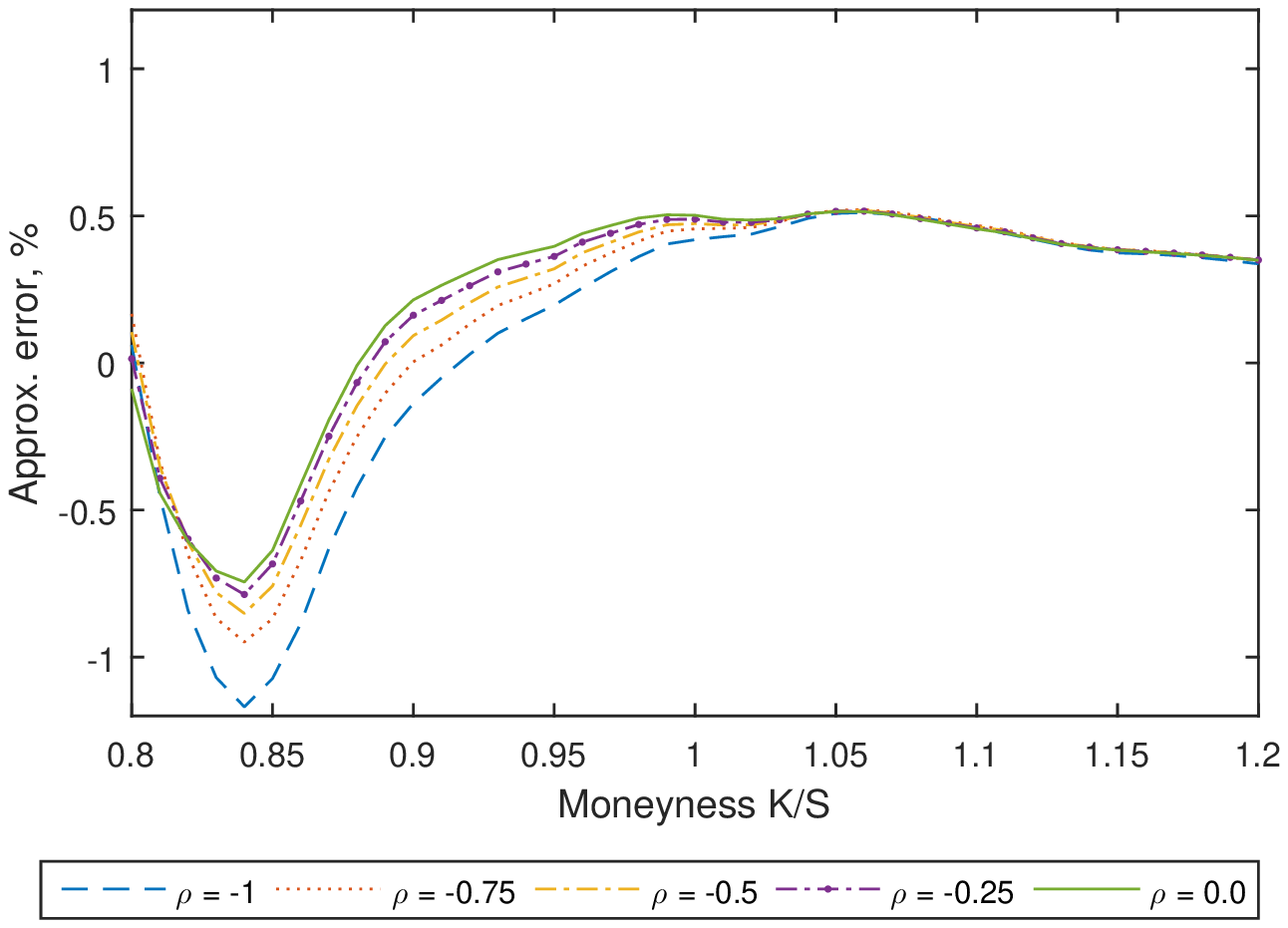}
		\caption{$T - t = 0.5$}
		\label{CEVcorr1gam06}
	\end{subfigure}\\
	\begin{subfigure}{.7\textwidth}
		\centering
		\includegraphics[width=.9\linewidth]{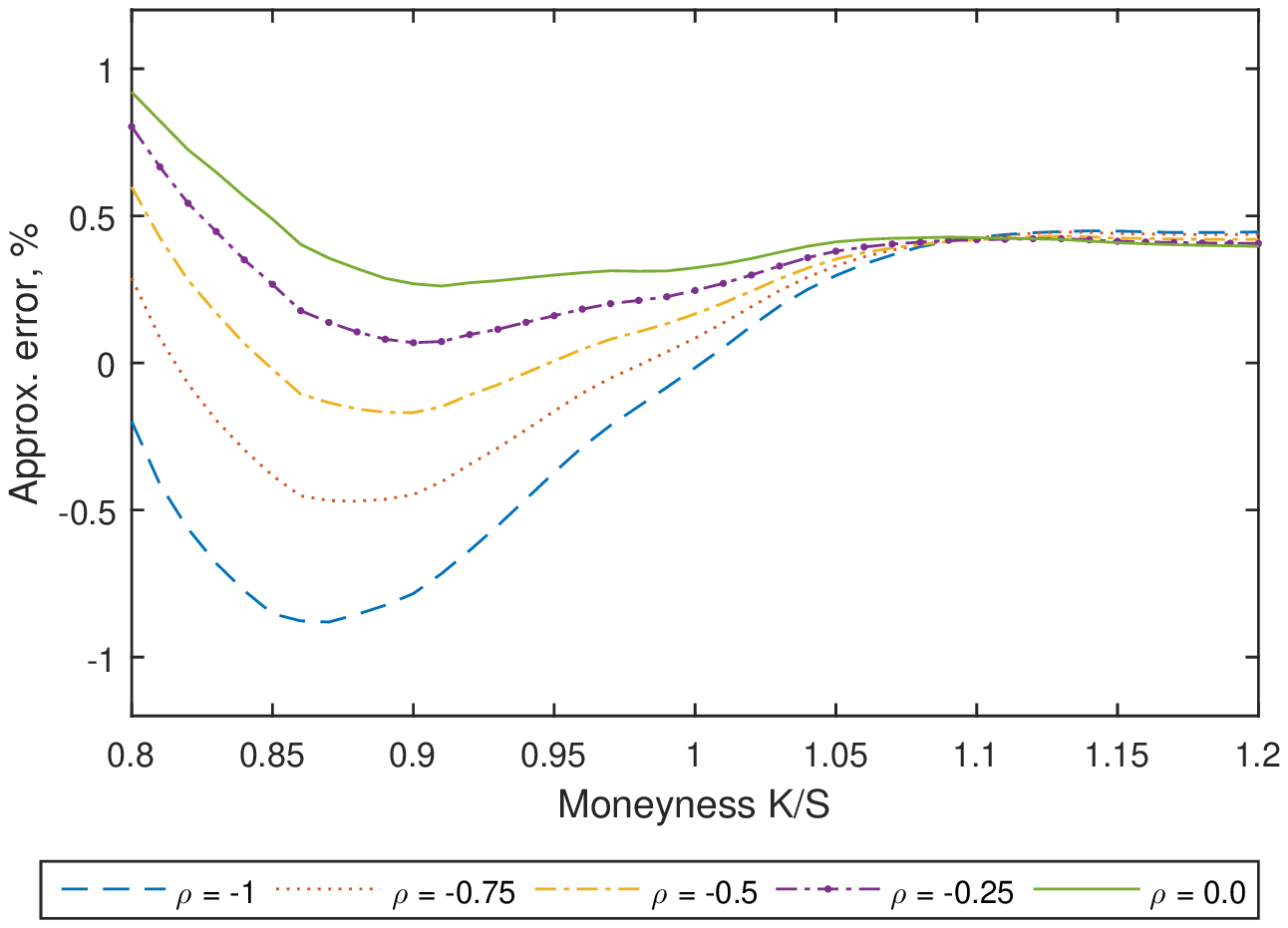}
		\caption{$T - t = 1.0$}
		\label{CEVcorr2gam06}
	\end{subfigure}
	\caption{CEV: Correlation and Moneyness for $\gamma = 0.6$}
	\label{fig.CEVCorrgam06}
	\floatfoot{\textit{Notes: Effect mooneyness and correlation on the KM approximation of the CEV model, using five corrective terms. Panel (a): Time to maturity is three months. Panel (b): Time to maturity is six months. $\gamma = 0.6$, $\theta = 0.04$, $\omega = 0.10$, $\rho = -0.5$, $r = 0.1$ and the spot variance $v = 0.05$. Black-Scholes volatility is set to $\sqrt{v}$. Reference values are obtained via Monte Carlo simulation using the specification in Appendix \ref{CEVMCRef}.}}
\end{figure}

Focusing on the $\gamma = 1.33$ case, Figure \ref{fig.CEVCorr} shows for different maturities that correlation seems to have no significant influence on the accuracy of the approximation. All parameters are the same as for Figure \ref{CEV1B}, despite the level correlation and the time to maturity. For the shorter time to maturity of six months in Figure \ref{CEVcorr1} KM's approximation achieves exactly the same accuracy for each level of correlation. For the longer time to maturity of one year in Figure \ref{CEVcorr2} the lines indicating the accuracy of the approximations move slightly apart with an increasing level of correlation, however the differences are extremely small. Hence, it seems that the effect of correlation on the accuracy of KM's approximation for the CEV model with $\gamma = 1.33$ is not significant. Since the level of correlation previously had an effect on the accuracy when approximating the Heston model, this effect now seems to be offset by the high elasticity of variance. Figure \ref{fig.CEVCorrgam06} performs the same analyzes for a CEV parameter of $\gamma = 0.6$, leaving all other parameters equal. The result supports the view that effect of correlation on the approximation is offset by a high elasticity of variance. For  $\gamma = 0.6$ correlation does have a visible effect, however the effect vanishes as the option moves deeper in the ITM region. As expected the influence of correlation is also increasing with increasing time to maturity.

\begin{figure}[H]
	\centering
	\includegraphics[scale=.9]{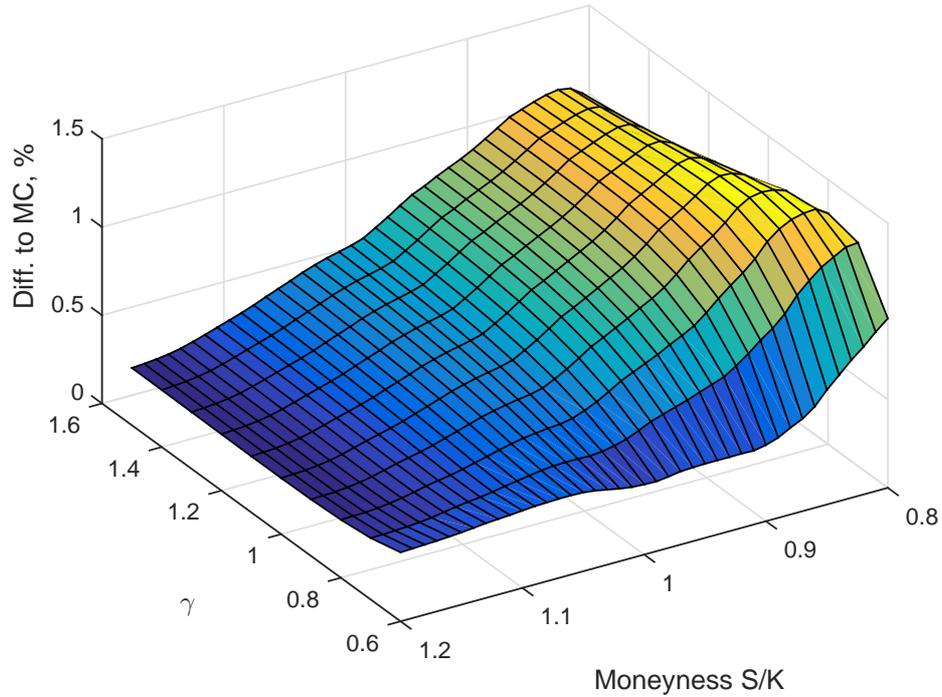}
	\caption{Level of elasticity of variance and moneyness}
	\floatfoot{\textit{Notes: Effect of different elasticity of variance parameters and moneyness on the KM approximation of the CEV model, using five corrective terms. Time to maturity is one year, $\theta = 0.04$, $\omega = 0.10$, $\rho = -0.5$, $r = 0.1$ and the spot variance $v = 0.05$. Black-Scholes volatility is set to $\sqrt{v}$. Reference values are obtained via Monte Carlo simulation using the specification in Appendix \ref{CEVMCRef}.}}
	\label{fig.CEVxi}
\end{figure}

Finally, it might be interesting to analyze the effect of $\gamma$ itself in a more focused way. Figure \ref{fig.CEVxi} attempts to do this. The parameters are the same as for Figure \ref{fig.CEV1}, but now letting $\gamma$ vary from $0.6$ to $1.6$. As in the previous graphs I use an order of $N = 4$ for the KM approximation. The figure suggests that there is no significant influence of the elasticity parameter on the accuracy of the approximation for $\gamma > 0.6$. For all values the series expansion shows the expected pattern of being less precise for OTM options. Nevertheless, the absolute maximum difference to the MC price never exceeds 1.5\%.

\clearpage

\section{Stochastic volatility with an OU - process} \label{SZModel}

In this section I will change the focus to another class of stochastic volatility models for option prices, in which stochastic volatility is modeled directly and follows an Ornstein - Uhlenbeck process. Consider the following system of SDEs  
\begin{align}
dS(t) ~=&~ r~dt + \sigma (t)~dW_1(t) \label{SZ1}\\
d\sigma (t) ~=&~ \kappa \left(\theta - \sigma (t)\right)~dt + \omega ~ dW_2(t) \label{SZ2} \\
dW_1(t) dW_2(t) ~=&~ \rho dt \nonumber
\end{align}
Where $\kappa$ denotes the speed of mean-reversion parameter,  $\theta$ the long-run mean and $\omega$ the instantaneous volatility of volatility. This model had been originally suggested by \citet{Stein91} (henceforth S\&S) for the case of zero correlation between the stock price and the volatility process, such that in the original formulation $\rho \equiv 0$. Since correlation between the volatility and the stock price process plays an important role in replicating volatility smiles observed in the data, zero correlation seems to be rather restrictive. However, the model had been extended by \citet{Schobel99}(henceforth SZ) to also cover arbitrary correlation between the two processes. I consider this model since it falls outside of the class of affine models and hence yields a more common non-affine model than the previous non-affine CEV example. \citet{Schobel99} provide a (semi-)closed-form solution via Fourier inversion techniques for this model, such that the accuracy and convergence behavior of KM's approximation can be checked against an analytic solution.\\
Note that the SZ model is not a special case of the Heston model or vice versa. For the parameter restriction $\kappa = \kappa_{Heston}/2,~\omega = \omega_{Heston}/2,~ \theta = 0$ and $\theta_{Heston} = \omega^2/\kappa_{Heston}$ the \citet{Schobel99} model is equivalent to the Heston model. However, these restrictions over-determine the Heston model such that the two models are not consistent for a wide range of parameter values.\footnote{See \citet{Schobel99}, p. 30.} This provides another advantage in assessing the robustness of KM's approximation. The non-affine and affine specifications of the CEV model shared the same general structure of the variance process, with the one specification arising as a special case of the other. 
While KM do not consider any non-affine option pricing model other than the CEV case, they do so in the context of interest rate models. Specifically, KM apply their approximation to the non-affine two-factor model suggested by \citet{Forani}.\footnote{See \citet{KM2011}, p. 407.} In this context KM found that the approximation errors only decreased with the number of corrective terms for maturities less than three years and conclude that for longer maturities probably many corrective terms are needed to achieve accurate approximations. Since the accuracy of KM's method also deteriorated with increasing time to maturity in the Heston, it might be worthwhile to assess the accuracy of the approximation for different times to maturity in the context of the \citet{Stein91} and \citet{Schobel99} models.  \\
\subsection{Approximation of the SZ model}
The steps to derive KM's approximation for the SZ/S\&S - model are essentially the same as for the Heston model. Again the Black-Scholes formula serves as the baseline model. The first step is to derive the covariance matrix of the SZ - model.\footnote{Since the SZ- and S\&S- model are equivalent, except for the choice of $\rho$ I will base the whole expansion on the SZ specification. An approximation of the S\&S-model is simply obtained by setting $\rho$ to zero.} By applying the same principle as in (\ref{CovHest}) in section \ref{HestonApproxKM} the covariance matrix is given by
\begin{align}
\begin{bmatrix}
\sigma(t)S(t) & 0 \\
\rho \omega & \sqrt{1 - \rho^2} \omega
\end{bmatrix} \times
\begin{bmatrix}
\sigma(t)S(t) & \rho \omega \\
0 & \sqrt{1 - \rho^2} \omega
\end{bmatrix} = 
\begin{bmatrix}
\sigma(t)^2S(t)^2 & \rho \omega \sigma(t) S(t) \\
\rho \omega \sigma(t) S(t) & \omega^2
\end{bmatrix}
\end{align}
Applying (\ref{initdel}) with the right-hand side in the above matrix equation yields the initial pricing error 
\begin{align}
\delta_0 = \frac{1}{2}\left(\sigma(t)^2 - \eta_0^2\right)S(t)^2\dfrac{\partial^2 C^{BS}}{\partial S(t)^2}
\end{align}
Note that $\delta_0$ is the same as in (\ref{CEVd0}) by just replacing $v(t)$ with $\sigma(t)^2$. Based on this initial pricing error all further corrective terms can be computed iteratively according to the rule in (\ref{It1}). The  iteration of the corrective terms is shown in Table \ref{tab.SZit} below. Note that for the case of the S\&S-model, i.e. with $\rho = 0$, all terms including the cross-derivatives $\partial^2 \delta_n / \partial S \partial \sigma(t)$ drop out. Using (\ref{CorrecHest}) together with the corrective terms in Table \ref{tab.SZit} the price of a plain vanilla call under the dynamics of the SZ-model can be expressed. For this model I developed the expansion for the SZ-model until $N = 5$, since this was the optimal order of the series expansion used by KM for their approximation of the \citet{Forani} model.\\
\clearpage
\begin{landscape}
	\newcolumntype{C}{>{\centering\arraybackslash} m{18cm} } 
	\renewcommand{\arraystretch}{2.0}
	\begin{table}[H]
		\centering
		\caption{Iterations of the pricing error for the SZ/S\&S-model}
		\begin{tabular}{m{1cm}|C}
		\hline
		\textbf{n} & \textbf{Pricing error $\delta_n(S,t)$} \\\hline\hline
		\textbf{1} & $\dfrac{\partial \delta_0}{\partial t} + rS\dfrac{\partial \delta_0}{\partial S} + \kappa\left(\theta - \sigma(t)\right)\dfrac{\partial \delta_0}{\partial \sigma(t)} + \frac{1}{2}\left(\sigma(t)^2S^2\dfrac{\partial^2 \delta_0}{\partial S^2} + \omega^2\dfrac{\partial^2 \delta_0}{\partial \sigma(t)^2} \right) + \rho \omega S \sigma(t)\dfrac{\partial^2\delta_0}{\partial S\partial \sigma(t)} - r\delta_0$\\\hline
		\textbf{2} & $\dfrac{\partial \delta_1}{\partial t} + rS\dfrac{\partial \delta_1}{\partial S} + \kappa\left(\theta - \sigma(t)\right)\dfrac{\partial \delta_1}{\partial \sigma(t)} + \frac{1}{2}\left(\sigma(t)^2S^2\dfrac{\partial^2 \delta_1}{\partial S^2} + \omega^2\dfrac{\partial^2\delta_1}{\partial \sigma(t)^2}\right) + \rho \omega S \sigma(t)\dfrac{\partial^2\delta_1}{\partial S\partial \sigma(t)} - r\delta_1$ \\\hline
		\textbf{$\vdots$} & $\vdots$ \\\hline
		\textbf{N} & $\dfrac{\partial \delta_{N-1}}{\partial t} + rS\dfrac{\partial \delta_{N-1}}{\partial S} + \kappa\left(\theta - \sigma(t)\right)\dfrac{\partial \delta_{N-1}}{\partial \sigma(t)} + \frac{1}{2}\left(\sigma(t)^2S^2\dfrac{\partial^2 \delta_{N-1}}{\partial S^2} + \omega^2\dfrac{\partial^2\delta_{N-1}}{\partial \sigma(t)^2}\right) + \rho \omega S \sigma(t)\dfrac{\partial^2\delta_{N-1}}{\partial S\partial \sigma(t)} - r\delta_{N-1}$  \\\hline
	\end{tabular} \\
	\label{tab.SZit}
\end{table}
\end{landscape}
\clearpage

\subsection{Accuracy of the approximation}
\begin{figure}[H]
	\begin{subfigure}{.6\textwidth}
		\includegraphics[width=.8\linewidth]{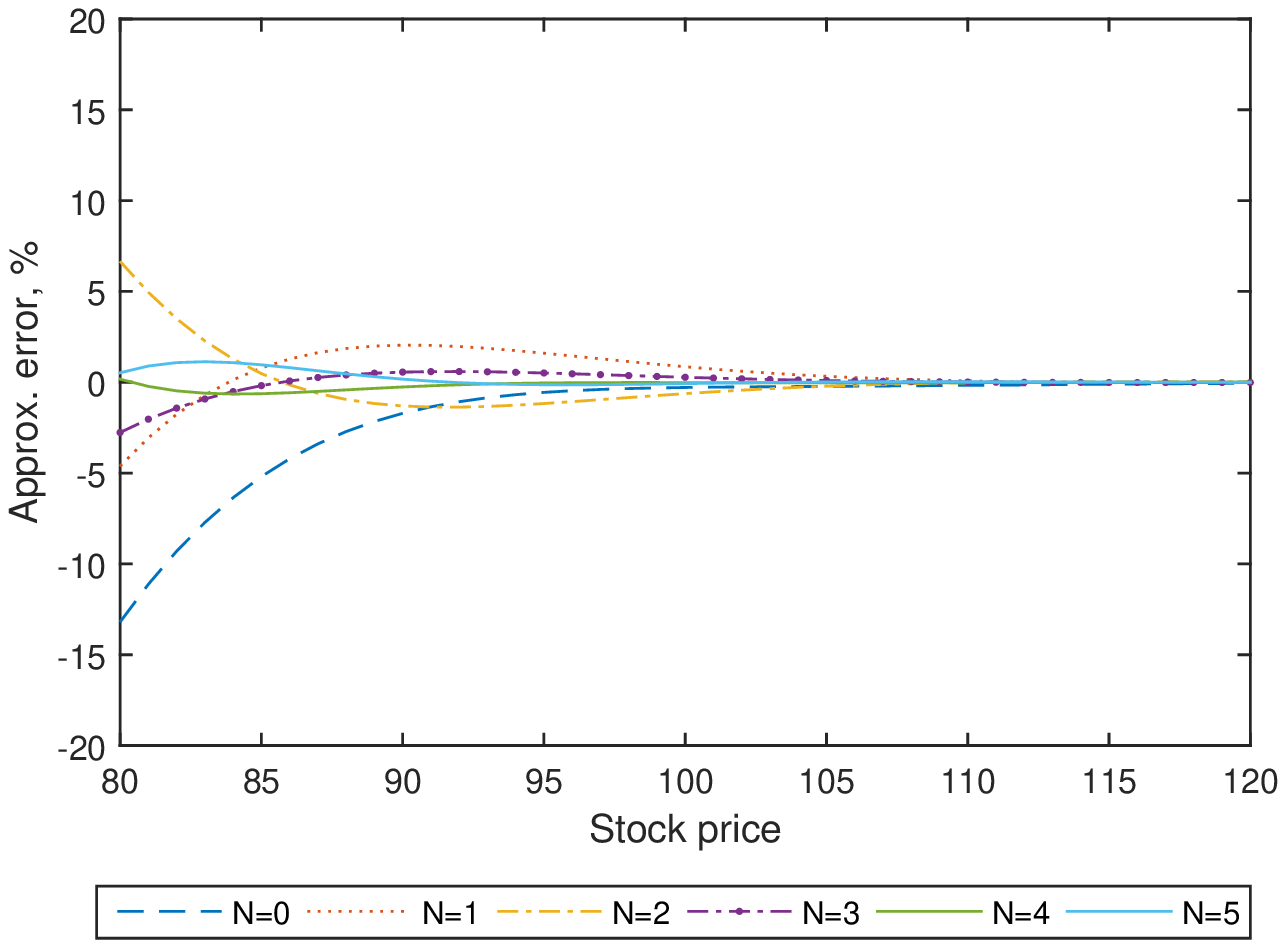}
		\caption{Time to maturity $T-t = 0.25$}
		\label{SS025}
	\end{subfigure} 
	\begin{subfigure}{.6\textwidth}
		\includegraphics[width=.8\linewidth]{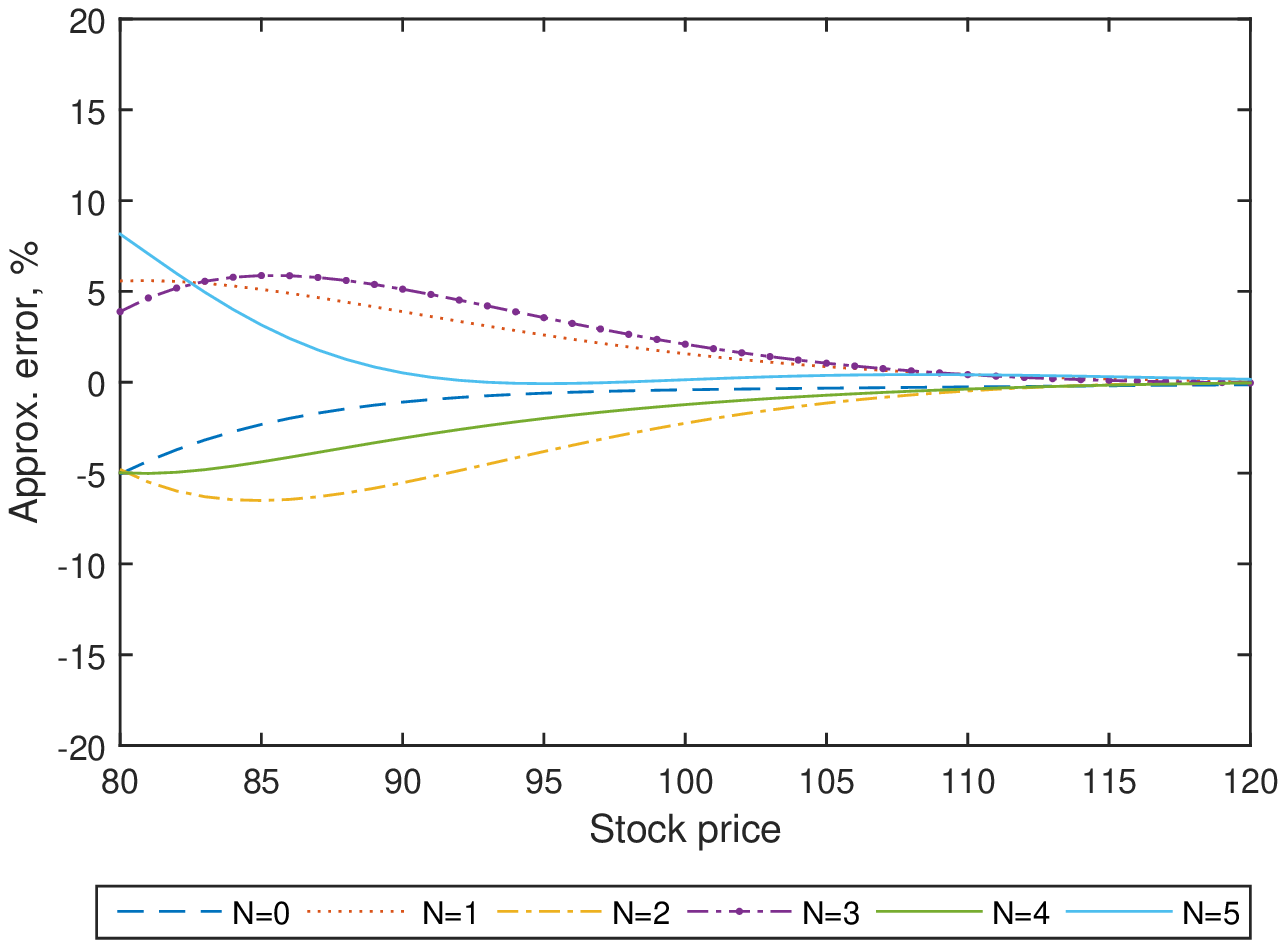}
		\caption{Time to maturity $T-t = 0.50$}
		\label{SS05}
	\end{subfigure}
	\begin{subfigure}{.6\textwidth}
		\includegraphics[width=.8\linewidth]{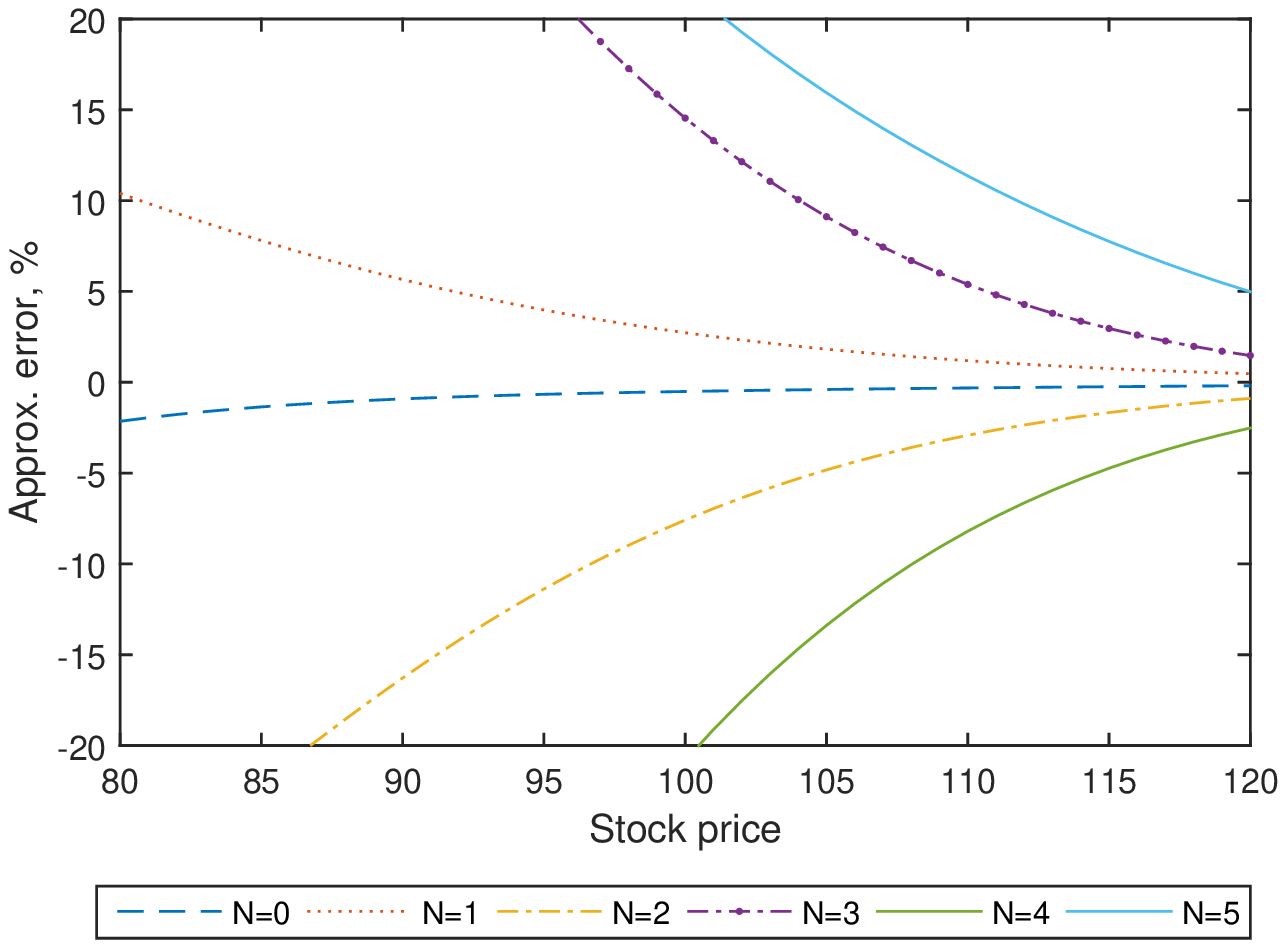}
		\caption{Time to maturity $T-t = 1.00$}
		\label{SS10}
	\end{subfigure}
	\caption{Approximation of the \citet{Stein91} model.}
	\label{fig.SS1}
	\floatfoot{\textit{Notes: For all Panels: Strike Price = 100, $\theta = 0.2$, $\kappa = 4.0$, $\omega = 0.1$, $r = 0.0953$, spot volatility $\sigma = 0.2$. Black-Scholes volatility $\sigma_0 = \sigma = 0.2$. Analytic prices via Fourier inversion solution of the SZ-model with $\rho = 0$.}}
\end{figure}

\begin{figure}[H]
	\centering
	\begin{subfigure}{.6\textwidth}
		\centering
		\includegraphics[width=.7\linewidth]{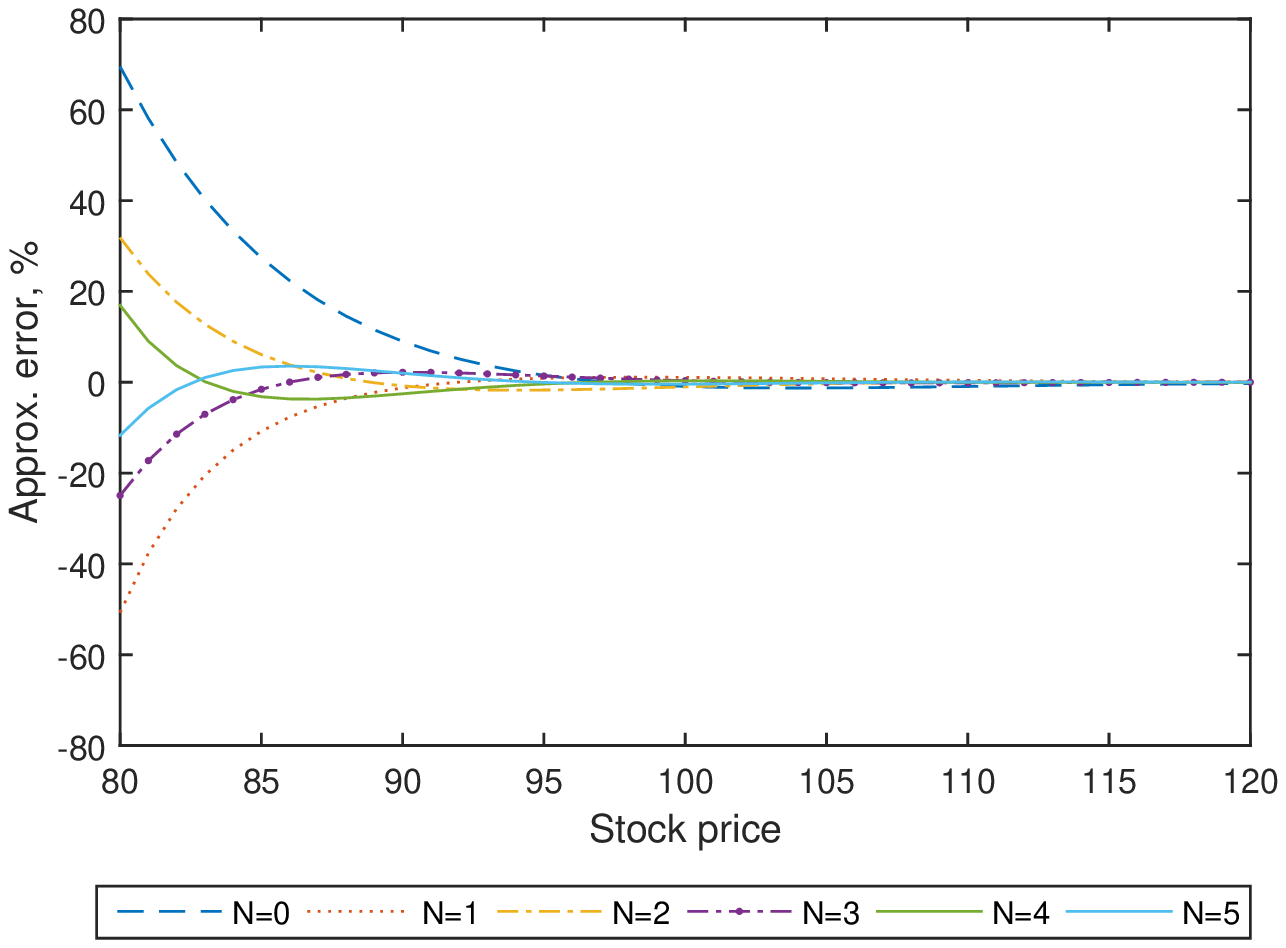}
		\caption{Time to maturity $T-t = 0.25$}
		\label{SZ025}
	\end{subfigure} 
	\begin{subfigure}{.6\textwidth}
		\centering
		\includegraphics[width=.7\linewidth]{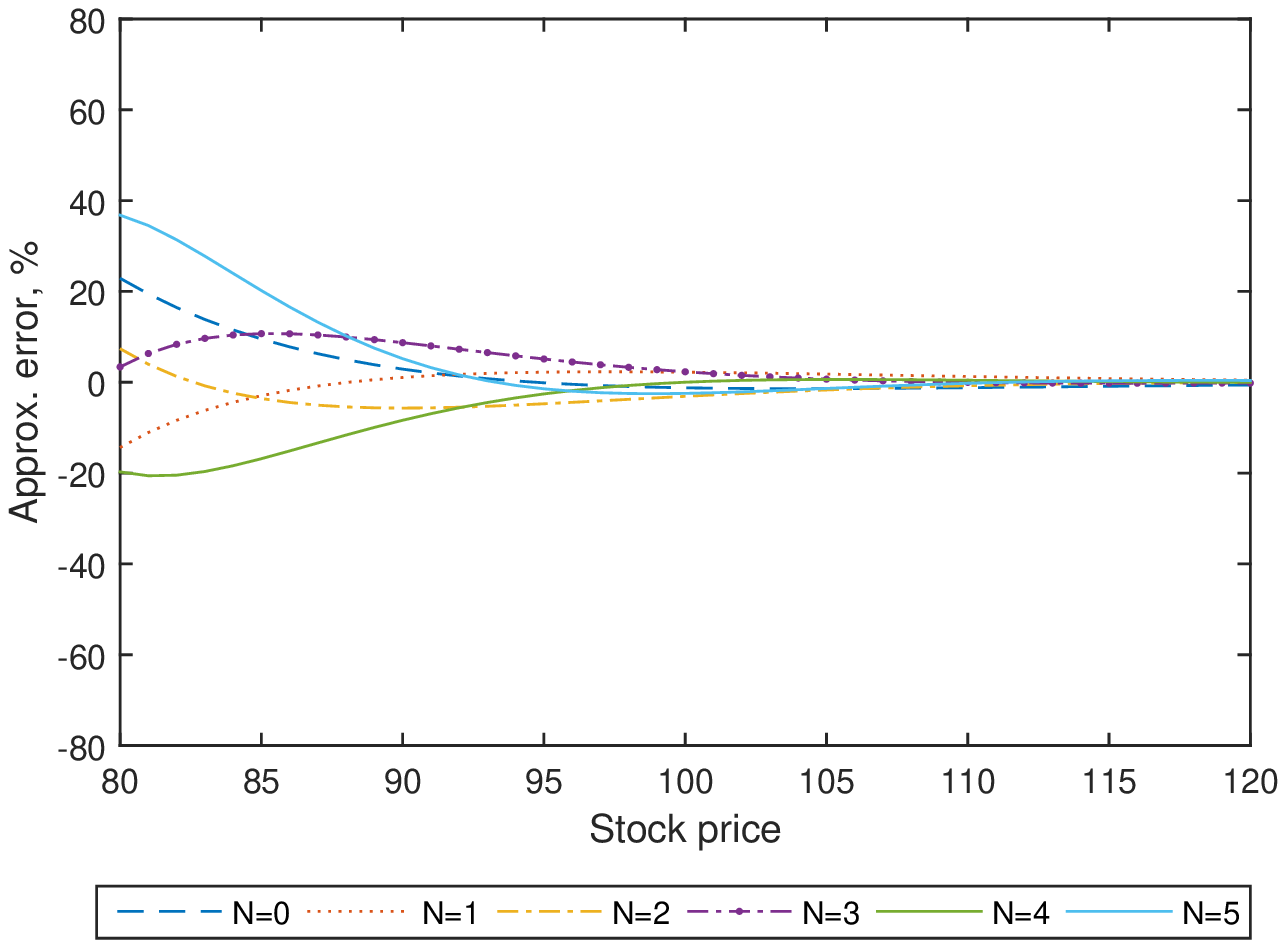}
		\caption{Time to maturity $T-t = 0.50$}
		\label{SZ05}
	\end{subfigure}
	\begin{subfigure}{.6\textwidth}
		\centering
		\includegraphics[width=.7\linewidth]{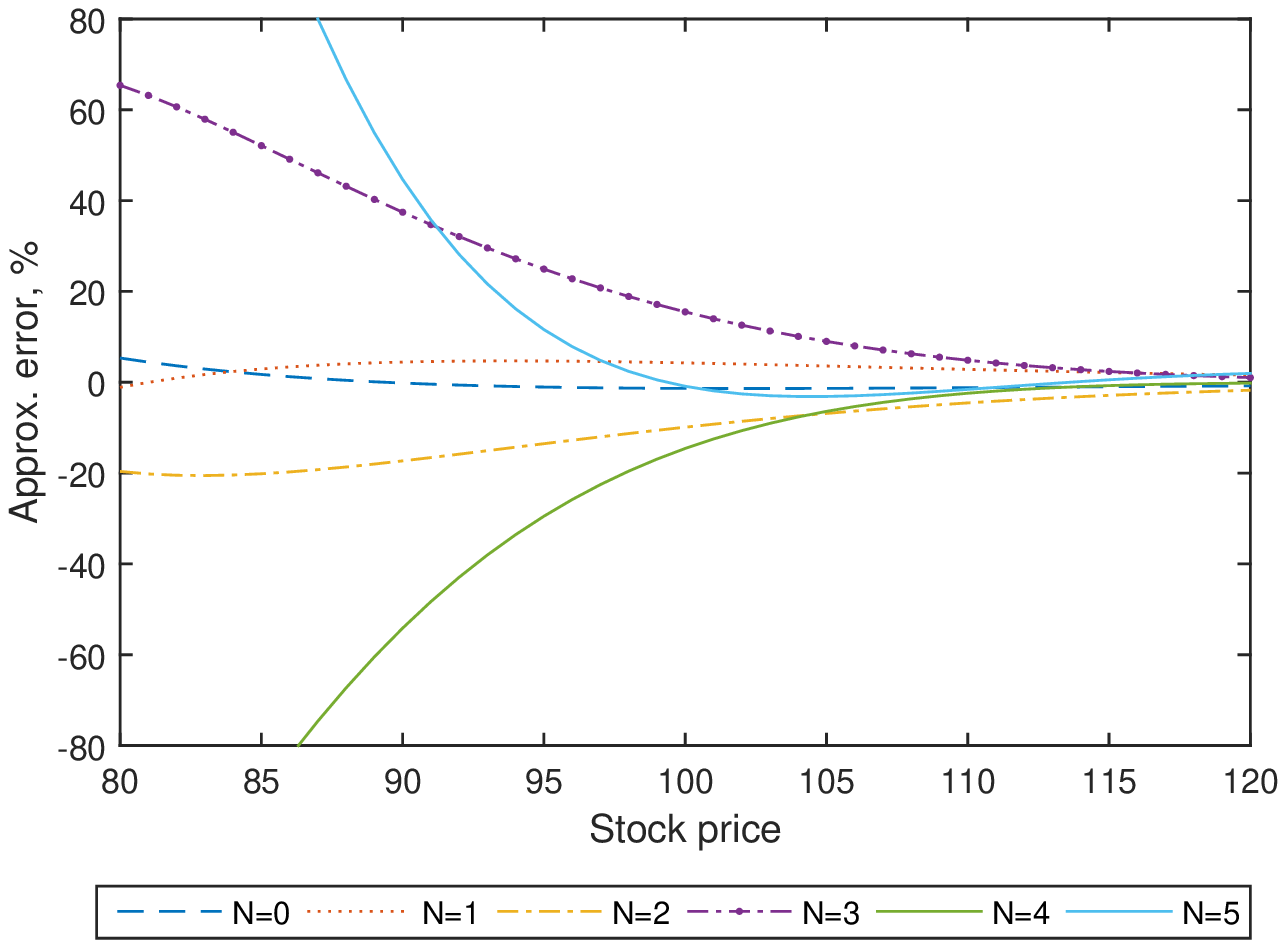}
		\caption{Time to maturity $T-t = 1.00$}
		\label{SZ10}
	\end{subfigure}
	\caption{Approximation of the \citet{Schobel99} model.}
	\label{fig.SZ1}
	\floatfoot{\textit{Notes: For all Panels: Strike Price = 100, $\theta = 0.2$, $\kappa = 4.0$, $\omega = 0.1$, $r = 0.0953$, spot volatility $\sigma = 0.2$, $\rho = -0.5$. Black-Scholes volatility $\sigma_0 = \sigma = 0.2$. Analytic prices via Fourier inversion.}}
\end{figure}

Figure \ref{fig.SS1} follows the same idea as the initial figures in the previous sections, showing the behavior of the approximation errors for different stock prices, when approximating the S\&S-model, i.e. the zero correlation case of the model. The parameter values are taken from \citet{Schobel99}. Figure \ref{SS025}, with a maturity of $0.25$ years, shows the expected convergence pattern. For $N = 4$ or $N = 5$ corrective terms KM's approximation yields quite precise results with percentage errors not exceeding $0.6458 \%$ for $N = 4$ and $1.1282 \%$ for $N = 5$. That the maximum absolute error for $N = 4$ is slightly less then for $N = 5$ indicates that a higher number of corrective terms do not necessarily for all models leads to greater precession. A feature that makes it difficult to determine the optimal order of the approximation if an analytic solution is indeed unknown. \\
With increasing time to maturity accuracy and also the convergence behavior deteriorates fast. Figures \ref{SS05} and \ref{SS10} show the same approximation for times to maturity of $0.5$ years and one year. While in Figure \ref{SS05} no obvious convergence pattern seems to be present, the last two Panels even seem to hint towards divergence instead of convergences, at least for the considered number of corrective terms. This indicates that the approximation becomes instable much faster with increasing time to maturity when using an Ornstein-Uhlenbeck instead of a square-root like process as in the previous sections. There instability occurred for maturities well above one year only.\\
Figure \ref{fig.SZ1} shows the result of approximating the more general SZ-model by KM's approximation. I assume a correlation coefficient of $\rho = -0.5$, while all other parameters are the same as before. The convergence pattern appears to be the same as for the S\&S model. The approximation converges for a short time to maturity of $0.25$ years as can be seen in Panel \ref{SZ025}. However, the Panels \ref{SZ05} and \ref{SZ10} show that for times to maturity of half a year and one year the convergence seems to turn into divergence. Note that even for the short maturity case the approximation error for $N = 5$ remains substantial for OTM calls. The approximation error dies out fast and more uniformly with an increasing stock price compared to the zero correlation case. Additionally, in this case the approximation indeed unambiguously gets more precise with each new corrective term, while in the zero correlation case for the same maturity the sixth corrective term slightly reduced the accuracy. When time to maturity increases, the approximation of the SZ models shows the same behavior as the S\&S model. For a maturity of half a year the convergence pattern starts to dissolve and turns into divergence for a maturity of one year. Overall the compatibility of KM's approximation approach seems to be lower for this class of models compared to the CEV or the Heston model. For the Heston model a pattern of convergence remained for maturities of up to two years. \citet{Kimmel08} finds that such divergence behavior for some maturities occurs if the series expansions used for pricing has singularities at some complex maturities. In such a case the series expansion will start to diverge for maturities of the same absolute magnitude as the complex maturity at which the singularity occurred.\footnote{See \citet{Kimmel08}, p. 1.} More specifically \citet{Kimmel08} argues that there exists a circle in the complex plane, which radius is defined by distance from zero to singularity closest to zero, such the series expansion converges if the maturity lies within the circle and diverges if it lies outside. This suggests that for the SZ model maturities over 0.5 years already lie outside this circle.\\   
Note that both, Figure \ref{fig.SS1} and \ref{fig.SZ1}, show that the most precise KM expansion is the one only including the first corrective terms, i.e. $N = 0$, when time to maturity is $0.5$ years or longer. For long maturities of one year these expansions seem to be strikingly precise. However, this can simply be explained by the specific choice of the nuisance parameter and the general behavior of the volatility smile. By choosing $\eta_0 = \sigma(t)$ one effectively sets the first corrective term equal to zero, such that the $N = 0$ expansion just uses the Black-Scholes price as an approximation. As I argued already in the discussion of the approximation of the Heston model, for long maturities the volatility smile tends to flatten and thus the Black-Scholes price will be closer to the true price.

\begin{figure}[H]
	\centering
	\includegraphics[scale=.9]{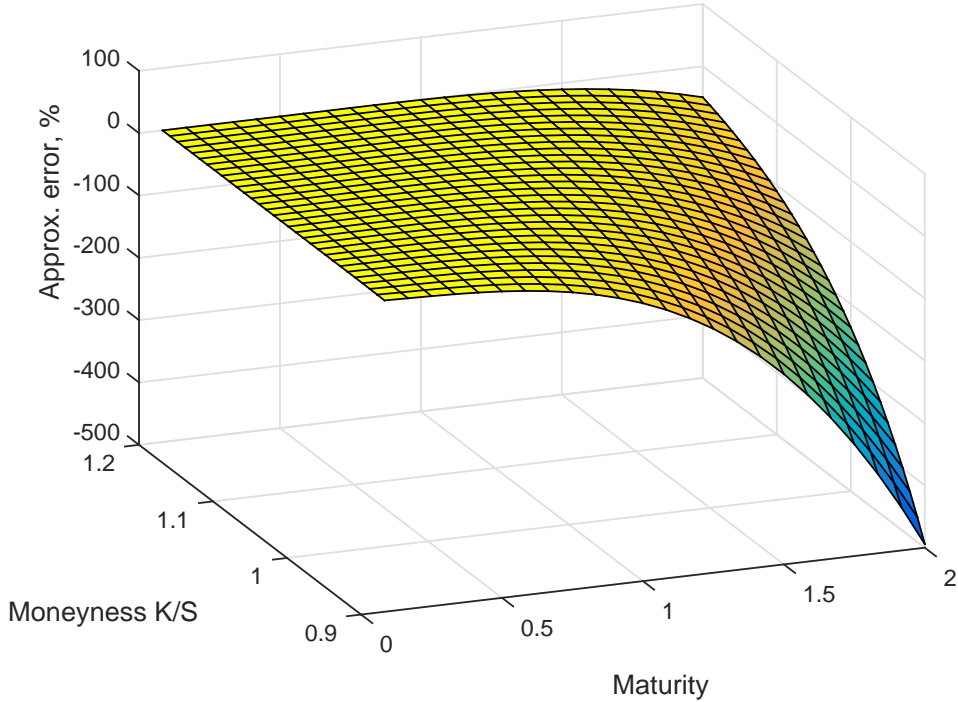}
	\caption{Sensitivity towards increasing time to maturity}
	\floatfoot{\textit{Notes: Instability of KM's approximation of the SZ model for increasing time to maturity and six corrective terms. Strike Price = 100, $\theta = 0.2$, $\kappa = 4.0$, $\omega = 0.1$, $r = 0.0953$, spot volatility $\sigma = 0.2$, $\rho = -0.5$. Black-Scholes volatility $\sigma_0 = \sigma = 0.2$. Analytic prices via Fourier inversion.}}
	\label{fig.SZTimeMon}
\end{figure}

It appears like the approximation only yields results with sufficient accuracy if time to maturity is below six month. This is summarized by Figure \ref{fig.SZTimeMon}, which clearly shows that the accuracy of the approximation deteriorates fast with increasing time to maturity. Hence, I will first focus on short maturities of three month and one month to investigate the effect of different levels of correlation on the performance of the approximation. Figures \ref{SS025} and \ref{SZ025} indicate significantly lower accuracy for OTM calls than for ITM calls when correlation is at a level of $-0.5$. \\

\begin{figure}[H]
	\centering
	\begin{subfigure}{.9\textwidth}
		\centering
		\includegraphics[width=.9\linewidth]{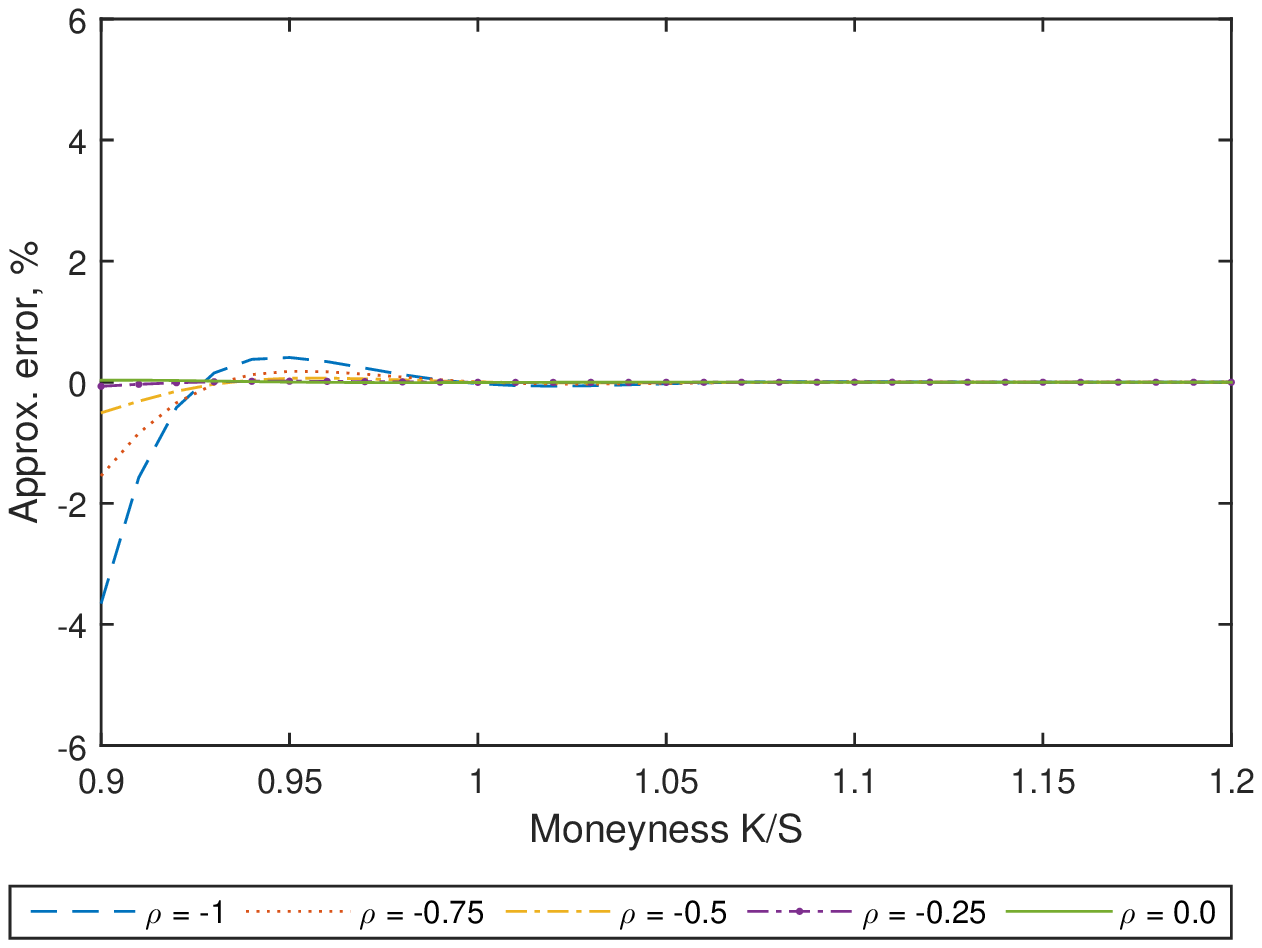}
		\caption{Time to maturity equals one month}
		\label{SZcorr1}
	\end{subfigure} 
	\begin{subfigure}{.9\textwidth}
		\centering
		\includegraphics[width=.9\linewidth]{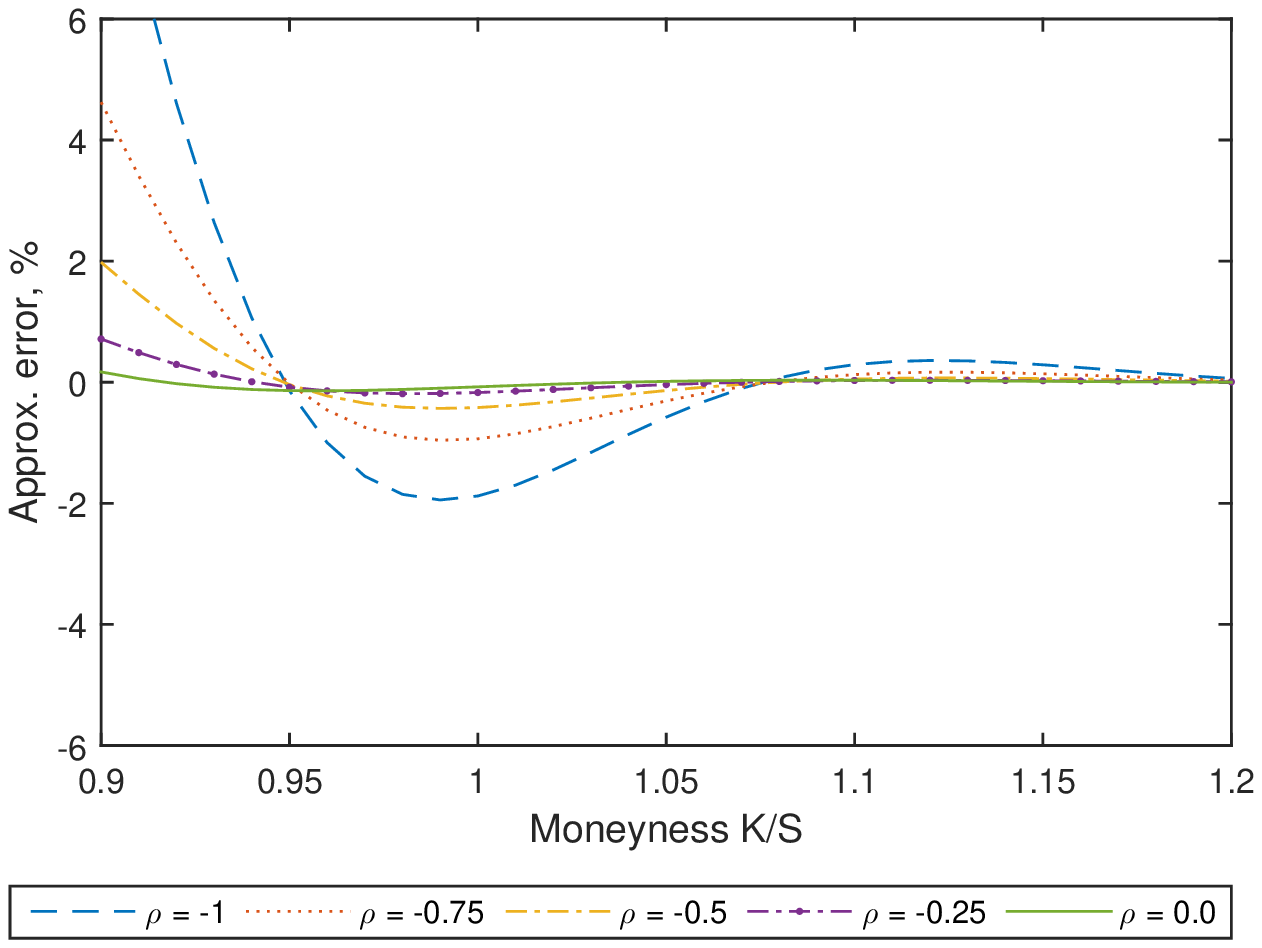}
		\caption{Time to maturity equals three months}
		\label{SZcorr2}
	\end{subfigure}
	\caption{SZ-model with six corrective terms}
	\label{fig.SZcorr}
	\floatfoot{\textit{Note: Analytic prices via Fourier transform. Strike $=~100$, $\kappa = 4.0$, $\theta = 0.2$, $\omega = 0.1$, $r = 0.0953$, and spot volatility $\sigma(t) = \theta$. Black-Scholes volatility equal spot volatility. All approximations are of order $N = 5$.}}
\end{figure}

Figure \ref{fig.SZcorr} shows the behavior of the approximation for maturities of one month and three months when using six corrective terms (i.e. $N = 5$) for different levels of correlation and moneyness. The figure indicates a behavior similar as in the case of the Heston model (see Figure \ref{fig.HestCorr3d} for a comparison). For the case of a time to maturity of one month in Figure \ref{SZcorr1} only levels correlation of $-0.5$ to $-1$ have an influence on the accuracy of the approximation, whereby this influence is also restricted to OTM options. ITM options are approximated with extremely high accuracy for all levels of correlation. If a longer time to maturity of three month, as in Figure \ref{SZcorr2}, is considered the influence of correlation increases significantly. For the zero correlation case the approximation remains completely stable over the whole moneyness range. An increasing level of correlation reduces the accuracy of the approximation visibly, whereby the effect is now not only restricted to OTM options but also appears for ITM options. Even though it is less pronounced for the latter. Longer maturities are not considered for the SZ model due to the instability of the approximation in these cases.

\subsubsection{Approximating Greeks in the SZ model}

In order to obtain KM's approximation of hedge ratios for a Call option under the dynamics of the SZ-model I again use equation (\ref{GrApprox}). As in the case of the Heston model I will also again use central finite differences to obtain derivatives of the corrective terms. Analytic reference values can be obtained via Fourier transforms following the same principle as for the Heston model, which I outlined in appendix \ref{HestGreek}.\footnote{For a detailed description of the Greeks in the SZ-model see  \citet{Schobel99}, p. 29.}\\
Tables \ref{tab.SZgreeks} and \ref{tab.SZgreeksGam} show KM's approximation of $\Delta$ and $\Gamma$ of a call option under the dynamics of the SZ-model. As before the greeks are reported in percent and thus the approximation error is given in percentage points. The chosen parameters are the same as before and time to maturity is $0.25$. The accuracy of the approximation is, as expected, lower as it was for the case of Heston model. Table \ref{tab.SZVega} shows the approximation of the options $\mathscr{V}$. Approximation errors are reported as absolute deviations. Note that also here the Black-Scholes $\mathscr{V}$ is equal to zero. Panel A in Table \ref{tab.SZVega} shows the approximation for varying moneyness. The approximation errors remain below one percent and show magnitudes similar to the other approximations. However, note that the spot volatility is set to $0.2$. Panel B of the same table indicates that the approximation is highly sensitive to the choice of the spot volatility parameter. The approximation yields accurate results for spot volatilities of $0.1$ and $0.2$. With increasing spot volatility the approximation of $\mathscr{V}$ becomes instable, yielding extremely high negative values for the hedge ratio. Despite the time to maturity this reveals another source of instability of the approximation procedure in case of the SZ model.    
  
\begin{table}[H]
	\centering
	\caption{Approximating $\Delta$ for the SZ-model}
	\begin{tabular}{c|c c c}
		\hline
		\multicolumn{4}{c}{\textbf{Panel A:}} \\
		\textbf{Stock price}&\textbf{Fourier Transform, \%}&\textbf{KM approx., \%}&\textbf{Diff., pp.}\\\hline
		\textbf{ 80}&1.5781&1.9609&-0.38278\\
		\textbf{ 85}&7.6057&8.6632&-1.0575\\
		\textbf{ 90}&22.6168&22.4057&0.21117\\
		\textbf{ 95}&44.2531&42.1098&2.1433\\
		\textbf{100}&64.6821&62.9238&1.7583\\
		\textbf{105}&79.5248&79.2665&0.2583\\
		\textbf{110}&88.791&89.2885&-0.4975\\
		\textbf{115}&94.0791&94.6592&-0.58008\\
		\textbf{120}&96.9379&97.473&-0.53503\\\hline
	\end{tabular}
	\begin{tabular}{c|c c c}
		\multicolumn{4}{c}{\textbf{Panel B:}} \\
		~~~\textbf{$\sigma(t = 0)$}~~~~~~~&\textbf{Fourier Transform, \%}&\textbf{KM approx., \%}&\textbf{Diff., pp.}\\\hline
	\textbf{0.1}&69.0697&69.594&-0.52425\\
	\textbf{0.2}&64.6821&62.9238&1.7583\\
	\textbf{0.3}&62.3638&61.0968&1.267\\
	\textbf{0.4}&61.121&60.7674&0.35363\\
	\textbf{0.5}&60.4721&60.6394&-0.16725\\
	\textbf{0.6}&60.1788&60.5942&-0.41539\\
	\textbf{0.7}&60.1131&60.6083&-0.49521\\
	\textbf{0.8}&60.2009&60.6637&-0.46274\\
	\textbf{0.9}&60.3968&60.7424&-0.34556\\
	\textbf{  1}&60.6711&60.8272&-0.15604\\
	\textbf{1.1}&61.004&60.9022&0.10177\\\hline
	\end{tabular}
	\floatfoot{\textit{Notes: Time to maturity $0.25$, $\theta = 0.2$, $\kappa = 4.0$, $\omega = 0.1$, $r = 0.0953$, $\rho = -0.5$. Black-Scholes volatility $\sigma_0 = \sigma = 0.2$. Panel A: Strike Price = 100 and spot volatility $\sigma = 0.2$. Panel B: Stock price = strike price = 100.}} \label{tab.SZgreeks}
\end{table}

\begin{table}[H]
	\centering
	\caption{Approximating $\Gamma$ for the SZ-model}
	\begin{tabular}{c|c c c}
		\hline
		\multicolumn{4}{c}{\textbf{Panel A:}} \\
		\textbf{Stock price}&\textbf{Fourier Transform, \%}&\textbf{KM approx., \%}&\textbf{Diff., pp.}\\\hline
		\textbf{ 80}&0.57373&0.7438&-0.17007\\
		\textbf{ 85}&2.0189&2.0166&0.0022704\\
		\textbf{ 90}&3.9028&3.4479&0.45492\\
		\textbf{ 95}&4.4477&4.2612&0.18651\\
		\textbf{100}&3.581&3.8587&-0.27766\\
		\textbf{105}&2.3676&2.6194&-0.25175\\
		\textbf{110}&1.3973&1.4584&-0.061154\\
		\textbf{115}&0.77065&0.7635&0.0071461\\
		\textbf{120}&0.40811&0.3993&0.0088082\\\hline
	\end{tabular}
	\begin{tabular}{c|c c c}
		\multicolumn{4}{c}{\textbf{Panel B:}} \\
		~~~\textbf{$\sigma(t = 0)$}~~~~~~~&\textbf{Fourier Transform, \%}&\textbf{KM approx., \%}&\textbf{Diff., pp.}\\\hline
		\textbf{0.1}&0.046477&-0.077595&0.12407\\
		\textbf{0.2}&0.03581&0.038587&-0.0027766\\
		\textbf{0.3}&0.028296&0.028236&6.0567e-05\\
		\textbf{0.4}&0.023168&0.023232&-6.4091e-05\\
		\textbf{0.5}&0.019529&0.019976&-0.00044687\\
		\textbf{0.6}&0.016836&0.017531&-0.00069402\\
		\textbf{0.7}&0.014772&0.015584&-0.00081176\\
		\textbf{0.8}&0.013142&0.013988&-0.00084634\\
		\textbf{0.9}&0.011823&0.012652&-0.0008288\\
		\textbf{  1}&0.010735&0.011519&-0.00078471\\
		\textbf{1.1}&0.0098209&0.010544&-0.00072277\\\hline
	\end{tabular}
	\floatfoot{\textit{Notes: Time to maturity $0.25$, $\theta = 0.2$, $\kappa = 4.0$, $\omega = 0.1$, $r = 0.0953$, $\rho = -0.5$. Black-Scholes volatility $\sigma_0 = \sigma = 0.2$. Panel A: Strike Price = 100 and spot volatility $\sigma = 0.2$. Panel B: Stock price = strike price = 100.}} \label{tab.SZgreeksGam}
\end{table}

	\begin{table}
		\centering
		\caption{Approximating SZ's $\mathscr{V}$}
		\begin{tabular}{c|c c c}
			\hline
			\multicolumn{4}{c}{\textbf{Panel A:}} \\
			\textbf{Stock price}&\textbf{Analytic $\mathscr{V}$}&\textbf{KM approx.}&\textbf{Diff., abs.}\\
			\textbf{ 80}&0.86888&0.49785&0.37103\\
			\textbf{ 85}&3.5542&3.5823&-0.028132\\
			\textbf{ 90}&8.2878&9.6389&-1.3511\\
			\textbf{ 95}&11.7159&12.9976&-1.2817\\
			\textbf{100}&11.6291&11.2967&0.33233\\
			\textbf{105}&9.2815&8.3014&0.98002\\
			\textbf{110}&6.4823&6.2093&0.27295\\
			\textbf{115}&4.163&4.3411&-0.1781\\
			\textbf{120}&2.5337&2.4634&0.070352\\\hline
		\end{tabular}
		\begin{tabular}{c|c c c}
			\multicolumn{4}{c}{\textbf{Panel B:}} \\
			~~~\textbf{$\sigma(t = 0)$}~~~~~~~&\textbf{Fourier Transform, \%}&\textbf{KM approx., \%}&\textbf{Diff., abs.}\\
			\textbf{0.1}&10.438&11.2173&-0.77928\\
			\textbf{0.2}&11.6291&11.2967&0.33233\\
			\textbf{0.3}&12.1307&87.6963&-75.5656\\
			\textbf{0.4}&12.3724&-1048.4084&1060.7808\\
			\textbf{0.5}&12.4988&-35868.8908&35881.3896\\
			\textbf{0.6}&12.5661&-372544.5672&372557.1333\\
			\textbf{0.7}&12.5994&-2388099.2386&2388111.8381\\
			\textbf{0.8}&12.6113&-11380315.5007&11380328.112\\
			\textbf{0.9}&12.6084&-44060486.9252&44060499.5336\\
			\textbf{  1}&12.5946&-145970936.3363&145970948.931\\
			\textbf{1.1}&12.5724&-428003736.4368&428003749.0092\\\hline
			\end{tabular}
		\floatfoot{\textit{Notes: Time to maturity $0.25$, $\theta = 0.2$, $\kappa = 4.0$, $\omega = 0.1$, $r = 0.0953$, $\rho = -0.5$. Black-Scholes volatility $\sigma_0 = \sigma = 0.2$. Panel A: Strike Price = 100 and spot volatility $\sigma = 0.2$. Panel B: Stock price = strike price = 100.}} \label{tab.SZVega}
	\end{table}

\clearpage
\section{Commodity futures with stochastic volatility} \label{CommModel}

All of the models considered so far had in common that the underlying itself was assumed to follow a geometric Brownian motion. Only stochastic volatility was assumed to be mean-reverting. While this is a fairly standard assumption in the modeling of financial derivatives like options or futures e.g. on stocks, the prices of commodities are usually modeled by mean-reverting processes. Such mean-reversion of the underlying is well documented for a wide range of commodities. \citet[Chapter 2]{Lutz09} provides a good discussion on the empirical evidence and theoretical justifications of the mean-reversion property of commodity prices.\\
In the following, I consider a model due to \citet{Lutz09} in which both, the price of the underlying as well as the stochastic variance, perform mean-reversion.
\begin{align} 
dX(t) =& \left(\eta\left(\alpha - X(t)\right) - \frac{1}{2}v(t)\right) + \sqrt{v(t)}dW_1(t) \label{Lutz1} \\
dv(t) =& \kappa\left(\theta - v(t)\right) + \omega \sqrt{v(t)}dW_2(t)  \label{Lutz2} \\
dW_1(t) dW_2(t) =& \rho dt \nonumber \\
S(t) =& e^{X(t)} \nonumber
\end{align}  
Stochastic variance is modeled by a Heston-liked square-root process, which in the previous experiments had been the most compatible with KM's approximation. The commodity price itself follows an Ornstein-Uhlenbeck process, where the above representation results from setting $X(t)= ln(S(t))$ and then applying Ito's lemma. This combination of processes is especially interesting since KM's approximation yielded quite good results for the square-root variance process but was unstable for Ornstein-Uhlenbeck volatility. The model above combines both processes.  While in the previous sections I always considered plain vanilla call options, I will instead consider Futures contracts in this section. In the absence of market frictions futures and forward prices are equal if the interest rate is deterministic.\footnote{See \citet[Proposition 3]{CIR81}, p. 325.} As in the previous section I will assume a constant interest rate, and hence future and forward prices will be considered the same in the following.


\subsection{A baseline model for commodity futures}
Since I will focus on Futures in this section, the Black-Scholes model can not serve as the baseline model here. However, there are several choices of possible models of commodity futures prices available in the literature. A suitable baseline to approximate the above futures price would be \textit{Model 1} from \citet{Schwartz97}, which I will refer to as Schwartz model in the following. The model assumes that the log of the commodity price follows an Ornstein-Uhlenbeck process with constant volatility
\begin{align}
dx(t) &= \kappa \left(\alpha - x(t)\right)dt + \sigma_0 dW(t) \label{Schwartz} \\
\alpha &= \mu - \frac{\sigma_0^2}{2\kappa} \nonumber \\
S(t) &= e^{x(t)} \nonumber
\end{align} 
Note that here it is assumed that the $x(t) = ln(S(t))$ follows an Ornstein-Uhlenbeck process, not $S(t)$ itself. Hence, the processes for the underlying differs slightly between the true and the baseline model. It is well known that a variable that follows an Ornstein-Uhlenbeck process has a Gaussian distribution. Hence, by using standard results from statistics it is easily verified that $x(t)$ has the following distribution\footnote{See \citet{Schwartz97}, p. 926.}
\begin{align}
x \sim N\left(\underbrace{e^{-\kappa T}x + \left(1 - e^{-\kappa T}\right)\alpha}_{= \mathbb{E}(x)}, \underbrace{\frac{\sigma_0^2}{2\kappa}\left(1 - e^{-2\kappa T}\right)}_{Var(x)}\right)
\end{align}
Since the price of a futures contract with maturity $T$ equals the expectation of the price of the underlying at that maturity date, the price is given by\footnote{See for this model \citet{Schwartz97}, p. 927, or \citet{CIR81}, p. 339 in general.}
\begin{align}
F(S,T) = \mathbb{E}(S(T)) = exp\left[e^{-\kappa T}x + \left(1 - e^{-\kappa T}\right)\alpha + \frac{1}{2}\frac{\sigma_0^2}{2\kappa}\left(1 - e^{-2\kappa T}\right)\right] \label{BaseSol}
\end{align}

\subsection{KM expansion}
KM's series expansion can again be computed according the principles outlined in section \ref{AssetApprox}. It is straightforward to derive drift vector and the covariance matrix of the commodity model in (\ref{Lutz1}) and (\ref{Lutz2})
\begin{align}
\mu(X,t) &= \left[\eta\left(\alpha - X(t)\right) - \frac{1}{2}v(t) ~~ \kappa\left(\theta - v(t)\right)\right]' \text{ ,} \\
\sigma^2(X,t) &=
\begin{bmatrix}
v(t) & \rho \omega v(t) \\
\rho \omega v(t) & \omega^2v(t)
\end{bmatrix} \label{FutureCov}
\end{align}
Whereas the model's PDE is
\begin{align}
\dfrac{\partial F}{\partial t} + \left(\eta\left(\alpha - X(t)\right) - \frac{1}{2}v(t)\right)\dfrac{\partial F}{\partial X} + \kappa\left(\theta - v(t)\right)\dfrac{\partial F}{\partial v} + \\ + \frac{1}{2}\left(v(t)\dfrac{\partial^2 F}{\partial X^2} + \omega^2v(t)\dfrac{\partial^2 F}{\partial v^2}\right) + \rho\omega v(t)\dfrac{\partial^2 F}{\partial X \partial v} = 0 \nonumber
\end{align}
Or, in terms of the operator $\mathscr{L}$ given in (\ref{L}), simply
\begin{align}
\mathscr{L} F(X,t) = 0 \label{CommPDEshort}
\end{align}
With the boundary condition $F(X,T) = X(T)$. Note that the PDE in (\ref{CommPDEshort})  lacks the $rF(X,t)$ term usually appearing in the PDEs of other derivatives.\\
It is well established in the financial literature that the price of every derivative should be determined by the fundamental PDE (see also (\ref{fundaPDE}))
\begin{align}
\mathscr{L} F(X,t) + c(X,t) = rF(X,t) \label{fundamentalPDE}
\end{align}  
Where $c(X,t)$ denotes a cash flow from the underlying asset. In all of the previous sections I assumed $c(X,t) \equiv 0$, as this was also done by KM. However, doing so in case of Futures might be misleading. \citet[Propsoition 7]{CIR81} states that the price of a Futures contract equals the value of an asset that receives continual payments of $c(X,t) = rX(t)$ and additionally $F(X,T) = X(T)$.\footnote{See \citet{CIR81}, p. 338.} Consequently (\ref{fundamentalPDE}) reduces to (\ref{CommPDEshort}). \citet[Theorem 6.1]{Friedman75} shows that a direct solution to (\ref{CommPDEshort}) would be given by $F(X,t) = \mathbb{E}_t\left[X(T)\right]$. By applying Fourier transform techniques \citet{Lutz09} provides an analytic (semi-)closed-form solution to this expectation. However, KM's closed-form approximation approach instead is based on the Feynman-Kac representation of a PDE describing the difference between the true and the baseline model. By denoting with $\Delta F(X,t) =  F(X,t) -  F_0(X,t)$ the difference between the Futures price under the true and the Schwartz model, the relevant PDE can be written as
\begin{align}
\mathscr{L}\Delta F(X,t) + \delta(X,t) &= 0 \label{FutureDiffPDE}\\
\text{s.t. }\quad \quad \quad \quad \quad \quad d(X) &= 0 \nonumber
\end{align} 
Where $\delta(X,t)$ is the same as in (\ref{Dif1}), and thus denoting the difference in the driving forces between the true and the baseline market. The boundary condition $d(X)$ is equal to zero, since both models, the true and the baseline, have the same final payoff. 
In order to find a solution to (\ref{FutureDiffPDE}) similar to (\ref{Theo1}) recall the outline of KM's approach in section \ref{AssetApprox}, there it was mentioned that this representation follows from \citet[Theorem 7.6]{Shreve}. Further note that (\ref{FutureDiffPDE}) differs from the general pricing bias PDE in (\ref{DiffPDE}) only by the term $R(X,t)\Delta F(X,t)$. Such that effectively in (\ref{FutureDiffPDE}) the instantaneous short-term rate is $R(X,t) = 0$. Hence, setting $R(X,t) = 0$ and $d(X) = 0$ in (\ref{Theo7.6}) yields 
\begin{align}
F(X,t) = F_0(X,t) + \int_{t}^{T}\mathbb{E}_{t,X}\left[\delta(X,s)\right]dt \label{FeynmanKacKMFut}
\end{align}
The case shown here is analogous to the case described in \citet[p. 397]{Shreve}. Therein  $R(X,t) = \delta(X,t) = 0$ whereas $d(X) \neq 0$. Then the solution to (\ref{FutureDiffPDE}) could be found again through \citet[Theorem 6.1]{Friedman75} to be $\Delta F(X,t) = \mathbb{E}_t\left[d(X)\right]$. Analogous to that case (\ref{FeynmanKacKMFut}) is the solution to (\ref{FutureDiffPDE}). Based on this representation of the solution KM's series expansion is given as (see section \ref{AssetApprox})
\begin{align}
F(X,t) = F_0(X,t) + \sum_{n = 0}^{N}\dfrac{\delta_n(z,t)\left(T-t\right)^{n+1}}{(n+1)!}
\end{align}
Since $R(X,t)$ was set to zero to derive (\ref{FeynmanKacKMFut}) the corrective terms develop according to
\begin{align}
\delta_n(X,t) = \mathscr{L}\delta_{n-1}(X,t) \text{ for } n > 0 \label{FutureCorrec}
\end{align}
Whereby the initial pricing error $\delta_0$ is still given by (\ref{initdel}). Note that the expectation in (\ref{FeynmanKacKMFut}) alternatively could have been evaluated by Monte Carlo integration. However, in such a case the obtained solution would not have a closed-form and hence there would be no advantage compared to Monte Carlo integration to evaluate the direct solution mentioned before.\\
Using the covariance matrix in (\ref{FutureCov})  together with the definition in (\ref{initdel}) it is straightforward to derive the initial pricing error as
\begin{align}
\delta_0 = \frac{1}{2}\left(v(t) - \sigma_0^2\right)\dfrac{\partial^2 F_0}{\partial X^2} - \frac{1}{2}v(t)\dfrac{\partial F_0}{\partial X}
\end{align}
While the first element in the initial pricing error is the same convexity adjustment as in the previous experiments, there is second element which accounts for the different drift terms of the SDE of the underlying in the baseline and the true model. Table \ref{tab.CommFut} below shows how the series of corrective terms developed according to the rule in (\ref{FutureCorrec}).

\clearpage
\begin{landscape}
	\newcolumntype{C}{>{\centering\arraybackslash} m{18cm} } 
	\renewcommand{\arraystretch}{2.0}
	\begin{table}[H]
		\centering
		\caption{Iterations of the pricing error for the Commodity Futures model}
		\begin{tabular}{m{1cm}|C}
			\hline
			\textbf{n} & \textbf{Pricing error $\delta_n(S,t)$} \\\hline\hline
			\textbf{0} & $\frac{1}{2}\left(v(t) - \sigma_0^2\right)\dfrac{\partial^2 F_0}{\partial X^2} - \frac{1}{2}v(t)\dfrac{\partial F_0}{\partial X}$ \\\hline
			\textbf{1} & $\dfrac{\partial \delta_0}{\partial t} + \left(\eta\left(\alpha - X\right) - \frac{1}{2}v(t)\right)\dfrac{\partial \delta_0}{\partial X} + \kappa\left(\theta - v(t)\right)\dfrac{\partial \delta_0}{\partial v} + \frac{1}{2}v(t)\dfrac{\partial^2 \delta_0}{\partial X^2} + \rho \omega v(t)\dfrac{\partial^2 \delta_0}{\partial X \partial v}$ \\\hline
			\textbf{2} & $\dfrac{\partial \delta_1}{\partial t} + \left(\eta\left(\alpha - X\right) - \frac{1}{2}v(t)\right)\dfrac{\partial \delta_1}{\partial X} + \kappa\left(\theta - v(t)\right)\dfrac{\partial \delta_1}{\partial v} + \frac{1}{2} \left( v(t)\dfrac{\partial^2 \delta_1}{\partial X^2} + \omega^2v(t)\dfrac{\partial^2 \delta_1}{\partial v^2} \right) + \rho \omega v(t)\dfrac{\partial^2 \delta_1}{\partial X \partial v}$ \\\hline
			\textbf{$\vdots$} & $\vdots$ \\\hline
			\textbf{N} & $\dfrac{\partial \delta_{N-1}}{\partial t} + \left(\eta\left(\alpha - X\right) - \frac{1}{2}v(t)\right)\dfrac{\partial \delta_{N-1}}{\partial X} + \kappa\left(\theta - v(t)\right)\dfrac{\partial \delta_{N-1}}{\partial v} + \frac{1}{2} \left( v(t)\dfrac{\partial^2 \delta_{N-1}}{\partial X^2} + \omega^2v(t)\dfrac{\partial^2 \delta_{N-1}}{\partial v^2} \right) + \rho \omega v(t)\dfrac{\partial^2 \delta_{N-1}}{\partial X \partial v}$ \\\hline
		\end{tabular} \\
		\floatfoot{\textit{Notes: $\partial^2 \delta_0/\partial v^2 = 0$}}
		\label{tab.CommFut}
	\end{table}
\end{landscape}
\clearpage

\subsection{Numerical accuracy}
\begin{figure}[H]
	\centering
	\includegraphics[scale=.4]{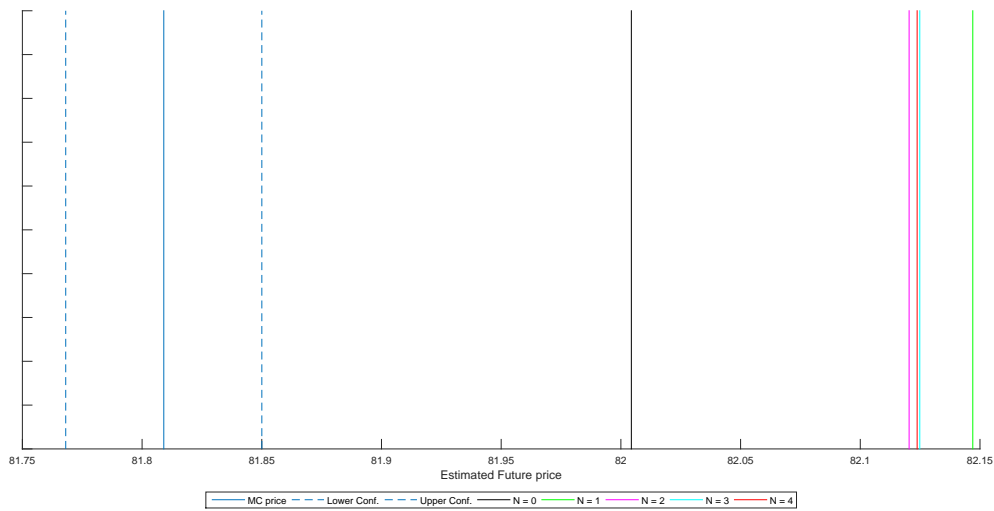}
	\caption{MC result vs. KM approximation - maturity 0.5 years}
	\floatfoot{\textit{Notes: KM's approximation compared to MC result for the stochastic variance commodity futures model. A discription of the MC approach is provided in Appendix \ref{CommRefSol}. Parameter values: $S = 80$, $\bar{S} = 85$ (i.e. $\alpha = log(85)$), $\eta = 1.0$, $T - t = 0.5$, $\omega = 0.2$, $v(t) = 0.04$, $\kappa = 1.0$, $\theta = 0.05$, and $\rho = -0.5$. Baseline model is Model I from \citet{Schwartz97}.}}
	\label{fig.Future1}
\end{figure}

\begin{figure}[H]
	\centering
	\includegraphics[scale=.4]{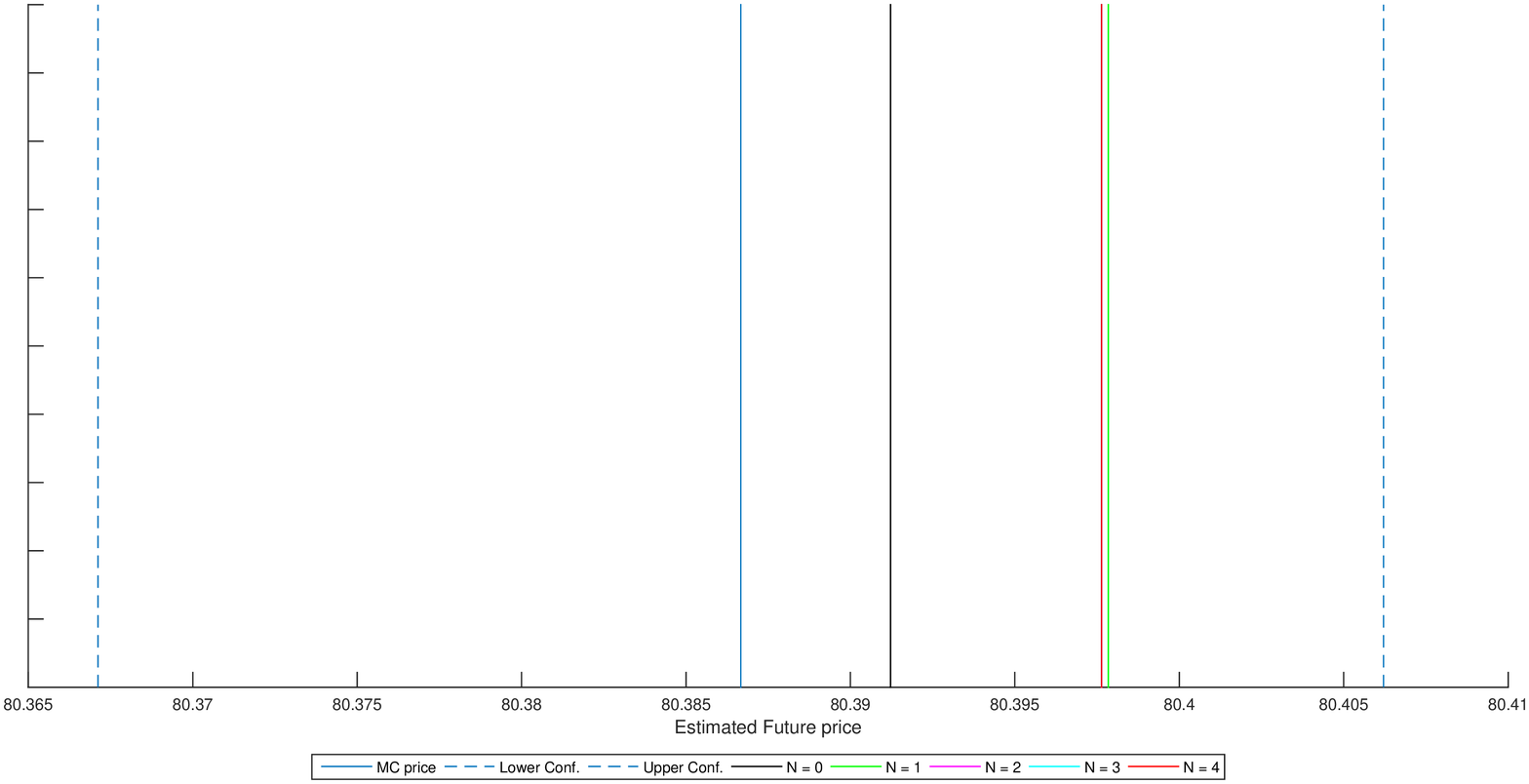}
	\caption{MC result vs. KM approximation - maturity 0.25 years}
	\floatfoot{\textit{Notes: KM's approximation compared to MC result for the stochastic variance commodity futures model. A discription of the MC approach is provided in Appendix \ref{CommRefSol}. Parameter values: $S = 80$, $\bar{S} = 85$ (i.e. $\alpha = log(85)$), $\eta = 1.0$, $T - t = 0.25$, $\omega = 0.2$, $v(t) = 0.04$, $\kappa = 1.0$, $\theta = 0.05$, and $\rho = -0.5$. Baseline model is Model I from \citet{Schwartz97}.}}
	\label{fig.Future2}
\end{figure}

\begin{figure}[H]
	\centering
	\includegraphics[scale=.4]{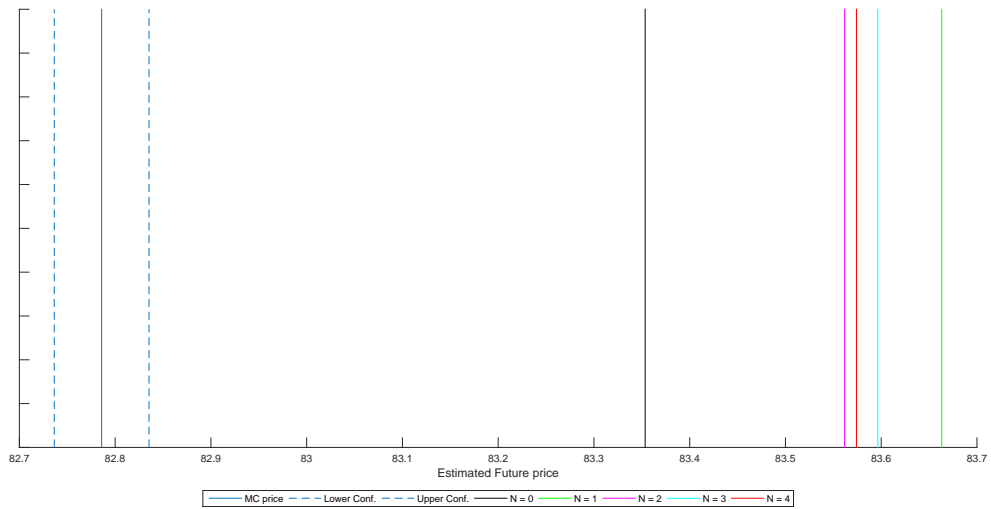}
	\caption{MC result vs. KM approximation - maturity 1.0 years}
	\floatfoot{\textit{Notes: KM's approximation compared to MC result for the stochastic variance commodity futures model. A discription of the MC approach is provided in Appendix \ref{CommRefSol}. Parameter values: $S = 80$, $\bar{S} = 85$ (i.e. $\alpha = log(85)$), $\eta = 1.0$, $T - t = 1.0$, $\omega = 0.2$, $v(t) = 0.04$, $\kappa = 1.0$, $\theta = 0.05$, and $\rho = -0.5$. Baseline model is Model I from \citet{Schwartz97}.}}
	\label{fig.Future3}
\end{figure}

To assess the accuracy of KM's approximation for this model I use the parameter values suggested by \citet{Lutz09}: $S = 80$, $\bar{S} = 85$ (i.e. $\alpha = log(85)$), $\eta = 1.0$, $\omega = 0.2$, $v(t) = 0.04$, $\kappa = 1.0$, $\theta = 0.05$, and $\rho = -0.5$ together with times to maturity of 12, 6, and 3 months. Following the same argumentation as in the previous sections the nuisance parameter of the baseline model is set to $\sigma_0 = \sqrt{v(t)}$. Figure \ref{fig.Future1} visualizes the MC result as well as its 95\% confidence interval together with the results of the KM approximation for $N = 0$ to $N = 4$ corrective terms for these parameter values. The simulated value of $\mathbb{E}_t\left[e^{X(T)}\right]$ is 81.8091 with 95\% confidence interval of $\left[81.7681~;~81.8500\right]$.\footnote{For the same set of parameters \citet{Lutz09} reports a MC result of 81.8016 and a 95\% confidence interval of $\left[81.7941~;~81.8090\right]$. \citet{Lutz09} also reports an analytic price for theses parameters of 81.8008. This means that my own MC results overestimates the analytic price by 0.0101\%. Hence, I believe that the use of my MC procedure won't have any significant influence on the drawn conclusions.} The results of KM's approximation overestimate the MC for each of the considered orders of the approximation. Also non of the KM approximations lies inside the estimated confidence interval. Nevertheless, the approximation errors are rather small, ranging between 0.2387\% and 0.4131\% depending on the order of the approximation. However, the series expansion fails to converge to the reference value after five corrective terms. An approximation using only the initial corrective term would lead to the most accurate result (Approximation error: 0.2387\%). The precision is then reduced to it's lowest level by including one more element in KM's expansion (Approximation error: 0.4131\%). However, with including more corrective terms there seems be slight improvements. Overall, the behavior of the series expansion for the first five elements suggests that there is a significantly larger number of corrective terms required to achieve clear convergence. While the approximation errors are such low that the accuracy seems to be sufficient for most practical applications, this hints at a problematic aspect of KM's approach in general: It is not always clear which number of corrective terms yields an optimal approximation.\\
Figure \ref{fig.Future2} shows the result of the approximation when reducing the time maturity to one month. The overall pattern is the same as before. An approximation only using the initial corrective term would yield the most precise approximation. Adding only one additional corrective term yields the least precise result. Adding more corrective terms then improves the approximation slightly again, whereas the results for $N = 2$ to $N = 4$ are identical. For a maturity of one month the results of KM's approximation lie within the 95\% confidence interval for each considered order of the approximation. Note that even though the use of all corrective terms would not be optimal the accuracy is nevertheless extremely high, with an approximation error of just 0.0136\% for the $N = 4$ approximation. However, using only the initial corrective term yields an approximation error as small as 0.0057\%. \\
Figure \ref{fig.Future3} depicts the result of KM's approximation for a time to maturity of one year. The series expansion behaves as expected in that case. The accuracy is lower than both previous examples. Again the first order approximation yields the most and the second order approximation yields the least precise result. Non of the approximations falls inside the 95\% confidence interval of the MC simulation. However, also with a maturity of one year all approximation errors remain remarkably low. For the approximation of order $N = 4$ the error is just 0.9521\% and for the expansion of order $N = 0$ the error is 0.6856\%. 


\clearpage
\section{Conclusion} \label{Conclude} 
 
The main theme of this thesis was the analysis of the behavior of the closed-form approximation approach developed by \citet{KM2011}. In terms of derivatives pricing models I focused on stochastic volatility models as they yield a convenient set-up to test the accuracy and convergence behavior of the approximations. I applied KM's approach to obtain closed-form approximation of derivative prices as well as Greeks. \\  
KM's approximation yields remarkably precise approximations for the Heston and the CEV model over a wide range of parameter choices. The series expansion converges fast, such that five corrective terms are sufficient to approximate asset prices in these models. Since the approximation has closed-form it is straightforward to obtain greeks, where the performance in the approximation of greeks in the Heston is equally well as the approximation of option prices.\\
Especially in the case of the CEV model KM's approximation reveals it strength in terms of computational efficiency. While computation of reference values via Monte Carlo simulation requires substantial computation time, KM's closed-form approximation yield results virtually immediately.\\
Both models, CEV and Heston, were approximated for two different sets of parameters. For the Heston model the approximation mostly performs better for ITM than for OTM options. Whereby this difference in accuracy becomes more pronounced the fewer corrective terms in the series expansion are considered. Additionally the effect of moneyness on the accuracy of the approximation decreases with the level correlation and with time to maturity, whereas the latter had a more significant impact on put than on call options. The approximation of the CEV model appears to be less robust against changes in the model parameter, especially for high levels of elasticity of variance. An exception seems to be the sensitivity to correlation, which is almost offset by a high elasticity of variance.\\  
The approximation of the SZ model performed significantly worse. Specifically, time to maturity influences the accuracy of the approximation much stronger than it was the case for the Heston model. While in the Heston model the approximation remains stable for maturities of up to two years, in the SZ model instability already occurs for maturities longer than six month. Since the only difference between the two models is the variance/volatility process, the instability is most likely related to the usage of the Ornstein-Uhlenbeck process in the SZ model. The Heston model is affine in the sense of exponentially linear characteristic functions, while the SZ model as well as the CEV model are non-affine in this sense. The analyzes of the approximations for the CEV model indicate that accuracy of KM's approximation is lower for non-affine models as it is for affine models. Nevertheless, the convergence behavior of the series expansion is similar in the CEV and the Heston case, suggesting the instability might be not related to the non-affine structure of the SZ model. These results suggest that the convergence of KM's series expansion might be not uniform over time. A possible explanation of this could be found in \citet{Kimmel08}, who links diverging behavior of series expansion solutions to PDEs to singularities for some maturities. \citet{Kimmel08} suggests to extend the area of convergence of such series expansion by replacing the time variable by a non-linear function of time. KM mention that the approach developed by \citet{Kimmel08} could be used to improve convergence of their approximations, but do not describe how this can be done. \citet{Kimmel08} applies his \textit{time-change} approach only to univariate diffusions and indicates that an extension to multivariate diffusion might be not straightforward in many cases.\footnote{See \citet{Kimmel08}, p. 38.} Hence, the combination of the approaches of \citet{Kimmel08} and KM might be a promising possibility for future research. If only maturities below six months are considered, the approximation of the SZ model is overall more precise if correlation is zero.\footnote{As already noted, in this case the SZ model is identical to the S\&S model.} However, while in the zero correlation case an approximation using five corrective terms yields the most precise results, an approximation using six corrective terms is superior if correlation is non-zero. Similar to the Heston and the CEV model cases the approximation performs better for ITM than for OTM options, whereby again this effect increases with the level of correlation. The accuracy in the approximation of greeks in the SZ model mimics the accuracy in the approximation of option prices. However, the approximation seems to be unstable with respect to spot volatility.\\
Finally, I applied KM's approximation to a model of commodity future prices which models the price of the underlying through an Ornstein-Uhlenbeck process and stochastic variance through an Heston-like square-root process. The series expansion was developed for up to five corrective terms, generally showing neither clear convergence nor clear divergence for maturities of up to one year. It seems like the approximation is converging to a slightly biased futures price instead to the true price.  However, approximation errors remained below one percent. Most noteworthy, for the approximation of the commodity futures model the approximation using only the first corrective terms always yields the highest accuracy. Considering that also for the SZ model with zero correlation the approximation using six terms is less precise than the approximation using five terms and that for the CEV and the Heston model in some cases also the approximation using four terms was slightly more accurate than the one using five terms, this indicates that it might be difficult to determine the optimal order of approximation without solving the model under investigation also through another procedure to obtain reference values.\\
Nevertheless, KM's approach yields a viable alternative to Monte Carlo solutions or Runge-Kutta solutions for the futures model due to significantly higher computational efficiency. The closed-form approximation could also be considered as an alternative to the analytic solution of the model, which is only available in terms of Kummer functions of the first and second kind.\footnote{See \citet{Lutz09}, p. 64.} While these functions might be difficult to implement in some programming languages, KM's series expansion is straightforward to implement in most languages.\\
Overall, KM's series expansion yields satisfying results for all models considered in this thesis except for the SZ model. Due to the very abstract nature of the underlying assumptions it appears to be difficult to point at specific reason for differences in the convergence behavior or sensitivities of the approximation towards specific parameter values. Even though the implementation of the series expansion is straightforward, the derivation is tedious and prone to error if it is done manually. However, even if symbolic mathematical software such as Maple is used, the computation is restricted to only five terms, since the corrective terms are increasing in length too fast exceeding the computers memory. A major issue in this matter might be the structure of the chosen baseline model. For most of the examples in this thesis I used the Black-Scholes model as baseline. As mentioned in one of the previous sections if only five corrective terms are used, this already requires the computation of derivatives of the Black-Scholes model up to an order of ten, which effectively means taking derivatives and cross-derivatives of a normal distribution up this order. Although if symbolic mathematical software is used, the series expansion usually can only be developed for a view terms. Approaches like  of finite difference approximation for derivatives of the baseline model, as it as done by \citet{Catalan}, ease mathematical difficulties in developing the approximation but significantly slow down the computation speed. Additionally, the method then loses its feature of yielding pricing formulas in closed-form and thus the approach would lose its two main advantages. Hence, I would suggest three areas of future research regarding KM's approach. (i) The combination of \citet{Kimmel08}'s time-change approach with KM's series expansion to improve robustness regarding the time to maturity. (ii) A closer investigation of the iterative procedure of computing the corrective terms to derive shorter expressions for the pricing bias and thus enabling the computation of more terms in the series expansion. (iii) The development of a tractable test procedure of the underlying assumptions for specific models. This could be connected with an investigation of criteria to determine an optimal order of the approximation for specific models. (iv) Finally, a closer investigation of the interaction between true model and baseline model to determine optimal choices of the baseline.

\addcontentsline{toc}{section}{Appendix}

\addtocontents{toc}{\vspace{2em}}
\appendix

\clearpage
\section{Reference solutions}
This appendix provides descriptions on the Fourier transform solutions of the \citet{Heston93} model and the \citet{Schobel99} model as well as on the Monte Carlo solutions used for the CEV model and the commodity futures model.

\subsection{Fourier transform solution of the Heston model}
The analytic prices of the Heston model are obtained via Fourier transforms. The Matlab code which I use to compute these prices is provided by \citet{Ricardo}, who uses a formulation of the models characteristic functions that had been suggested by \citet{Gartheral06} instead of Heston's original formulation. One particular drawback of Heston's original formulation of the characteristic functions is the possibility of branch-cuts when evaluating the complex valued logarithm appearing in the characteristic functions. These branch-cuts might be problematic in the present context as I will compare the performance of KM's approximation under a variety of different parameter values. Hence, if issue of possible branch-cuts would be ignored, the validity of the analytic reference values could not be ensured for all parameter choices and thus potentially spoil the conclusions drawn from the analysis of the approximation errors. However, by using the characteristic function from \citet{Gartheral06} the argument of the logarithm in the characteristic function never crosses the negative real axis, such that branch cuts do not occur.\footnote{See \citet{Gartheral06}, p. 20.} Nevertheless, Gartherals formulation of the model is equivalent to Heston's original formulation.\footnote{See e.g. \citet{Zhu10}, p. 55.} \citet{Ricardo} uses the following pricing framework for the Heston model
\begin{align}
C_{Analytic}(S,t) &= S_0P_1 - e^{-r(T-t)}KP_2 \label{HestonCallEq}\\
&\text{with} \nonumber \\
P_1 &= \frac{1}{2} + \frac{1}{\pi}\int_{\infty}^{0}real\left(\frac{e^{-i\phi log(K)}\varphi(\phi - i)}{i\phi\varphi(-i)}\right)d\phi \nonumber \\
P_2 &= \frac{1}{2} + \frac{1}{\pi}\int_{\infty}^{0}real\left(\frac{e^{-i\phi log(K)}\varphi(\phi)}{i\phi}\right)d\phi \nonumber 
\end{align}
Where the characteristic function $\varphi(\cdot)$ is given in \citet{Ricardo}. Note that this approach slightly deviates from the one in \citet{Gartheral06}. Both \citet{Gartheral06} and \citet{Heston93} derive two distinct characteristic functions for the probabilities $P_1$ and $P_2$, while \citet{Ricardo} uses only one characteristic function for both probabilities. Also \citet{Ricardo} applies Gatherals method of deriving the characteristic function to $log(S(t))$, instead of $log(S(t)/K)$ as in \citet{Gartheral06}. However, \citet[Appendix B]{Ricardo} provides a prove for the equivalence of the two approaches.\\
In order to evaluate the real integrals in (\ref{HestonCallEq}) direct integration via Matlabs \texttt{integral()} function is used, which employs global adaptive quadrature to numerically evaluate integrals.\footnote{See Matlab help file for more details. \citet[Ch. 4.3 to 4.5]{Zhu10} provide a good overview on integration algorithms in the application of Fourier transform methods.}

It is straightforward to obtain a (semi-)closed-form solution via Fourier transforms for European puts in the Heston framework\footnote{See e.g. \citet{ZhuDiss}, p. 35.}
\begin{align}
	Put(S,t) &= e^{-r(T-t)}KP_2 - S_0P_1 \label{HestonPutEq}\\
	\text{with } \nonumber \\
	P_1 &= \frac{1}{2} - \frac{1}{\pi}\int_{\infty}^{0}real\left(\frac{e^{-i\phi log(K)}\varphi(\phi - i)}{i\phi\varphi(-i)}\right)d\phi \nonumber\\
	P_2 &= \frac{1}{2} - \frac{1}{\pi}\int_{\infty}^{0}real\left(\frac{e^{-i\phi log(K)}\varphi(\phi)}{i\phi}\right)d\phi \nonumber
\end{align}
Where the characteristic function $\varphi$ is the same in (\ref{HestonCallEq}). I implemented the Heston Puts by adapting the Matlab code from \citet{Ricardo}.

\subsubsection{Analytic Greeks in the Heston model} \label{HestGreek}
The computation of these so called Greeks for the Heston model is straightforward, since integration and differentiation of the characteristic function can be interchanged. Note that I again use the formulation of the characteristic function developed by \citet{Ricardo}. Recalling the general structure of call option prices $C(S,t) = SP_1 - e^{-r(T-t)}KP_2$ and the definitions 1.) to 4.) in section \ref{sec.ApproxGreeks} the analytic greeks of the Heston model are defined as \footnote{See e.g. \citet{ZhuDiss}, p. 36.}
\begin{align}
\Delta_S ~=&~ P_1 \\
\Gamma_S ~=&~ \dfrac{\partial P_1}{\partial S} \\
\mathscr{V} ~=&~ S\dfrac{\partial P_1}{\partial v} - e^{-r(T-t)}K\dfrac{\partial P_2}{\partial v} \\
~~~~~ \nonumber \\
\text{with } \dfrac{\partial P_j}{\partial h} ~=&~ \frac{1}{\pi}\int_{0}^{\infty}\dfrac{\partial \varPsi_j(\phi)}{\partial h}d\phi \nonumber \\
\varPsi_1 ~&=~ real\left(\frac{e^{-i\phi log(K)}\varphi(\phi - i)}{i\phi\varphi(-i)}\right) \nonumber \\
\varPsi_2 ~&=~ real\left(\frac{e^{-i\phi log(K)}\varphi(\phi)}{i\phi}\right) \nonumber
\end{align}
Where $h$ equals either $S$ or $v$, and $j = 1,2$.

\subsection{Fourier transform solution of the SZ model}
The analytic prices for both, the S\&S- and the SZ-model had been obtained via Fourier transforms, whereas the S\&S prices can be computed by simply setting $\rho = 0$ in the characteristic function. The solution has the same structure as the Fourier solution of the Heston model in (\ref{HestonCallEq}). However, \citet{Schobel99} use two distinct characteristic functions to derive the probabilities $P_1$ and $P_2$. In order to implement the Fourier solution I adapted Matlab code that I used during the \textit{Numerical Methods in Finance} PC-Lab classes at University of Tuebingen. The Matlab program simply implements the two characteristic functions as given in \citet{Schobel99} and uses global adaptive quadrature\footnote{Numerical integration via Matlab's \texttt{integral()} function. The original PC-lab code used a simple trapezoidal algorithm via Matlab's \texttt{trapz()} function. However, while yielding numerically identical results, I found \texttt{integral()} to be significantly faster.} to numerically evaluate the involved integrals. Since the characteristic functions of the SZ-model also include a complex valued logarithm the same issue of possible branch-cuts arises as in the Heston model. While there I could conveniently use a reformulation of the characteristic functions to ensure stability of the reference solution, such a reformulation is, to my knowledge, not readily available for the SZ-model. Hence, I use a simple correction algorithm to adjust the complex logarithm whenever it's argument crosses the negative real axis. Appendix \ref{ComplexLog} provides a description of the used algorithm as well as the corresponding Matlab code.

\subsection{Digression on complex logarithms} \label{ComplexLog}
\begin{figure}[H]
	\centering
	\begin{tikzpicture}[scale=2]
	\draw[decoration={markings, mark=at position 0.10 with {\arrow[thick]{>}}}, 
	decoration={markings, mark=at position 0.35 with {\arrow[thick]{>}}}, 
	decoration={markings, mark=at position 0.65 with {\arrow[thick]{>}}}, 
	decoration={markings, mark=at position 0.85 with {\arrow[thick]{>}}}, 
	postaction={decorate}] (0,0) circle (1cm);
	\draw[thick, ->] (-1.5,0) -- (1.5,0) node[below=1pt,fill=white] {$Re(z)$};
	\draw[thick, ->] (0,-1.5) -- (0,1.5) node[above,fill=white] {$Im(z)$};
	\coordinate(A)at(1.5,0);
	\coordinate(O)at(0,0);
	\coordinate(Z)at(60:1);
	\coordinate(AA)at(1.0,1.0);
	\coordinate(AB)at(-1.0,1.0);
	\coordinate(AC)at(-1.0,-1.0);
	\coordinate(AD)at(1.0,-1.0);
	\draw [ultra thick, red](-1.5,0) --(O);
	\draw [ultra thick, blue] (O) -- 
	(Z)node[circle,fill=black,inner sep=0.025mm,transform shape,label=above:$Z_0$]{};
	\pic[draw,fill=orange,fill opacity=0.3,angle radius=4mm,"$\varphi$" opacity=1]
	{angle=A--O--Z};
	\draw (AA) node[above]{I};
	\draw (AB) node[above]{II};
	\draw (AC) node[above]{III};
	\draw (AD) node[above]{IV};
	\end{tikzpicture}
	\caption{$z_0$ path around the origin} \label{CompLogAxis}
\end{figure}
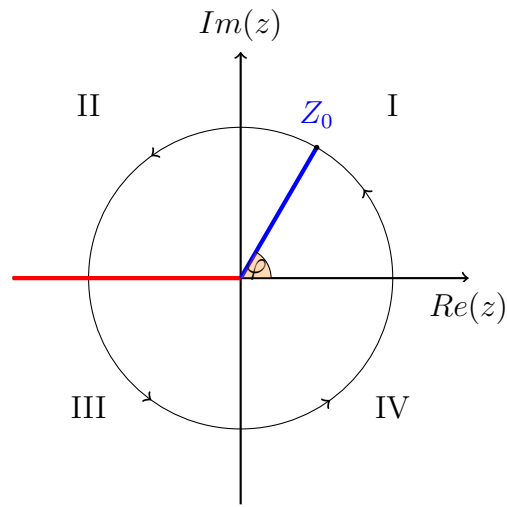

This appendix is an adapted excerpt from two assignments of the 2015 \textit{'Numerical Methods in Finance'}-class, which I wrote together with F. Slezak and A. Berg.\\
The issue of branch-cuts arose in the context of the Fourier transform solution to the Heston model in section \ref{HestonModel} as well as the SZ model in section \ref{SZModel}, since in both cases the computation of the option price includes the evaluation of a complex valued logarithm. It can be shown that the logarithm of a complex number $z$ has the following form
\begin{align} \label{eq:log2}
w &= log(z) = log(|z|) +  i (Arg(z) + 2k\pi) \\ \nonumber \\
&\textit{with } k = \pm 0,\pm 1, \pm 2,... \nonumber
\end{align}
Where $Arg(z)$ denotes the principal argument of $z$, i.e. $-\pi \leq Arg(z) < \pi$. Most commercial software packages like Matlab restrict calculation of complex logarithms to their principal values, i.e. $k$ is set to zero in equation (\ref{eq:log2}). As any value of $k$ would suffice to recover $e^w=z$, the choice of $k=0$ would not be a problem in an isolated computation.\footnote{See \citet{Zhu10}, p. 100.} However, \citet{Schobel99} were among the first to observe that this leads, at least for extreme parameter values, to discontinuities in the characteristic functions of their and other stochastic volatility models and thus to a wrong integration.\footnote{See \citet{Schobel99}, p. 28 and \citet{Lord2010}, p. 672.} The main issue in implementing equation (\ref{eq:log2}) in order to correctly compute the complex logarithm is that the appropriate $k$ value can only be found if the path of the complex number is known.\footnote{See \citet{Zhu10}, p. 100.} Every time $z$ crosses the negative part of the real axis in Figure \ref{CompLogAxis} the value of $k$ needs to be adjusted.\\
While \citet{Gartheral06} provided a reformulation of Heston's characteristic function such that the argument of the complex logarithm never crosses the red line in \ref{CompLogAxis}, I am unaware of any such reformulation for the SZ model. In order to obtain a numerically stable reference solution to the SZ model I used a correction algorithm which adjusts the number $k$ in (\ref{eq:log2}) every time the complex argument of the logarithm crosses the red line in Figure \ref{CompLogAxis}.\footnote{The used Matlab code was taken from the same Assignment as the content of this appendix section and also slightly adapted.} \\







\subsection{MC simulation for the CEV model} \label{CEVMCRef}
MC requires the discretization of the processes in (\ref{CEV1}) and (\ref{CEV2}), which may lead to the occurrence of negative variances in the sample path. Such values are problematic since in the process of generating a stock price sample path the square-root as well other roots of the variance need to be computed. This issue had been studied extensively in the financial literature, mostly in the context of the Heston model. Introducing the absolute value to the variance process as in (\ref{CEV2}) resembles the so called \textit{reflective assumption} which sets the rule: if $v < 0$, then $-v = v$. However, to implement this it is necessary to apply this rule to all roots of $v$. Another possibility to avoid negative variances would be to use the so called \textit{absorbing assumption} which sets the rule: if $v < 0$, then $v = 0$.\footnote{See e.g. \citet{Gartheral06}, p. 21.} If a sufficiently large number of trajectories and sample paths is chosen neither of the two procedures is likely to have no significant influence on the result, because the true variance is likely to be small if the discretization error leads to a negative simulated variance.\footnote{See \citet{Lee14}, p. 8.} KM do not point out which discretization scheme they use to obtain their MC results, neither they mention the number of trajectories nor the number of sample paths used in their simulation.\\
In order to discretize the processes in (\ref{CEV1}) and (\ref{CEV2}) I use the Milstein scheme\footnote{According to \citet{Glass03} the name Milstein scheme might be misleading since there are several methods due to Milstein. However, throughout this thesis I use the term to name the discretization scheme shown below.} which simply adds the next element of the Ito-Taylor expansion to the Euler scheme\footnote{i.e., the term $\frac{1}{2}Diff(h)Diff(h)'\left(Z^2_h(t_i) - 1\right)\Delta_i$ with $h = S~ or~ V$ is added. Where $Diff(h)$ denotes the diffusion term of the respective SDE and $Diff(h)'$ its derivative with respect to the variable denoted by $h$. (See e.g. \citet{Glass03}, p. 343)}
\begin{align}
S(t_{i+1}) &= S(t_i)\left[1 + r\Delta_i + \sqrt{|V(t_i)|}\sqrt{\Delta_i}Z_1(t_i) + \frac{1}{2}|V(t_i)|\left(Z^2_1(t_i) - 1\right)\Delta_i \right]  \\
V(t_{i+1}) &= V(t_i) + \kappa \left(\theta - |V(t_i)|\right)\Delta_i + \omega |V(t_i)|^{\gamma} \sqrt{\Delta_i}Z_2(t_i) \\ 
& \quad \quad + \frac{1}{2}\omega^2\gamma |V(t_i)|^{2\gamma - 1}\left(Z^2_2(t_i) - 1\right)\Delta_i \nonumber \\ \nonumber \\
\text{with } & Z_1(t_i) = Ran_1, \quad Ran_1 \sim N(0,1) \nonumber \\
& Z_2(t_i) = \rho Ran_1 + \sqrt{1 - \rho^2}Ran_2, \quad Ran_2 \sim N(0,1) \nonumber
\end{align}
Where $\Delta_i$ denotes the length of time step $i$. Note that I used the \textit{reflective assumption} to handle negative simulated variances if they occur. Applying the Milstein discretization instead of the Euler discretization do not increase accuracy of the MC results since both schemes are of weak order one.\footnote{See \citet{Glass03}, p. 347.} Nevertheless, using the Milstein scheme do not significantly increase the computational burden, but reduces the frequency of negative variances in the sample path.\footnote{See \citet{Gartheral06}, p. 22.} Hence, \citet{Gartheral06} advocates for always using the Milstein discretization for the Heston model. Due to the similarity between the CEV and the Heston model the positive effect of applying the Milstein scheme should also apply in this case. This further mitigates the effect of the choice between \textit{reflective} and \textit{absorbing} assumption. Additionally, I use 500 equally sized time steps and 20,000 sample paths in all simulations below.\footnote{The pseudo-random numbers are obtained via Matlabs multiple recursive generator (\textit{mrg32k3a}).}

\subsection{MC simulation of the Commodity Future} \label{CommRefSol}
\begin{figure}[H]
	\centering
	\begin{subfigure}{.9\textwidth}
		\centering
		\includegraphics[width=.9\linewidth]{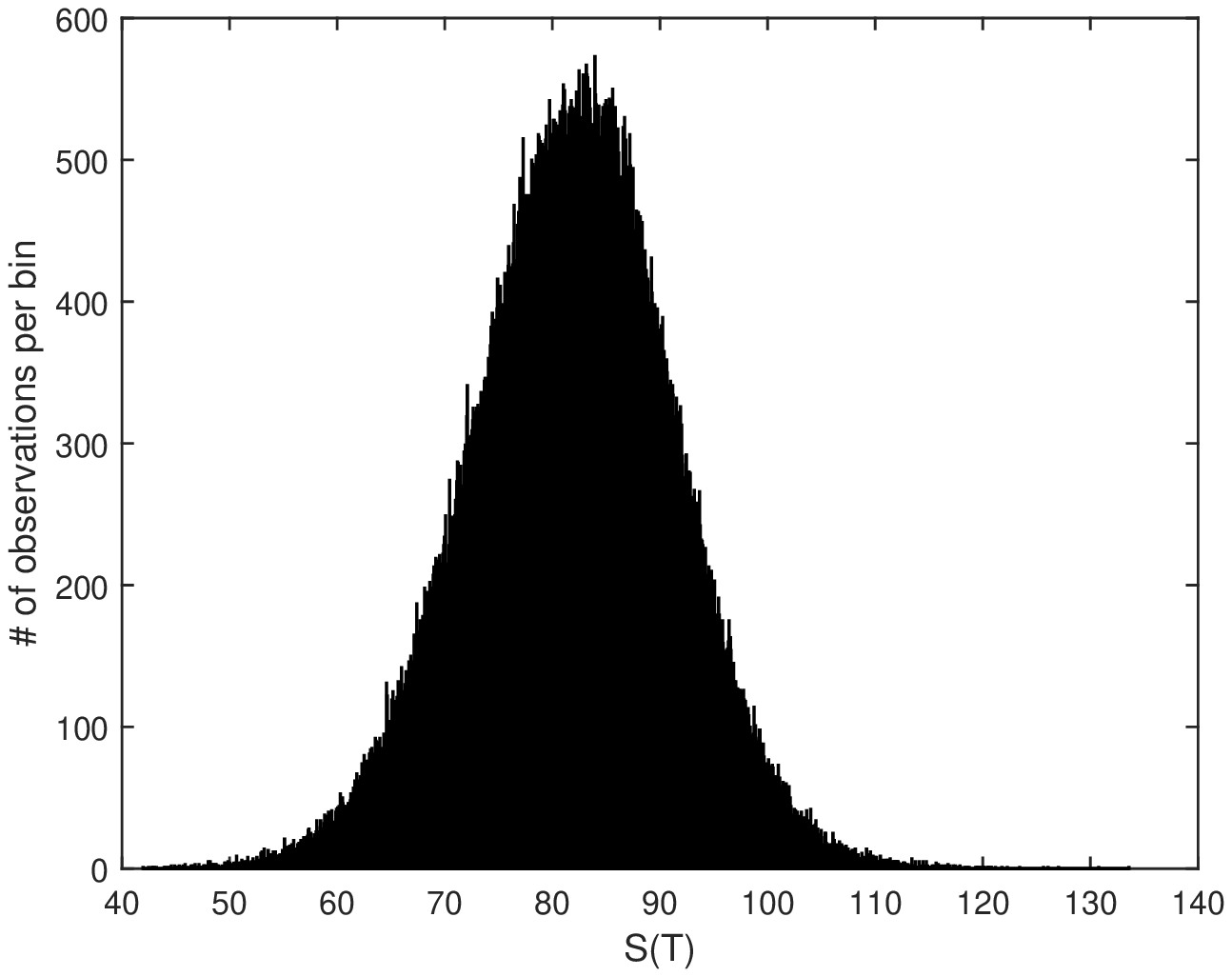}
		\caption{Histogram of simulated commodity prices at time $T$}
		\label{Comm1a}
	\end{subfigure} 
	\begin{subfigure}{.9\textwidth}
		\centering
		\includegraphics[width=.9\linewidth]{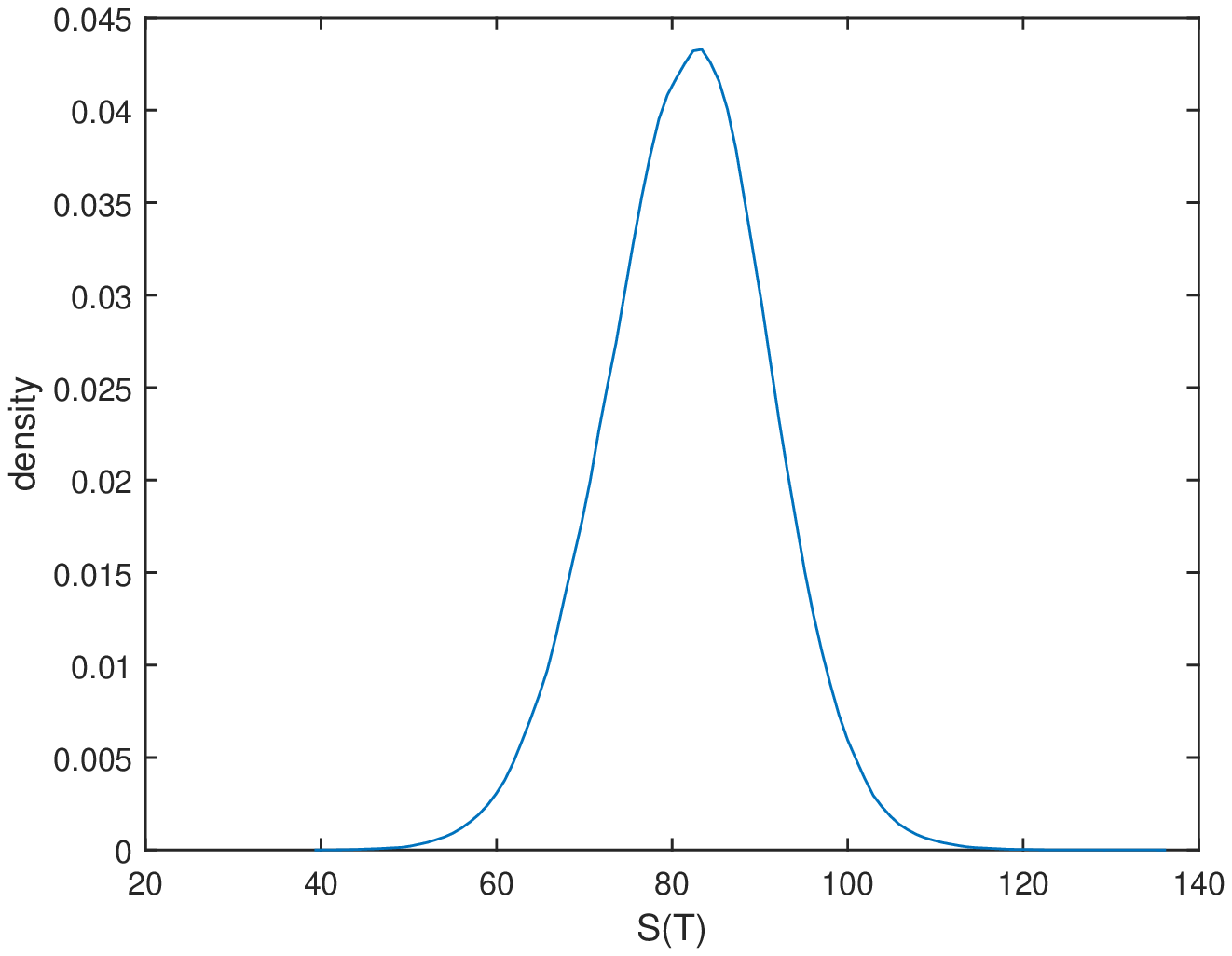}
		\caption{Density of commodity prices at time $T$}
		\label{Comm1b}
	\end{subfigure}
	\caption{Simulation of commodity prices}
	\label{fig.CommMC}
	\floatfoot{\textit{Notes: Panel (a): 1,000 bins had been used. Panel (b): Kernel density estimated from the data in Panel (a) by Matlab's \texttt{kdensity()} function. For both Panels: $S = 80$, $\bar{S} = 85$ (i.e. $\alpha = ln(85)$), $\eta = 1.0$, $T - t = 0.5$, $\omega = 0.2$, $v(t) = 0.04$, $\kappa = 1.0$, $\theta = 0.05$, and $\rho = -0.5$}}
\end{figure}

It was already mentioned before that \citet{Lutz09} provides an analytic solution for the model in (\ref{Lutz1}) and (\ref{Lutz2}).  However, this solution requires the evaluation of Kummer functions of the first and second kind.\footnote{The type of hypergeometric function that appears in the analytic solution depends on the parameter values. For the special case of perfect correlation between the underlying and the variance process the Bessel function of the first and second kind appears instead of the Kummer function (See \citet{Lutz09}, p. 64).} In order to implement the Kummer functions in the analytic solution \citet{Lutz09} uses Matlab code provided by B. Barrows via Matlabs central file exchange, but needs to adapt that code to handle integer arguments.\footnote{Note that since release \textit{R2014b} of Matlab, Kummer functions of the second kind are available as build-in functions via \texttt{kummerU(a,b,z)} (See Matlab online documentation). The function is only available via the Symbolic Math Toolbox.} Alternatively the analytic solution could be implemented by using Runge-Kutta algorithms.\footnote{See \citet{Lutz09}, pp. 77 - 79 for a comparison of these two implementation approaches.}\\
However, for simplicity I will use MC simulations instead of the analytic solution to obtain reference values for the futures price approximated by KM's series expansion. \citet{Lutz09} also compares the analytic solution to MC results, finding a very close match of the obtained prices. In order to generate the MC solution I use a similar approach as in section \ref{CEVMCRef}. Since the log of the price of the underlying is simulated instead of the underlying itself, I use a simple Euler discretization for the (\ref{Lutz1}). However, by the same arguments as in section \ref{CEVMCRef} I use a Milstein discretization for the variance process in (\ref{Lutz2}) to reduce the frequency of negative simulated variances in the sample paths.
\begin{align}
X(t_{i+1}) &= X(t_i) + \left[\eta \left(\alpha - X(t_i)\right) - \frac{1}{2}|V(t_i)|\right]\Delta_i + \sqrt{|V(t_i)|} \sqrt{\Delta_i}Z_1(t_i) \\
V(t_{i+1}) &= |V(t_i)| + \kappa \left(\theta - |V(t_i)|\right)\Delta_i + \omega \sqrt{|V(t_i)|\Delta_i}Z_2(t_i) \\ 
& \quad \quad + \frac{1}{4}\omega^2\left(Z^2_2(t_i) - 1\right)\Delta_i \nonumber \\ \nonumber \\
\text{with } & Z_1(t_i) = Ran_1, \quad Ran_1 \sim N(0,1) \nonumber \\
& Z_2(t_i) = \rho Ran_1 + \sqrt{1 - \rho^2}Ran_2, \quad Ran_2 \sim N(0,1) \nonumber
\end{align}
Where $\Delta_i$ denotes the length of time step $i$. As the above equations indicate I use again the reflective assumption to handle negative simulated variances in the sample path, if they occur. For his MC simulation \citet{Lutz09} uses a huge number of 2,500 time steps and 1.5 million sample paths.\footnote{Of which 375,000 were independent. The large number of 1.5 million paths results from the use of antithetic sampling.} Unfortunately the computational resources that had been available to me were not sufficient to use such numbers. Hence, I constraint myself to 1,000 time steps and 200,000 sample paths, which I believe should be sufficient to achieve satisfying results for the reference values. Below Figure \ref{Comm1a} and \ref{Comm1b} show the histogram and the density of the simulated price of the underlying at maturity of some Future contract. The simulation is based on the above shown discretization of the commodity model and the following set of parameter values: $S = 80$, $\bar{S} = 85$ (i.e. $\alpha = log(85)$), $\eta = 1.0$, $T - t = 0.5$, $\omega = 0.2$, $v(t) = 0.04$, $\kappa = 1.0$, $\theta = 0.05$, and $\rho = -0.5$. These parameters had been suggested by \citet{Lutz09}, since they  should yield a realistic set up.

\subsection{Definition of the scale measure of the CEV variance process} \label{AppendixScale}
Generally, the scale measure $\Omega(\cdot)$ of a diffusion $dx = \mu(x)dt + \sigma(x)dW$, with $dW$ denoting a Wiener process, is defined as\footnote{See \citet{Jones03}, p. 215.}
\begin{align}
	\Omega(x) &= \int_{m}^{n}\Theta(x)dx \\
	\text{with } \Theta(x) &= exp\left(- \int \frac{2\mu(x)}{\tilde{\sigma}^2(x)}dx\right)
\end{align} 
Where $\Theta(x)$ is called the \textit{scale density}. Note that $\tilde{\sigma}^2(x)$ denotes the variance of the process $x$ only and not of the whole covariance matrix, if e.g. $x$ would be part of a system of SDEs. Hence, in the case of the CEV variance process in (\ref{CEV2}) we have $\mu(v) = \kappa\left(\theta - v\right)$ and $\tilde{\sigma}^2(v) = \omega^2v^{2\gamma}$. Such that the scale density can be computed as 
\begin{align}
	\Theta(v) &= exp\left(- \int \frac{2\mu(v)}{\tilde{\sigma}^2(v)}dv\right) \nonumber\\
	\quad &= exp\left(- \int \frac{2\kappa\left(\theta - v\right)}{\omega^2v^{2\gamma}}dv\right) \nonumber\\
	\quad &= exp\left(- \frac{2\kappa}{\omega^2}\left(\theta\int\frac{1}{v^{2\gamma}}dv - \int\frac{1}{v^{2\gamma - 1}}dv\right)\right) \label{scale1}
\end{align}
Where the two integrals can be solved as 
\begin{align}
	\int\frac{1}{v^{2\gamma}}dv &= \label{int1}
	\begin{cases}
		-\frac{x^{1 - 2\gamma}}{2\gamma - 1}  & \quad \text{if } \gamma \neq 1/2 \\
		log(v)  & \quad \text{if } \gamma = 1/2 \\
	\end{cases} \\
	\int \frac{1}{v^{2\gamma - 1}}dv &= \label{int2}
	\begin{cases}
	-\frac{x^{2 - 2\gamma}}{2\gamma - 2}  & \quad \text{if } \gamma \neq 1 \\
	log(v)  & \quad \text{if } \gamma = 1 \\
	\end{cases}
\end{align}
Plugging the first solutions from (\ref{int1}) and (\ref{int2}) into (\ref{scale1}) and rearranging yields the scale density of the CEV variance process as it was given in (\ref{CEVscale})
\begin{align}
 \Theta(v) = exp\left(\frac{2\kappa\theta}{\omega^2\left(2\gamma - 1\right)}\frac{1}{v^{2\gamma - 1}} - \frac{\kappa}{\omega^2\left(\gamma - 1\right)}\frac{1}{v^{2\gamma - 2}}\right) \nonumber
\end{align}
Note that this expression is identical to the scale density for the CEV process given in \citet{Jones03} with $\alpha = \kappa\theta$ and $\beta = -\kappa$.

\addtocontents{toc}{\vspace{2em}}

\newpage


\addcontentsline{toc}{section}{References}  
\pagenumbering{roman}
\setcounter{page}{4}

\bibliographystyle{jf}
\bibliography{References}

\end{document}